\documentclass[structabstract]{aa}
\usepackage{graphicx}
\usepackage{subfigure}
\usepackage{rotating}
\usepackage{amsmath}
\usepackage{url}
\usepackage{xspace}
\usepackage{txfonts}
\usepackage{natbib}
\bibpunct{(}{)}{;}{a}{}{,}
\usepackage{comment}
\begin{document}

\newcommand{\dx}{\mathrm{d}x }
\newcommand{\dy}{\mathrm{d}y }
\newcommand{\m}{\mathrm{m}}
\newcommand{\pin}{``Pi~of~the~Sky''\xspace}

   \title{PSF modelling for very wide-field CCD astronomy}

   \author{L.~W.~Piotrowski\inst{1}\thanks{Presently at RIKEN Advanced Science Institute, Wako, Japan
} \and T.~Batsch\inst{2} \and H.~Czyrkowski\inst{1} \and M.~Cwiok\inst{1} \and R.~Dabrowski\inst{1} \and
G.~Kasprowicz\inst{3} \and A.~Majcher\inst{2} \and A.~Majczyna\inst{2} \and K.~Malek\inst{4,5} \and L.~Mankiewicz\inst{4} \and K.~Nawrocki\inst{2} \and R.~Opiela\inst{4} \and M. Siudek\inst{4} \and M.~Sokolowski\inst{2} \and R.~Wawrzaszek\inst{6} \and G.~Wrochna\inst{2} \and
M.~Zaremba\inst{1} \and A.~F.~\.Zarnecki\inst{1}}

   \institute{Faculty of Physics, University of Warsaw,
	Ho\.za 69, 00-681 Warsaw, Poland
	\and
	National Centre for Nuclear Research,
	Hoza 69, 00-681 Warsaw, Poland
	\and
	Institute of Electronic Systems, Warsaw University of Technology,
	Nowowiejska 15/19, 00-665 Warsaw, Poland
	\and
	Centre for Theoretical Physics, Polish Academy of Sciences,
	Al. Lotnikow 32/46, 02-668 Warsaw, Poland
	\and
	Institute for Advanced Research, Nagoya University, 
	Furo-cho, Chikusa-ku, 464-8601 Nagoya, Japan
	\and
	Space Research Center, 
	Polish Academy of Sciences, Bartycka 18A, 00-716 Warsaw, Poland
             }
 
\authorrunning{L. W. Piotrowski et al.}
\titlerunning{PSF modelling for very wide-field CCD astronomy}

\abstract
{
One of the possible approaches to detecting optical counterparts of GRBs requires monitoring large parts of the sky. This idea has gained some instrumental support in recent years, such as with the ``Pi of the Sky'' project. The broad sky coverage of the ``Pi of the Sky'' apparatus results from using cameras with wide-angle lenses ($20^\circ \times 20^\circ$ field of view). Optics of this kind introduce significant deformations of the point spread function (PSF), increasing with the distance from the frame centre. A deformed PSF results in additional uncertainties in data analysis. 
}
{Our aim was to create a model describing highly deformed PSF in optical astronomy, allowing uncertainties caused by image deformations to be reduced.}
{Detailed laboratory measurements of PSF, pixel sensitivity, and pixel response functions were performed. These data were used to create an effective high quality polynomial model of the PSF. Finally, tuning the model and tests in applications to the real sky data were performed.}
{We have developed a PSF model that accurately describes even very deformed stars in our wide-field experiment. The model is suitable for use in any other experiment with similar image deformation, with a simple tuning of its parameters. Applying this model to astrometric procedures results in a significant improvement over standard methods, while basic photometry precision performed with the model is comparable to the results of an optimised aperture algorithm. Additionally, the model was used to search for a weak signal -- namely a possible gamma ray burst optical precursor -- showing very promising results.}
{Precise modelling of the PSF function significantly improves the astrometric precision and enhances the discovery potential of a wide-field system with lens optics.}

   \maketitle
%

\section{Introduction}
\label{introduction}

The discovery of gamma ray bursts (GRBs) in 1969 \citep{Astroph1973} revealed a completely new type of astrophysical phenomena -- a transient event on time scales of seconds, coming from random directions. It was quickly shown that an effective study of these explosions requires satellite gamma-ray detectors capable of monitoring large portions of the sky. The discovery of prompt optical emission from GRBs has shown that similar transients in the optical regime also exist \citep{first_sim,second_sim}. Observations made for GRB080319B -- the naked-eye burst -- performed simultaneously in the $\gamma$ and optical regimes uncovered crucial information about the nature of the phenomenon \citep{nature}.

The case of the naked-eye burst leads to the conclusion that significant scientific information about the phenomenon is hidden in the very first seconds of the prompt optical emission. However, the current strategy is to perform follow-up optical observations of GRBs, which is almost always insufficient to study the very beginning of the transient in the optical regime. Additionally, there are predictions of orphan afterglows, which are GRB induced transients that are visible only in the optical regime \citep{orphan1,orphan2}. There is also a possibility of optical transients of still unknown origin. Therefore, a similar strategy to that of $\gamma$-ray observations has to be introduced for optical observations, namely the constant optical monitoring of a large part of the sky.

While the ideal solution for optical transient detections would be to use countless large telescopes to observe a large part of the sky with very short exposures, we have to restrict us to feasible solutions, such as observing with a few very wide-field cameras. This is the aim of the ``Pi~of~the~Sky'' experiment. The very short history of these observations limits the experience in this area, especially when compared to the observations with large telescopes. We have to deal with observational issues that have not been studied at all or only studied poorly. One of these problems is an image deformation due to the very significant spatial variance of the very wide-field cameras point spread function (PSF). 

Each linear optical imaging system (invariant under translation) is characterised by a PSF, a response of the detector (in this case lenses + a CCD sensor) to a point source of light. Thus a point-source image that normally would be contained in a single pixel is spread over several pixels, mainly owing to diffraction and optical aberrations. The standard PSF in optical experiments is a profile with approximately rotational symmetry\footnote{PSF elongations of a few percent are often found in wide-field experiments even in the central part of the frame. However, they are very small compared to the distortions described in this paper.}, often approximated well with a two-dimensional Gaussian in the core.

\begin{figure}
\begin{center}
\includegraphics[width=0.24\textwidth]{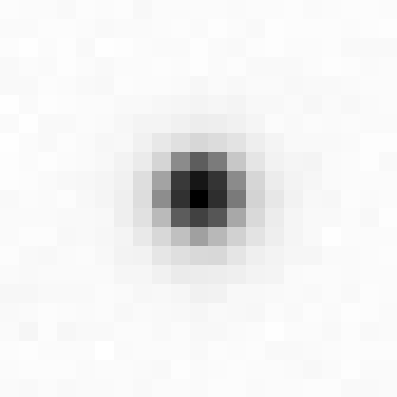}
\includegraphics[width=0.24\textwidth]{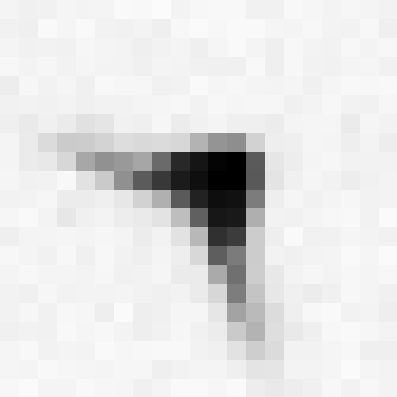}
\end{center}
\caption{Left: PSF of an object in the centre of a frame - Gaussian like profile, clear rotational symmetry. Right: PSF of the same object in the corner of a frame: no rotational symmetry, profile difficult to describe}
\label{star_samples}
\end{figure}

However, it has been noted that the star's PSF develops ``wings'' of signal extending far from the core, which deviate from the Gaussian by a few orders of magnitude \citep{King}. Thus more elaborate descriptions had to be developed, such as given by \cite{Moffat} or \cite{Kormendy}. Although these data and models have been obtained for plate sky images, they were easily adopted and extended for a modern, CCD experiments \citep{Bendinelli, Racine}. These representations are also satisfactory for the very wide-field experiments, for stars very close to the optical axis of the apparatus that are also approximately rotationally symmetric (fig. \ref{star_samples} left).

The rotational symmetry of the PSF becomes disrupted with the increasing distance from the optical axis, where off-axis optical aberrations make star shapes more elliptical. This is still quite a common deformation in astronomical experiments, easily treated with simple modifications to the standard, point-symmetric PSFs \citep{Stetson1990}. Additionally, such deformations (and in general, small deviations from the model shape) can often be described well by a combination of an analytical shape and a numerical table consisting of the residues between the measured star profile and the model. Extension of such an idea is a fully numerical, empirical PSF interpolated from the real measurements. While the last approach performs well in many cases, the analytical approach is preferred for the under-sampled star images, such as those of the ``Pi of the Sky'' experiment, where interpolation would introduce significant uncertainties \citep{Stetson1992}.

The approaches described above become insufficient for experiments, where very precise knowledge of the PSF is crucial to the quality of scientific results, such as weak lensing measurements. Therefore, in past two decades, a new approach has been developed that is an approximation of the PSF by a combination of multiple factors accounting for the profile deformations. There are multiple bases that are more or less suitable for this task, among other wavelets \citep{wavelets}, either Gauss-Hermite or Gauss-Laguerre polynomials (shapelets) \citep{shapelets1,shapelets2}, or cheblets \citep{cheblets}. That an infinite number of components from a complete set of basis vectors can describe any shape makes this approach suitable for approximating even the highly deformed PSF of very wide-field experiments, where much more complicated modifications of PSF develop with the distance from the frame centre as shown in fig. \ref{star_samples} (right). However, the PSF model has to account for the spatial variability of the PSF, therefore the best basis is the one yielding a good profile approximation with the fewest components. To minimise the number of components, a principal components analysis (PCA) method has been applied to PSF parametrisation, based on deriving basis vectors of the profile explicitly from star images, not involving any assumptions about the analytical form of the basis \citep{pca1,pca2}. The PCA method gives very satisfactory results for PSF parametrisation in weak lensing experiments.

\begin{figure}
\begin{center}
	\includegraphics[width=0.47\textwidth]{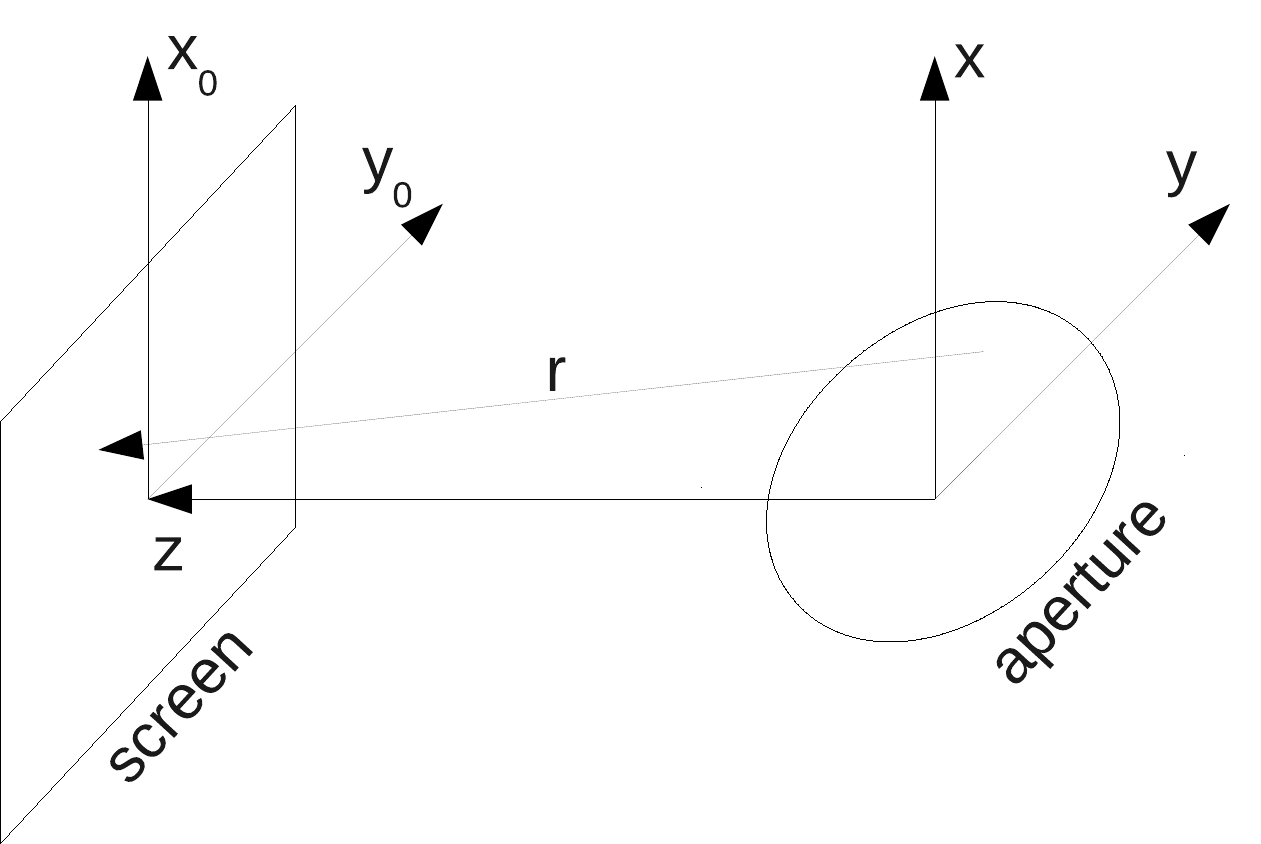}
\end{center}
\caption{Scheme of the aperture and screen coordinate systems, as used in the Rayleigh-Sommerfeld formula.}
\label{fig_diff_setup}
\end{figure}

The methods of PSF approximation by a combination of basis components mentioned above focus on describing the PSF of the star in the focal (image) plane. Parameters of such an effective model cannot be directly related to any physical parameter of the apparatus or the source, and as such are not universal. They cannot be easily adjusted to the change of focus or wavelength. Still, proper physical modelling should overcome these obstacles and may result in precise PSF reproduction in some cases, such as \cite{Jarvis}. In general, PSF can be derived from simplified equations describing the propagation of light through lenses:

\begin{equation}
PSF_L(x_0, y_0, z) = \left|\frac{1}{i\lambda}\iint\limits_{A}U(x,y,0)\frac{ze^{ikr}e^{W(x,y)}}{r^2}\dx \dy\right|^2
\label{eq_aber_diff}
\end{equation}

\noindent which is the Rayleigh-Sommerfeld formula with Kirchhoff approximation of finite aperture, introducing the deviations (aberrations, such as coma, astigmatism, etc.) of the wavefront $W(x,y)$ from sphericity. The Cartesian coordinates in the image plane, $(x_0, y_0)$, are at the distance $z$ from the aperture, $(x,y)$ are corresponding coordinates in the aperture plane, $r$ is the distance between point of the wavefront $(x,y,0)$ on the aperture and point of the image $(x_0, y_0, z)$ (as shown in fig. \ref{fig_diff_setup}), $\lambda$ is the wavelength, $k$ is the wavenumber, and $U(x,y,0)$ is the amplitude of the wavefront on the aperture. This formula is often simplified further into Fresnel or Fraunhofer formula, not applicable to very wide-field experiments due to large angles between the optical axis and outer parts of the ``screen'' (the CCD).

Calculations of this physical model are very demanding, even though the model is oversimplified. In the ``Pi of the Sky'' case it introduces a simple aperture, while in reality a number of thick lenses, including aspherical components, are present. According to the authors' knowledge, the physical approach to the parametrisation of the PSF has not been successfully applied to very wide-field experiments, where deformations of the PSF are very large. However, the model served as an inspiration for the approximation of the PSF by basis functions, the main topic of the work described in this article. Such an approximation to profiles with this level of deformation has also never been performed before, according to authors' knowledge.

\section{The ``Pi of the Sky'' project}

The ``Pi of the Sky'' experiment is designed for continuous monitoring of a large part of the sky with high time resolution \citep{pi1}. This will be achieved with the field of view of 1.5 steradians, obtained with 12 cameras, each covering $20^{\circ} \times 20^{\circ}$ of the sky, using Canon EF lenses with $f=85$ mm and $f/d=1.2$. Cameras are of a unique construction and design prepared by ``Pi of the Sky'' project members. One exception is a commercial CCD sensor with roughly $2000\times 2000$ pixels and a pixel size of $15\times15$ $\rm{\mu m}^2$, corresponding to an angular size of 36 arcseconds. The large solid angle covered by an individual pixel makes ``Pi of the Sky'' PSF virtually unaffected by atmospheric seeing.

This setup, with 10~s exposures and 2~s readout time gives estimated range of $11.5^{\rm{m}}$ on a single frame and $13-14^{\rm{m}}$ on 20 stacked frames. Parameters such as cooling, readout gain, etc. can be controlled by USB 2.0 or ethernet interfaces. Cooling is performed with a two-stage Peltier stack and allows reaching $40$ K below ambient temperature. 

A real-time analysis of the data stream, based on a multi-level triggering system, allows discoveries of GRB optical counterparts independently of satellite experiments \citep{pi1,pi2}. This approach resulted in the autonomous detection of the naked-eye burst GRB080319B at its very beginning \citep{nature}.

The very wide-field of view of each camera causes significant deformations of images far from the optical axis, which is much larger than in other astronomical experiments. This was also the case for GRB080319B, for which the position of the burst was in the corner of the frame up to $t_0+36$~s. Therefore, crucial information about this phenomenon was contained in a highly deformed, triangular-like, winged profile.

\section{Laboratory PSF measurement}

To model the shape of the PSF and its variation with the position on the frame, the shape itself has to be determined first. Images of the stars observed on the frames are a convolution of the PSF and the CCD pixel response. Since the pixel size is quite large in the case of a ``Pi~of~the~Sky'' camera, most single star images, even in the very corner of the frame, are under-sampled and consist of fewer than 30 pixels, thus giving fewer than 30 data points for shape analysis. While it is possible to fit a scaling factor and centre position of a well known profile to this amount of data, it would be very hard to derive a profile's shape itself. This derivation requires a significant spatial resolution, which is much higher than the pixel size. Assuming that a star is placed in a slightly different position on each frame, relative to pixel centre, a series of such images can result in a sub-pixel resolution. Additionally, one can assume that the PSF is invariant under the rotation around the frame's centre and take all the stars into account within a certain distance from it. This operation performed on a long series of frames should give an average profile with a good enough resolution. However, it turned out that the obtained profiles were blurred and with a rather irregular shape, which is insufficient for modelling purposes, especially for positions far from the frame centre.

There are many reasons for this method being not suitable for shape derivation purposes. It is important that star images used in the analysis are properly superimposed. This task is most difficult for peripheral positions, where the centre of the profile cannot be precisely defined, due to the lack of knowledge of the profile's shape. Additional blur comes from the fact that, instead of considering stars at a fixed radius, one has to sum the profiles from an annulus around the frame centre. Other uncertainties enter because we observe stars with different spectral types (which have different PSF), due to the blurring caused by cameras' vibrations when following the sky movement or simply because of atmospheric turbulence. 

Additionally, the behaviour of the CCD sensor itself may affect the PSF, for example due to different sensitivities to different wavelengths. All these uncertainties can be eliminated or at least vastly reduced when the data for PSF parametrisation is obtained from laboratory measurements in controlled conditions, when using an immutable source with a known spectra. Such measurements were performed for the profiles described in this article.

\begin{figure}[tb]
\begin{center}
	\includegraphics[width=0.12\textwidth,angle=90]{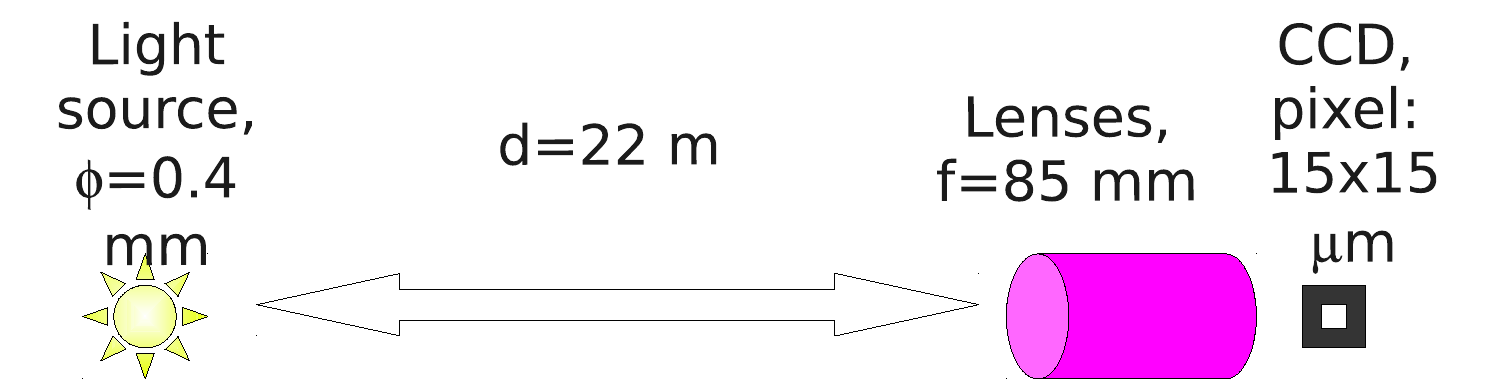}
\end{center}
\caption{The schematic layout of the setup used for laboratory measurements.}
\label{setup}
\end{figure}

\begin{figure*}
\begin{center}
\includegraphics[width=0.328\textwidth]{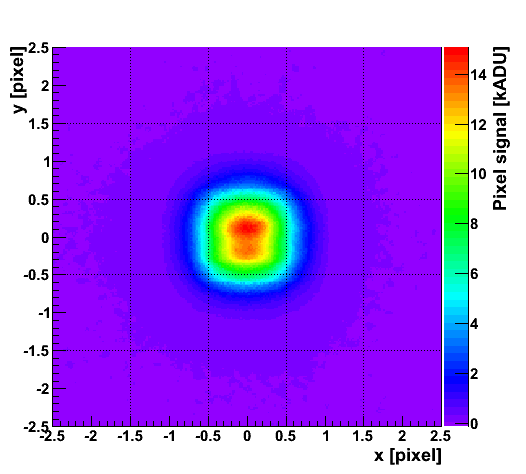}
\includegraphics[width=0.328\textwidth]{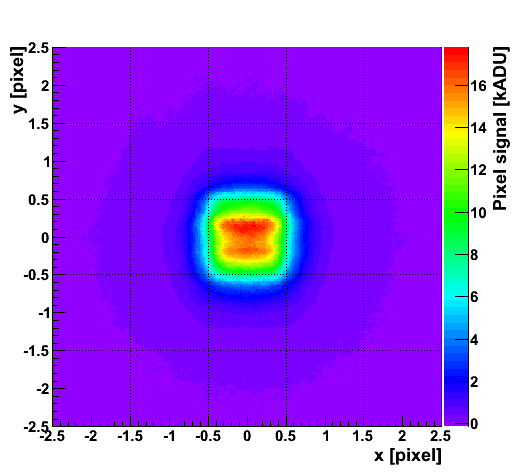}
\includegraphics[width=0.328\textwidth]{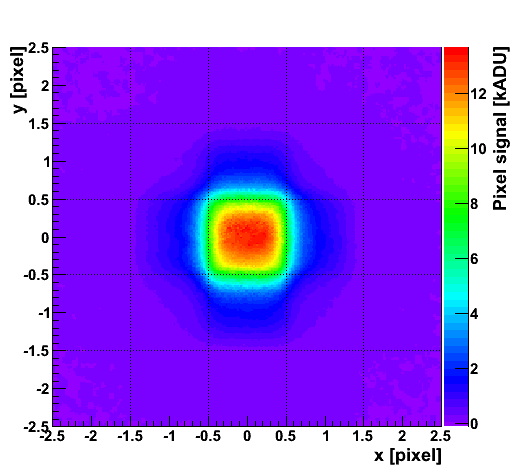}
\end{center}
\caption{Measurements used to extract pixel response function for the blue (left), red (centre) and white (right) diodes.}
\label{prf}
\end{figure*}

\begin{figure*}
\begin{center}
\includegraphics[width=0.328\textwidth]{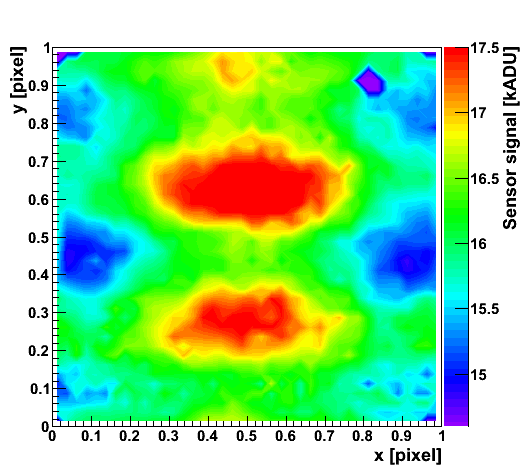}
\includegraphics[width=0.328\textwidth]{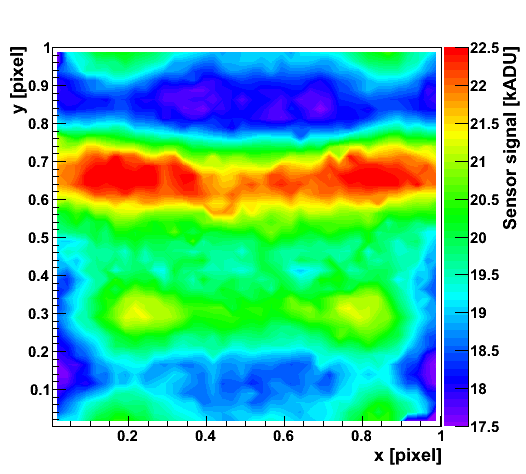}
\includegraphics[width=0.328\textwidth]{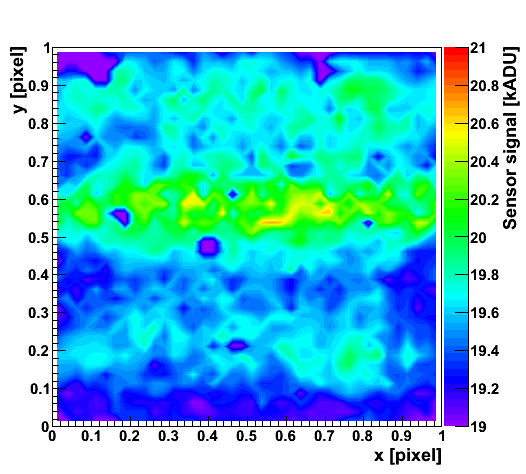}
\end{center}
\caption{Measurements used to extract pixel sensitivity function for the blue (left), red (centre) and white (right) diodes.}
\label{pixel_sens_func}
\end{figure*}

The experimental setup consisted of an LED diode (red, green, yellow, blue, or white) placed behind a pinhole of $0.4$ mm diameter at a $22$ m distance\footnote{The length of the corridor in the laboratory} from a CCD camera, the same as in the ``Pi of the Sky'' prototype and the final system (fig. \ref{setup}). The expected geometrical diameter of the image, when neglecting diffraction and assuming perfect optics, is 1.5 $\mathrm{\mu m}$ (0.1 pixel).

The diode was placed in a mechanic mount driven by two step motors that allowed a precise movement in the vertical and horizontal axes\footnote{Precision of the movement was about 0.1 mm}. A constantly shining source was much too bright for PSF measurements, even with the lowest possible voltage in the diode working range. Thus a pulse generator was used as a power source, so that the diode image brightness on the CCD could be adjusted by changing the pulse length. Exposures, the step motors and the generator pulse were controlled by a computer with self-written, dedicated software.

The focussing was set to the best focus for the red diode, which had the smallest difference between focus minima for central and peripheral diode images. An additional reason was that the sensor is most sensitive in the red band. All measurements were being performed more than 300 s after switching off the laboratory lights, when the background light became stable. The fluctuations of the source light intensity (after subtracting background) were on the level of $0.8\%$. Nearly $70\%$ of those fluctuations can be explained by photon statistics. Measurement of the light source stability showed drift of the image position on the frame that was smaller than $0.002$ pixels per minute. Therefore, the estimated laboratory set-up stability was perfectly sufficient for further measurements \citep{phd}.

\subsection{Intra and inter-pixel measurements}

The way the CCD sensor is designed causes spatial non-uniformity of a single-pixel light sensitivity. That is mainly due to the electrodes placed across the pixels and channel stops separating the sensor's columns. The non-uniformity can be measured with a source of light focussed on a spot that is smaller than the pixel size \citep{subpixel}. The geometrical size of the spot in the described apparatus setup is smaller than the pixel size, but the PSF causes the light to be spread over several pixels. The way to restrain the PSF is to put a circular aperture in the front of the lenses. The smaller the aperture, the smaller the illuminated lenses area and the smaller the PSF. However, a small aperture causes the spot size to be diffraction limited. As a compromise between PSF size and the diffraction size an aperture of $20$ mm diameter had been chosen for the inter-pixel measurements, resulting in 0.2 and 0.14 pixel spot size, respectively, for red and blue lights.

The dependence of detector response on light intensity has been tested and some departures from linearity for high signal values have been revealed. To minimise these effects, we have decided to keep the pixel readout in the range up to 20000 ADU\footnote{ADU (Analog/Digital Unit) is a unit of measure of charge stored in a pixel of a CCD sensor, resulting from an analogue to digital conversion. ADU can be converted to a number of photoelectrons if the gain of the CCD is known.}, where the response is almost perfectly linear.

Pixel response function (PRF) describes a single pixel signal value as a function of the spot position relative to the pixel edge. In an ideal case of infinitely small spot size and uniform pixel response, the PRF should have constant value inside the pixel and zero value outside. In the real case, it depends on the pixel sensitivity and on the finite spot size, so that the spot may be only partially contained inside the pixel. The latter causes a narrow but not-negligible transition of PRF from high to low values close to the pixel border (fig. \ref{prf}). 

However, the function is also non-zero for spot fully outside the pixel. This may be caused by a finite PSF size or diffraction of the spot, restrained by setup parameters, but still non-negligible. The more interesting possibility is that it is caused by a charge diffusion between pixels or photons reflecting from the sensor substrate, illuminating a single pixel causes some charge to be accumulated in a neighbouring pixels as well. In that case, the PRF ``tails'' also contribute to the observed PSF shape\footnote{The presented PRF is convoluted with a finite PSF of the diode of spot size 0.14-0.2 pixel (wavelength dependent). To obtain the most precise measurements of the PRF a proper deconvolution should be performed. We assume, however, that the uncertainty introduced in PSF modelling due to PRF convolution is negligible compared to other factors.}.

Pixel sensitivity function is defined similarly to the PRF, however instead of the single pixel signal an overall CCD signal is studied as a function of the spot position. Changes in pixel sensitivity are the main factor responsible for signal changes observed with the image movement across the CCD. With the knowledge of the pixel sensitivity function and the position of the source's centre on the pixel one should be able to compensate for this effect, performing more precise measurement of brightness.

\begin{figure}
\begin{center}
\includegraphics[width=0.4\textwidth]{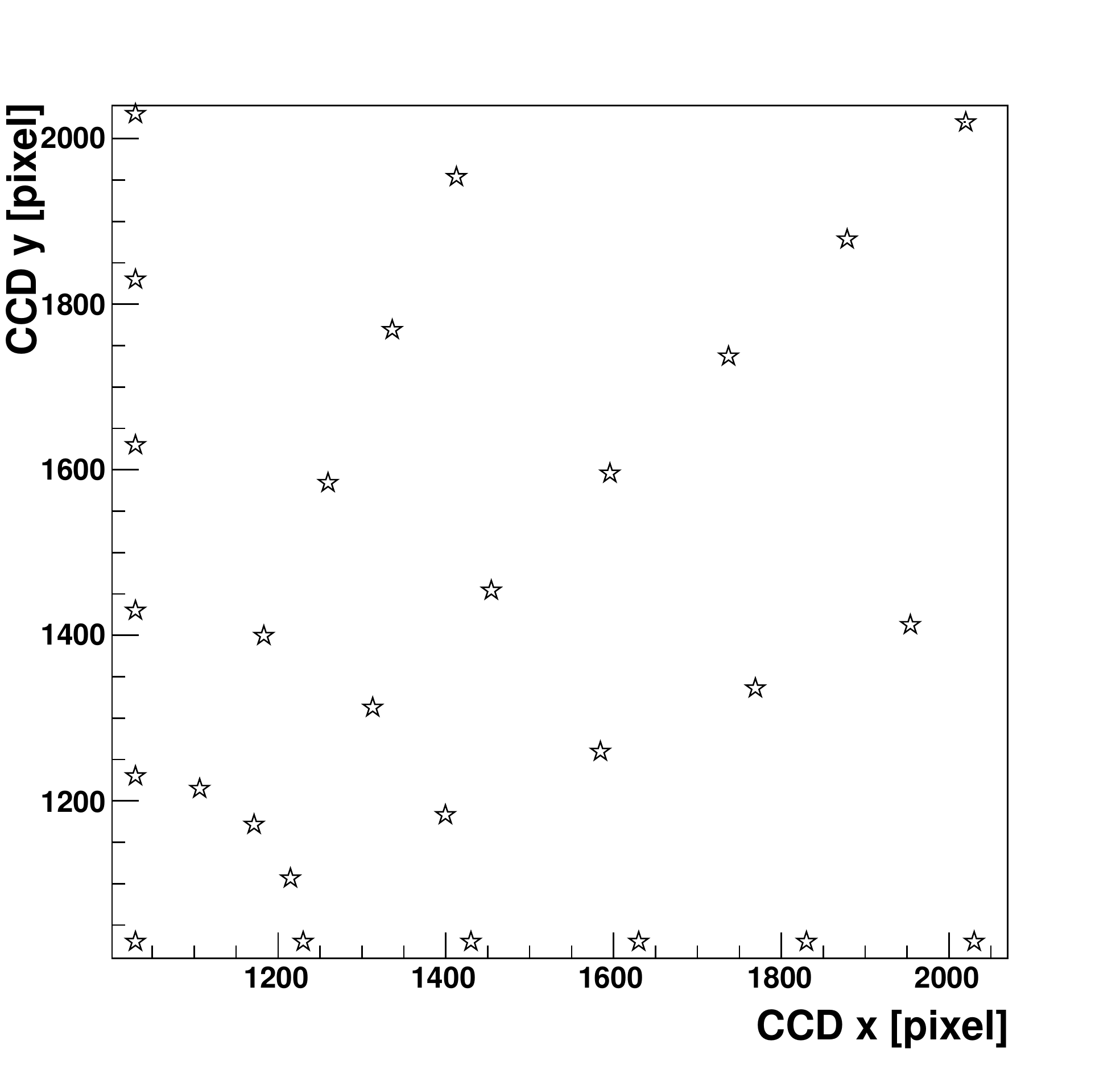}
\end{center}
\caption{Positions of the PSF measurements on the CCD surface in sensor coordinates. The measurements were performed for 5 angles (for $0^\circ$, $22.5^\circ$, $45^\circ$, $77.5^\circ$ and $90^\circ$ from the horizontal axis), and 5 or 8 distances from the frame's centre. For each angle, distances between considered positions were about 200 pixels.}
\label{meas_pos}
\end{figure}

The overall signal was estimated by a sum of $3\times 3$ pixels around the spot centre. Results of the measurement are shown in fig. \ref{pixel_sens_func}. The maximal observed changes in the signal due to the pixel sensitivity non-uniformity for a red diode are more than $30\%$, and for a blue diode more than $20\%$. However, for a normal PSF image (taken without aperture reducing PSF size), which is spread over more pixels, the amplitude of corresponding fluctuations is lower (around $6\%$).

A visible vertical structure similar for both red and blue diodes is probably caused by the electrodes. The horizontal structure is clearly colour dependent, probably due to different penetration depths of light of different wavelengths. Both structures are much less visible for the white diode, for which the effect is averaged all over the white diode spectrum, and thus much smaller.

\subsection{PSF measurements and reconstruction}
\label{sec_mes_rec}

A high resolution profile for selected coordinates on the frame was obtained using multiple images of a diode. Each exposure was taken for a specific position of the diode's centre, the full set of images covering $10\times 10$ points inside a single pixel. Additionally, 5 images with and 5 without a light pulse from a diode were taken in each position, for noise reduction and stability monitoring purposes. All the images were superimposed, taking into account coordinates of each image.

PSF measurements and reconstruction were performed for a white diode for 5 angles and 6 distances from the frame centre, covering 1/4 of the CCD, as shown on fig. \ref{meas_pos}. Each star represents a position on a CCD in which PSF was reconstructed. Due to time constraints and a very long time required to precisely measure a single PSF, only one quarter of the CCD chip was examined.

\begin{figure}
\begin{center}
\includegraphics[width=0.4\textwidth]{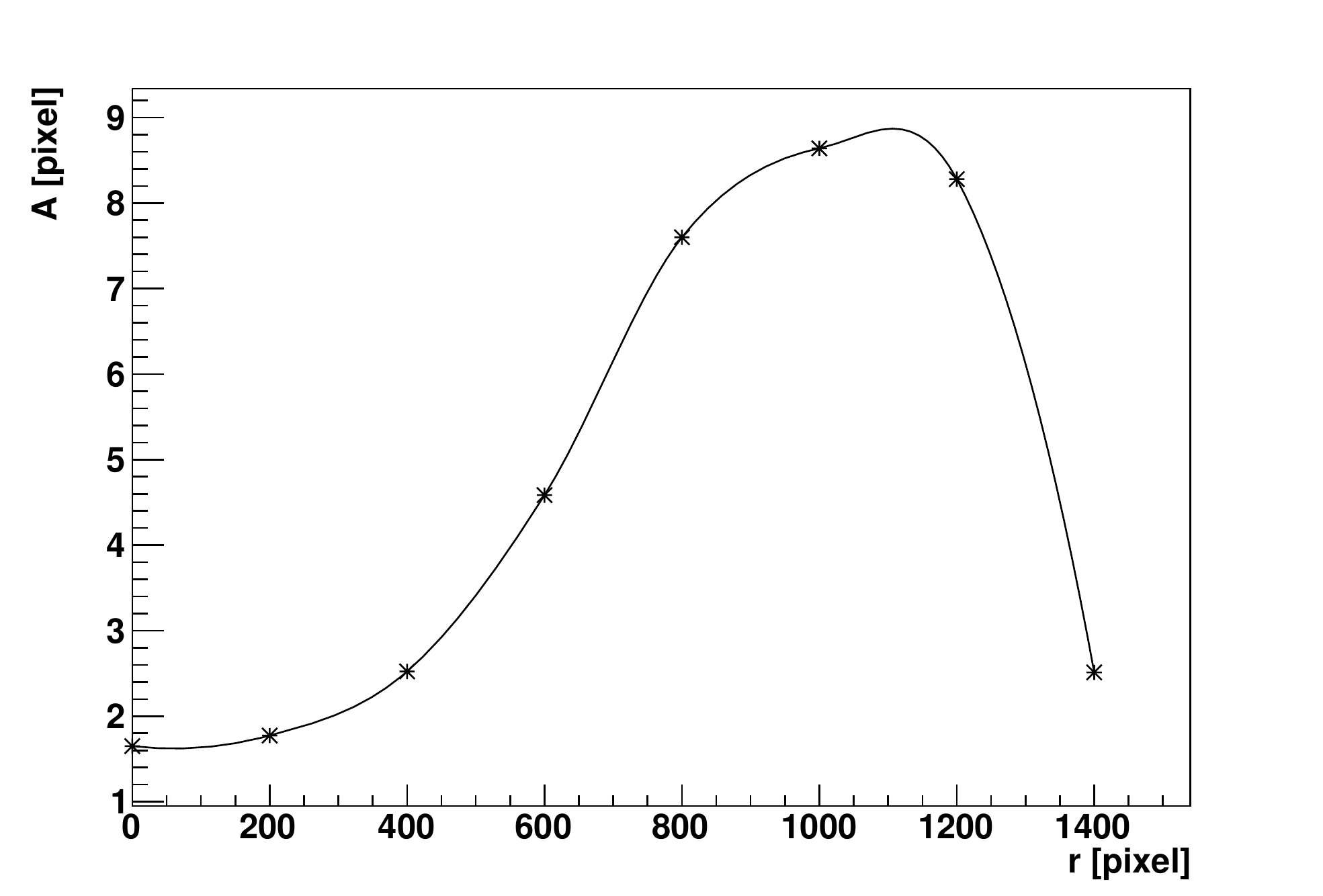}
\end{center}
\caption{The area A above half of the maximum height for PSFs measured along chip diagonal, as a function of the distance from the frame centre~r.}
\label{fig_fwhm}
\end{figure}

\begin{figure*}
\begin{center}
\includegraphics[width=0.328\textwidth]{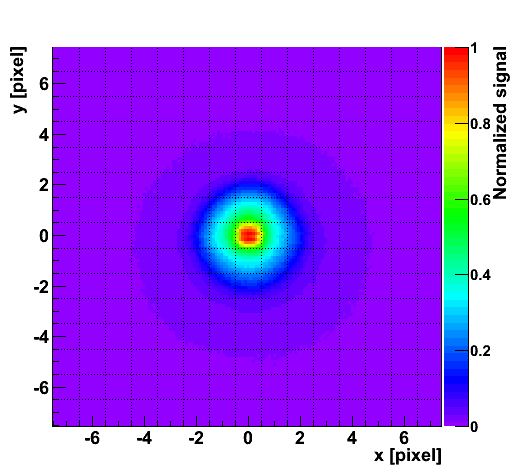}
\includegraphics[width=0.328\textwidth]{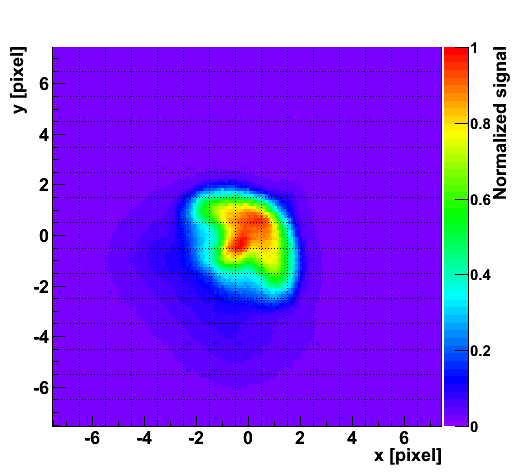}
\includegraphics[width=0.328\textwidth]{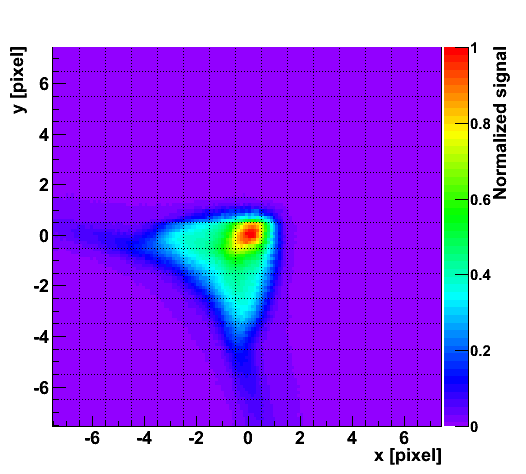}
\end{center}
\caption{Sample PSFs of the white diode measured along the diagonal, for 0, 800 and 1400 (from the left) pixels from the frame centre}
\label{diode_psf_white}
\end{figure*}

A significant deformation develops with the distance from the frame centre, causing not only the shape of the profile to change, but also the area containing the signal to grow. The area A covered by the signal greater than half of the maximal signal, as shown in fig. \ref{fig_fwhm}, increases from slightly more than 1.5 pixel for the central (r=0) profile to nearly 9 pixels for the profile r=1000 pixels from the frame centre. Even ignoring the shape deformation, such a growth may be a non-negligible factor increasing uncertainties of aperture photometry. For profile photometry the situation is even more difficult, because the PSF shape changes dramatically with radius, as shown on fig. \ref{diode_psf_white} (for measurements at angle of $45^\circ$).

\begin{figure*}
\begin{center}
\includegraphics[width=0.328\textwidth]{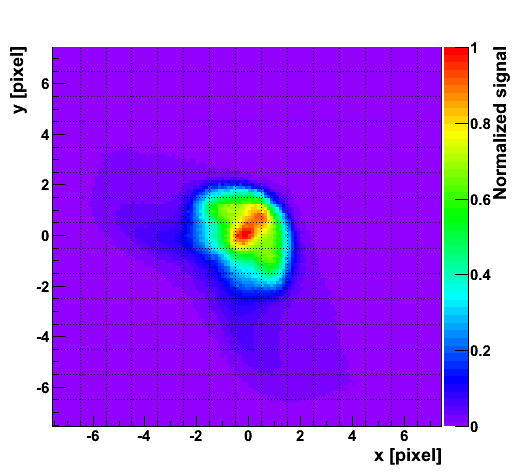}
\includegraphics[width=0.328\textwidth]{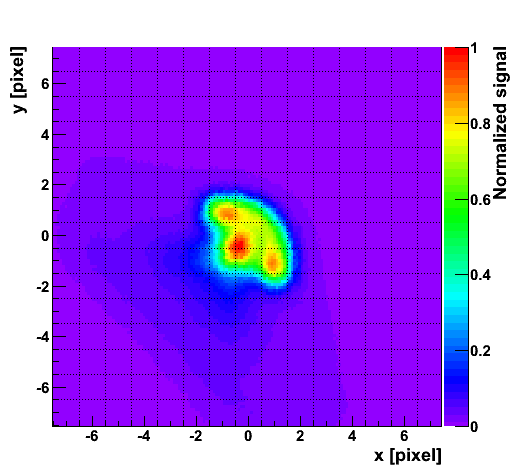}
\includegraphics[width=0.328\textwidth]{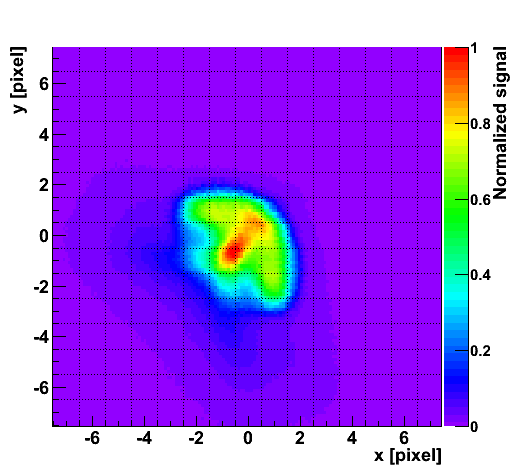}
\end{center}
\caption{Sample PSFs of the blue, red and white diodes (from the left) on the diagonal 1000 pixels from the frame centre.}
\label{diode_psf_blue}
\end{figure*}

\begin{figure*}
\begin{center}
\includegraphics[width=0.328\textwidth]{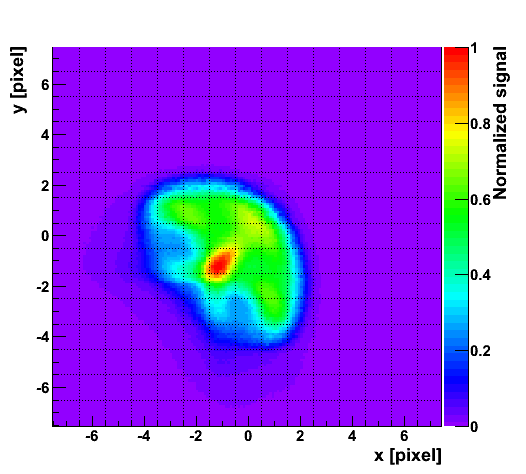}
\includegraphics[width=0.328\textwidth]{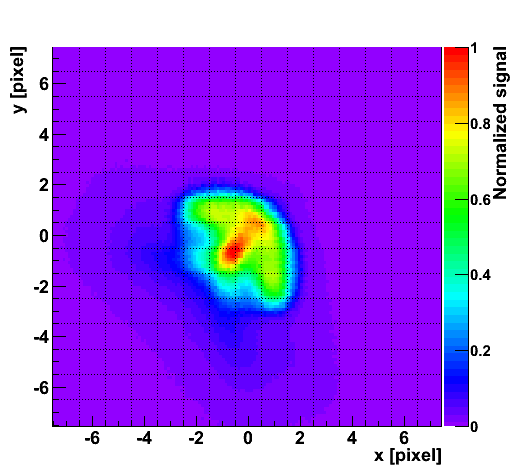}
\includegraphics[width=0.328\textwidth]{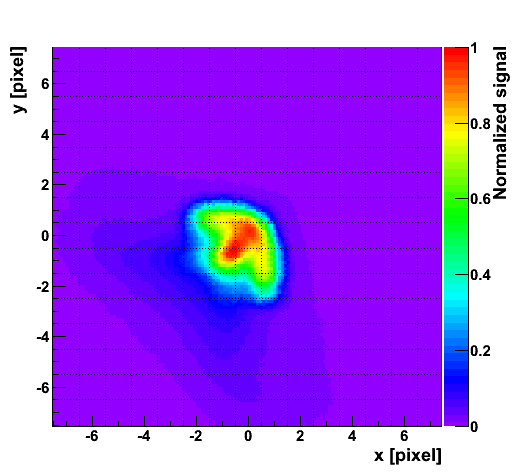}
\end{center}
\caption{Sample PSFs of the white diode measured for different focus settings (from the left): fs=1.2, 1.4 and 1.6 m, for 1000 pixels from the frame centre on the diagonal.}
\label{diode_psf_focus}
\end{figure*}

PSF reconstruction for the white diode is important from the point of view of studying a general image shape for real stars, which are hardly monochromatic. However, monochromatic PSFs may allow to study the chromatic dependence of the star shape. Thus a similar PSF reconstruction was performed for the blue and the red diode (fig. \ref{diode_psf_blue}), for 800, 1000, 1200 and 1400 pixels from the frame centre along its diagonal. It can be noticed, that PSFs of the white diode contain superimposed shape features of PSFs of the blue and the red diode, which is expected. However, polychromatic PSFs tend to be larger due to the fact, that different wavelengths are focused in a slightly different positions on the CCD \citep{phd}.

High quality photographic lenses of the same type should have PSFs in the same place close to identical, assuming the same focus is set. However, as mentioned before, focusing for the real stars is different than the focusing for the diode, and it is close to impossible to maintain exactly the same focus for different cameras. To study the possible influence of the focus setting on the PSF shape, measurements for three different focusing settings were repeated for the white diode, for 800, 1000, 1200 and 1400 pixels from the frame centre along its diagonal (fig. \ref{diode_psf_focus}).

The area covered by the PSF changes visibly with the focusing setting, but the general shape remains similar (blue and green parts). The very centre undergoes bigger changes, especially for 800 and 1000 pixels from the frame centre (red part). PSF seems to be best focused for the last setting, which was not the best focus for the central profile.

\section{Polynomial model of PSF}

Deriving the PSF description in the image plane is a task similar to finding a mathematical description of complicated shapes of an aberrated wavefront in the aperture plane -- $W(x,y)$ -- introduced in eq. \ref{eq_aber_diff}. This can be done using many bases, but perhaps the most popular is expansions to the set of Zernike polynomials \citep{zernike_coef}:

\begin{equation}
	Z^m_n(\rho, \phi) = R^m_n(\rho)\cos(m\phi),\ Z^{-m}_n(\rho, \phi) = R^m_n(\rho)\sin(m\phi)
\label{eq_zernike}
\end{equation}

the radial part being:

$$
	R^m_n = \sum^{(n-m)/2}_{k=0}\frac{(-1)^k(n-k)!}{k!((n+m)/2-k)!((n-m)/2-k)!}\rho^{n-2k}
$$

Inspired by the diffraction approach, we decided to consider slightly modified Zernike polynomials as a possible description in the image plane. First modification is due to the fact that the PSF is a function described on an infinite plane, while the wavefront described by the Zernike polynomials was bound to a finite, circular aperture. Therefore, a transformation of a PSF's radial coordinate $r$ to argument of Zernike polynomial, $u(r)$, where we assume that $u(0)=0$ and $u(\infty)=1$, has to be performed. Thus set of modified Zernike polynomials is now defined as:

\begin{equation}
Z^m_n:=Z^m_n(u, \phi),\ u=1-e^{-\frac{r}{\Lambda}}
\label{eq_zer_repar}
\end{equation} 

where $u$ and $\phi$ are standard radial and azimuthal coordinates of Zernike polynomials and $\Lambda$ is a parameter modifying the transformation. Zernike polynomials are defined as in eq. \ref{eq_zernike}.

Additional modification is needed, for the real PSF has a maximal value around its centre and asymptotically drops to zero in infinity, while the wavefront has a sharp cut-off at the border of the aperture. The asymptotic behaviour is introduced to the PSF by using Zernike polynomials to modify a gaussian-like profile:

\begin{equation}
PSF_L(r, \phi) = \exp(-\frac{1}{2}r^p\cdot \sum_{m,n}Z^m_n(u, \phi))
\label{eq_psf_lenses}
\end{equation} 

where $PSF_L$ stands for a PSF generated by lenses only (no convolution with the CCD) and $p$ is a parameter describing the asymptotic behaviour ($p=2$ for Gaussian shape). Additionally, we assume that the PSF pattern is axially symmetric (in respect to the axis connecting the centre of the PSF and the centre of the frame), thus only symmetric terms are allowed in Zernike polynomials (those with odd $n$ and $m<0$ or even $n$ and $m>=0$). The exact axial symmetry is expected only in the case of a perfect alignment of lenses and camera. However, our aim was to create a universal model able to describe the PSF on all cameras (with just a minor tuning of the model parameters, when required). Although some asymmetry is visible in the detailed examination of high resolution profiles, its effects are negligible compared to the signal fluctuations when standard image frames are considered. We attribute the asymmetry to the random errors in lenses and lenses-camera alignment, which should not be reflected in a general, universal model. We have decided to stick to a model with symmetric basis only as a general description of basic features of PSF, applicable to all our cameras.\footnote{Addition of asymmetric terms to the model, although in principle possible and leading to better description of single reconstructed PSF, would make the fine tuning of the model to each camera much more difficult. In fact, the dedicated model, with parameters selected and fitted for each camera separately, should be developed in such a case. This is an approach contrary to the one adopted in this paper, which would require an independent study.}

The obtained set of functions is not orthogonal, in contrast to the standard Zernike polynomials, but still is well suited for fitting purposes. We decided to use this basis hoping that the PSF would simply factorise to components similar as the aberrations from the spherical wavefront mentioned in the section \ref{introduction}. However, results show that hardly any physical meaning can be attributed to the parameters of the fit in this approach. On the other hand, tests showed that the above non-linear combination of parameters inspired by the diffraction formula \ref{eq_aber_diff} performs better than basis adopting the linear combinations, such as shapelets (sec. \ref{results_model}).

PSF visible as an image on the frame is a convolution of a PSF generated by lenses (eq. \ref{eq_psf_lenses}) and the CCD pixel structure:

\begin{equation}
	PSF(r', \phi') = \iint PRF(r, \phi, r', \phi')\cdot PSF_L(r, \phi)\ r\ dr d\phi
\label{eq_psf_conv}
\end{equation} 
 
\noindent where $PRF(r, \phi, r', \phi')$ stands for the PRF, $(r', \phi')$ are polar coordinates of the specific point of the final $PSF$ (corresponding to the CCD pixel centres), while $(r, \phi)$ are polar coordinates of the current point of integration. In the general case, the convolution (\ref{eq_psf_conv}) would require integrating over an infinite space. However, the rapid drop in the PRF close to the pixel edge allows restricting the integration space. An area of $1.6 \times 1.6$ square pixels around the pixel centre was used, the integration replaced by a simple sum of $PRF$ and $PSF_L$ products on the uniformly distributed grid of $20 \times 20$ point in this space.

\subsection{Optimal parameters for the basis}

Our main aim was to find a universal formula that describes the PSF all over the whole frame. Due to limited time, the function was measured only for a few chosen positions, so that deriving such a universal formula requires finding a local model for each of the measured PSFs first and then determining an interpolation method between these models.

We tried numerous approaches to selecting optimal parameters for the model. Most reproduced measured PSFs in great detail, but none showed a clear and simple dependence of polynomial coefficients on the distance from the frame centre. Therefore, for the effective approach, a linear interpolation in distance between parameters was chosen. The linear interpolation is not a perfect choice, for in some cases it leads to a sudden change in a parameter value, but does not cause suspicious behaviour -- false extrema -- as in all attempted functional interpolations.

The finally chosen model consisted of 17 polynomial terms: first three circular polynomial terms and 14 non-circular terms, which led to the smallest $\chi^2$ between model and measured PSFs. These parameters reproduced measurements best and were most successful in tests performed on PSFs measured in different positions on the frame.

\subsection{Results of the polynomial model}
\label{results_model}

\begin{figure*}
\begin{center}
	\includegraphics[width=0.328\textwidth]{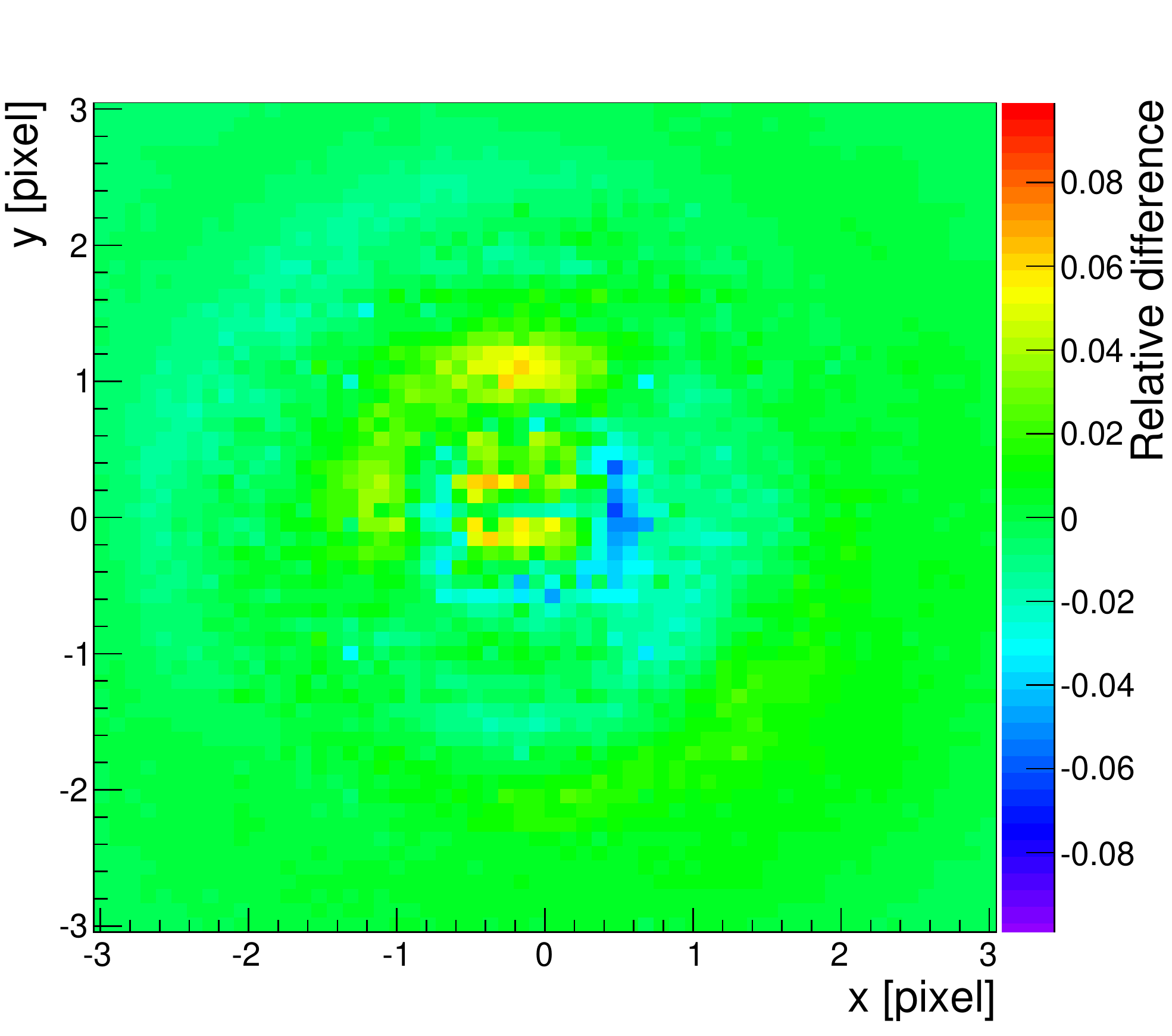}
	\includegraphics[width=0.328\textwidth]{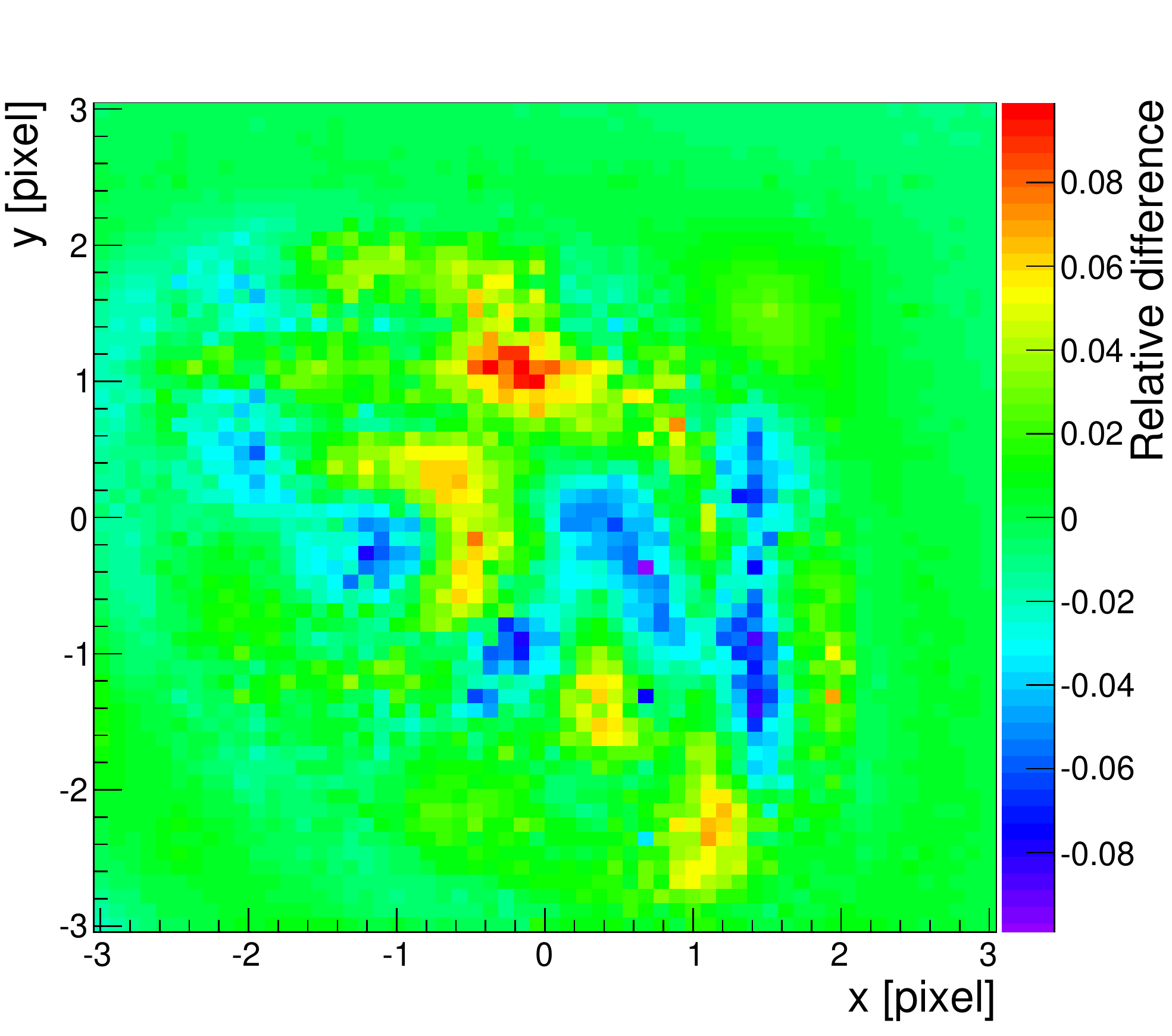}
	\includegraphics[width=0.328\textwidth]{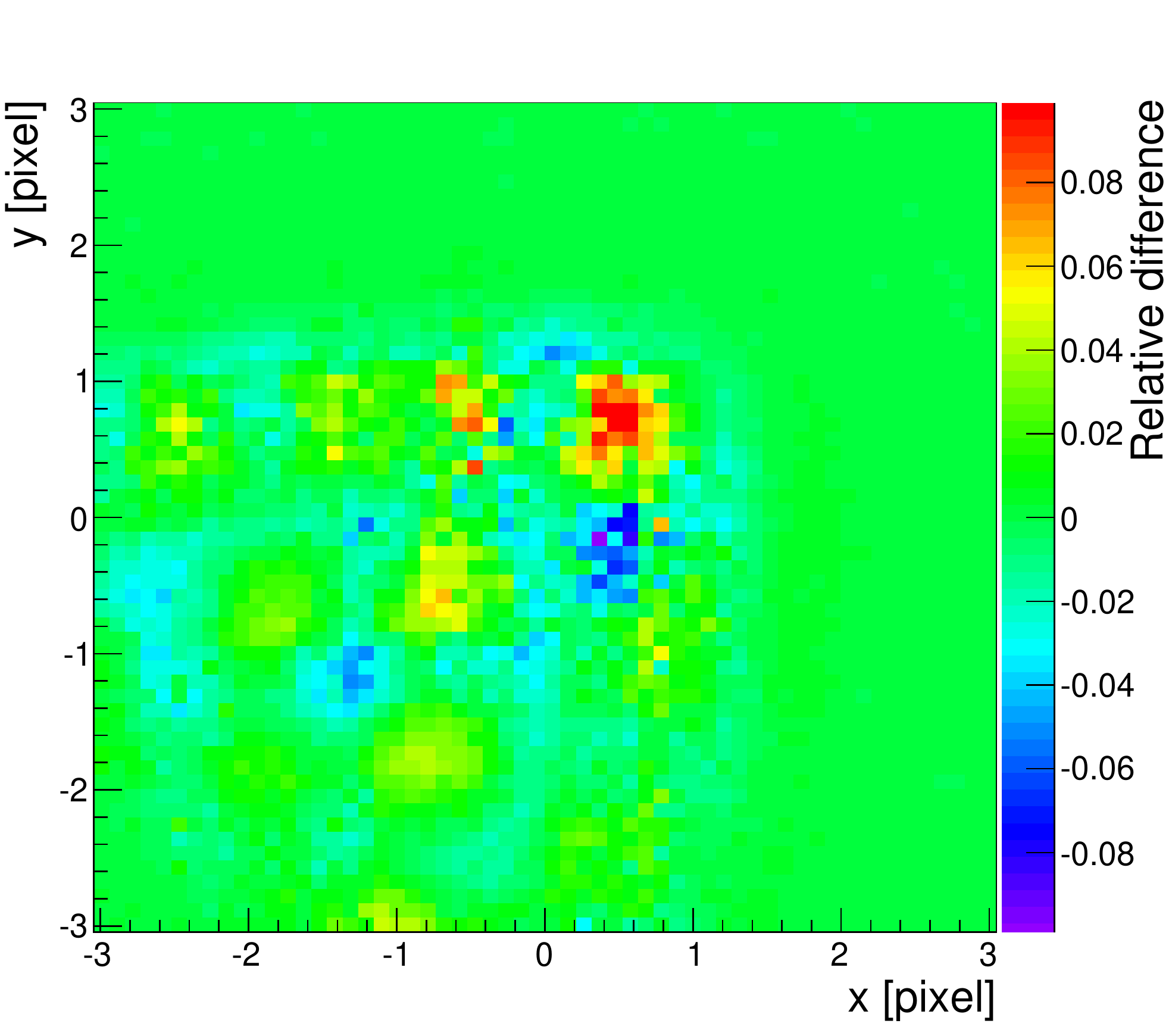} \\
	\includegraphics[width=0.328\textwidth]{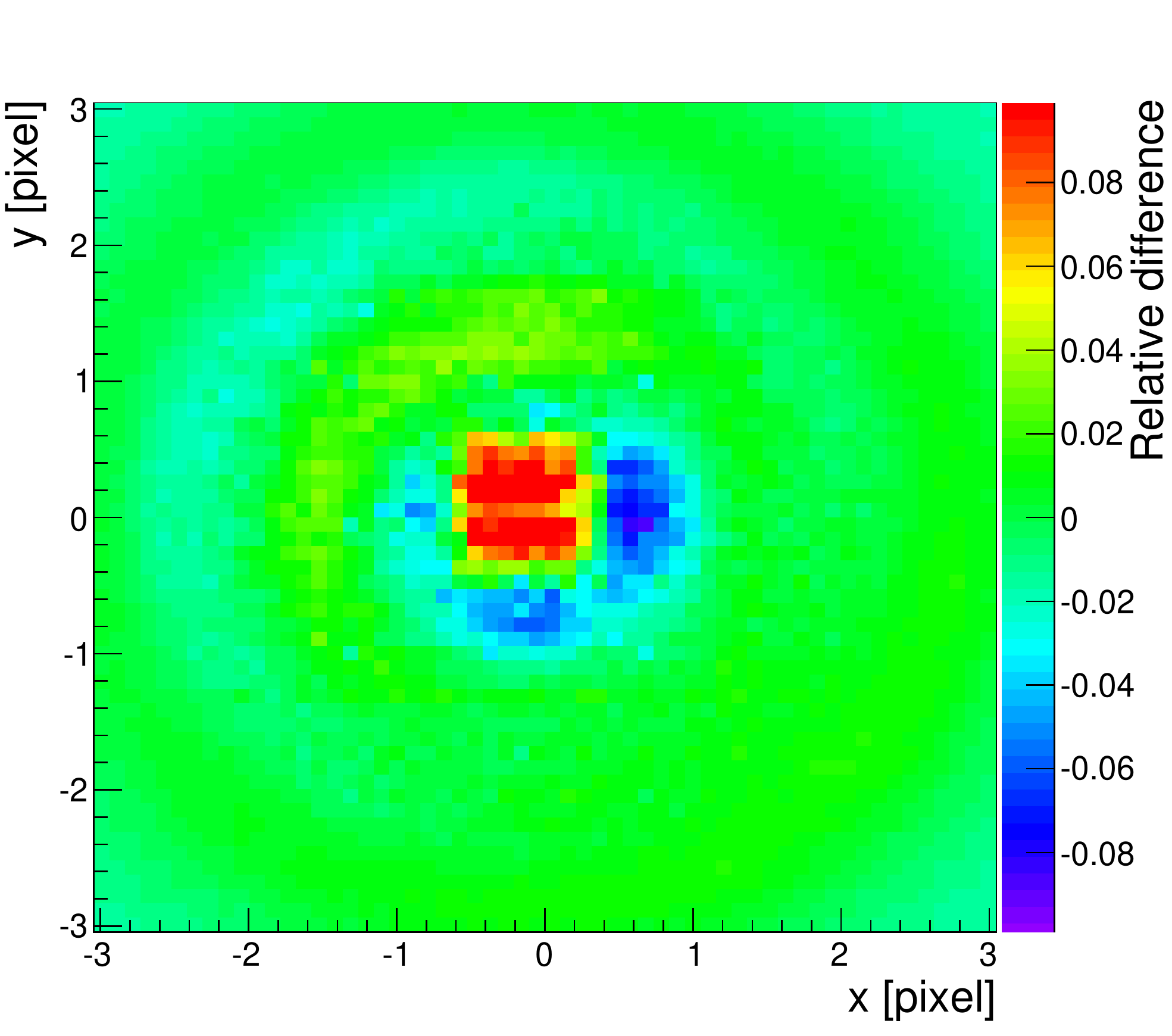}
	\includegraphics[width=0.328\textwidth]{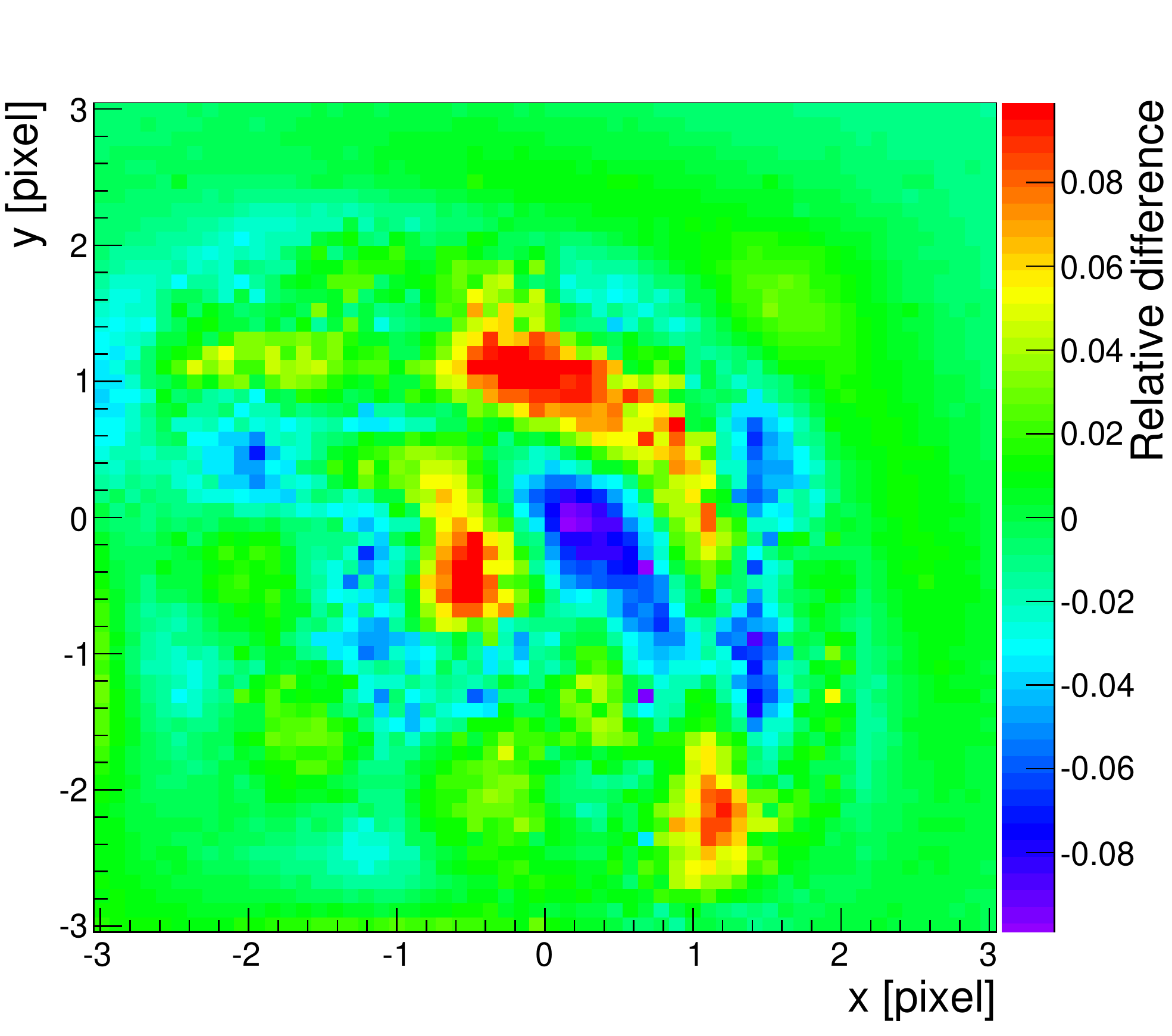}
	\includegraphics[width=0.328\textwidth]{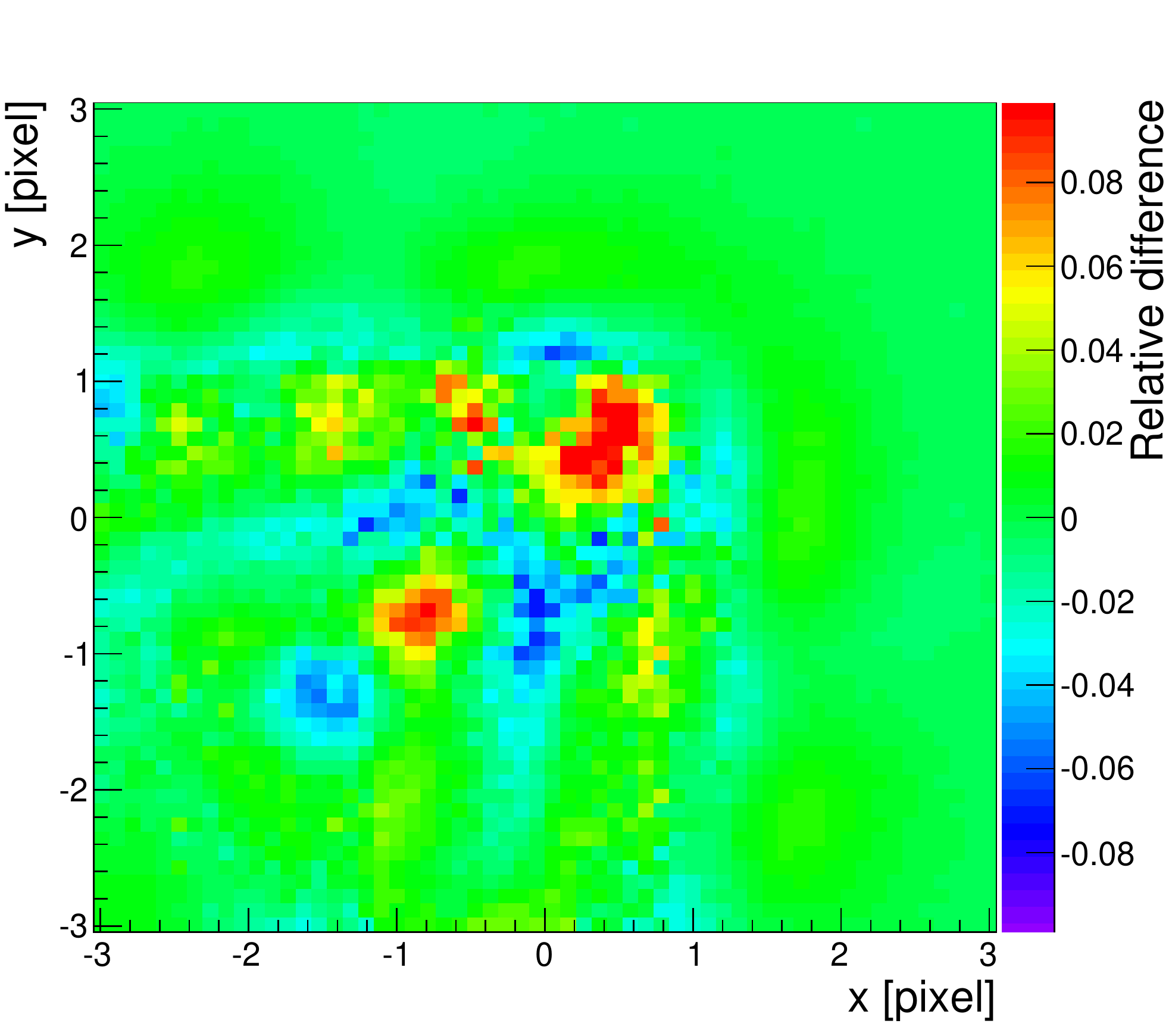}
\end{center}
\caption{Residuals (normalised to the maximum signal) of the high resolution PSF profile reconstructed from measurements and the model for (from the left) 0, 800, and 1400 pixels from the frame centre. On the top -- the polynomial model, on the bottom -- the shapelet model.}
\label{fig_res}
\end{figure*}

\begin{table}
\newcommand{\mc}[3]{\multicolumn{#1}{#2}{#3}}
\begin{center}
\begin{tabular}{|c|c|c|c|c|}
\hline
 & \mc{2}{c|}{\textbf{max. residue}} &  \mc{2}{c|}{\textbf{RRMSD}} \\
\hline
\textbf{r} & \textbf{polyn.} & \textbf{shap.} & \textbf{polyn.} & \textbf{shap.} \\
\hline
0 & $6.5\%$ & $15\%$ & $1.15\%$ & $1.91\%$ \\
200 & $5.7\%$ & $12.8\%$ & $0.95\%$ & $1.67\%$ \\
400 & $8.7\%$ & $14.5\%$ & $1.06\%$ & $1.74\%$ \\
600 & $7.4\%$ & $13.2\%$ & $1.55\%$ & $2.35\%$ \\
800 & $11.3\%$ & $14.8\%$ & $2.05\%$ & $2.55\%$ \\
1000 & $11.5\%$ & $13.1\%$ & $2.45\%$ & $3.23\%$ \\
1200 & $12.2\%$ & $17.9\%$ & $2.61\%$ & $4.21\%$ \\
1400 & $12.8\%$ & $13.2\%$ & $1.55\%$ & $1.96\%$ \\
\hline
\end{tabular}
\end{center}
\caption{Comparison of maximum residues and relative root mean square deviations (RRMSD) for the polynomial and the polar shapelet models, for distance r from the frame centre.}
\label{res_tab}
\end{table}

The model was fitted to the averaged, high resolution profiles of the PSFs, containing $10\times10$ points per pixel (sec. \ref{sec_mes_rec}). The uncertainty for measurements was determined with source stability tests, which are $0.8\%$ of the signal in the measured intensity range. The final uncertainty of the averaged profile was calculated from measurement uncertainty combined with the dark frames uncertainty, weighted by the number of measurement in the bin. The fit, performed on the area of 36 pixels surrounding the centre of the profile, was based on a standard $\chi^2$ estimation, performed with the Fumili2 algorithm from the ROOT framework \citep{ROOT}.

\begin{figure}
\begin{center}
\subfigure[0 pixels from the frame centre]{
	\includegraphics[width=0.32\textwidth]{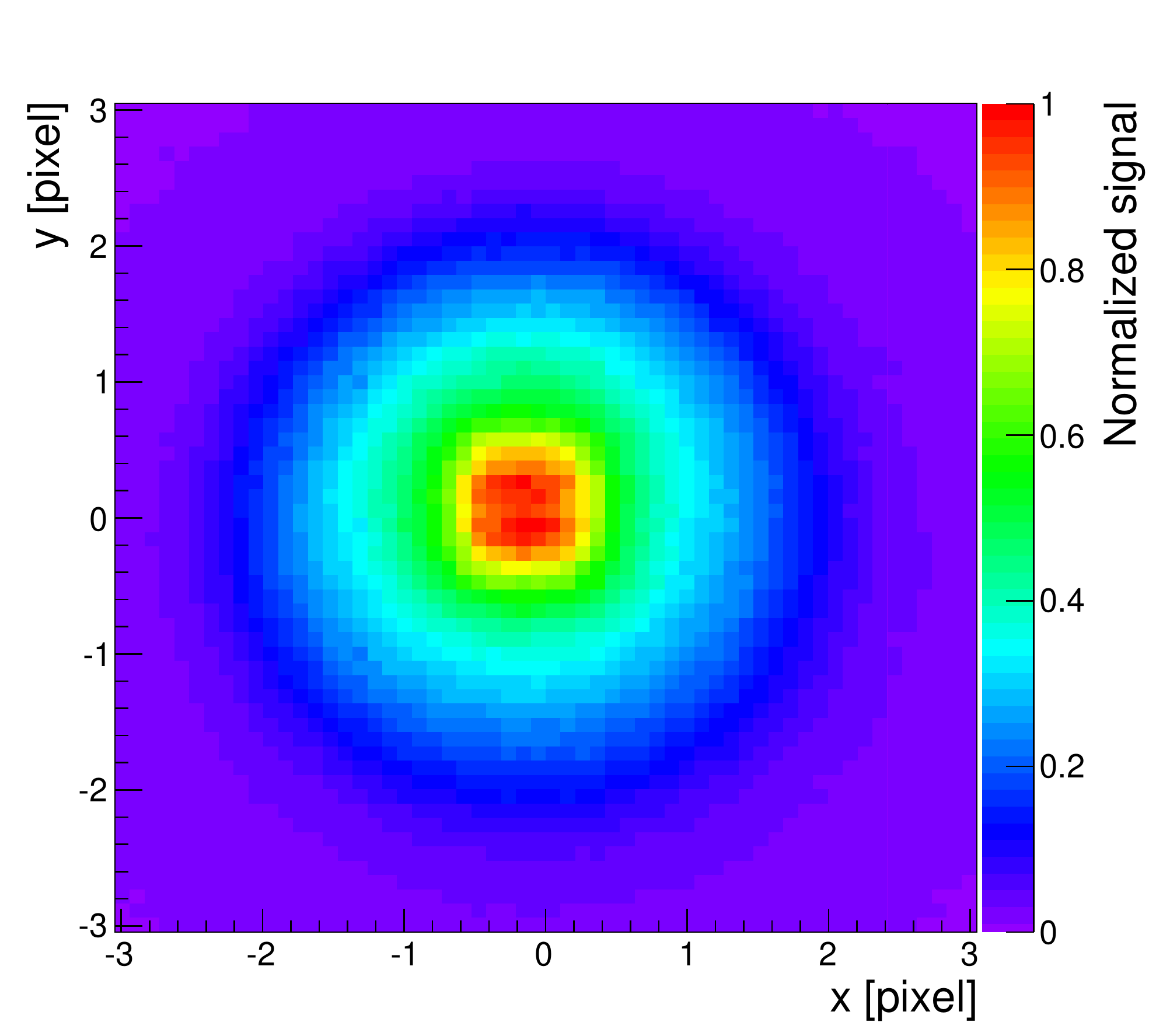}
	\includegraphics[width=0.32\textwidth]{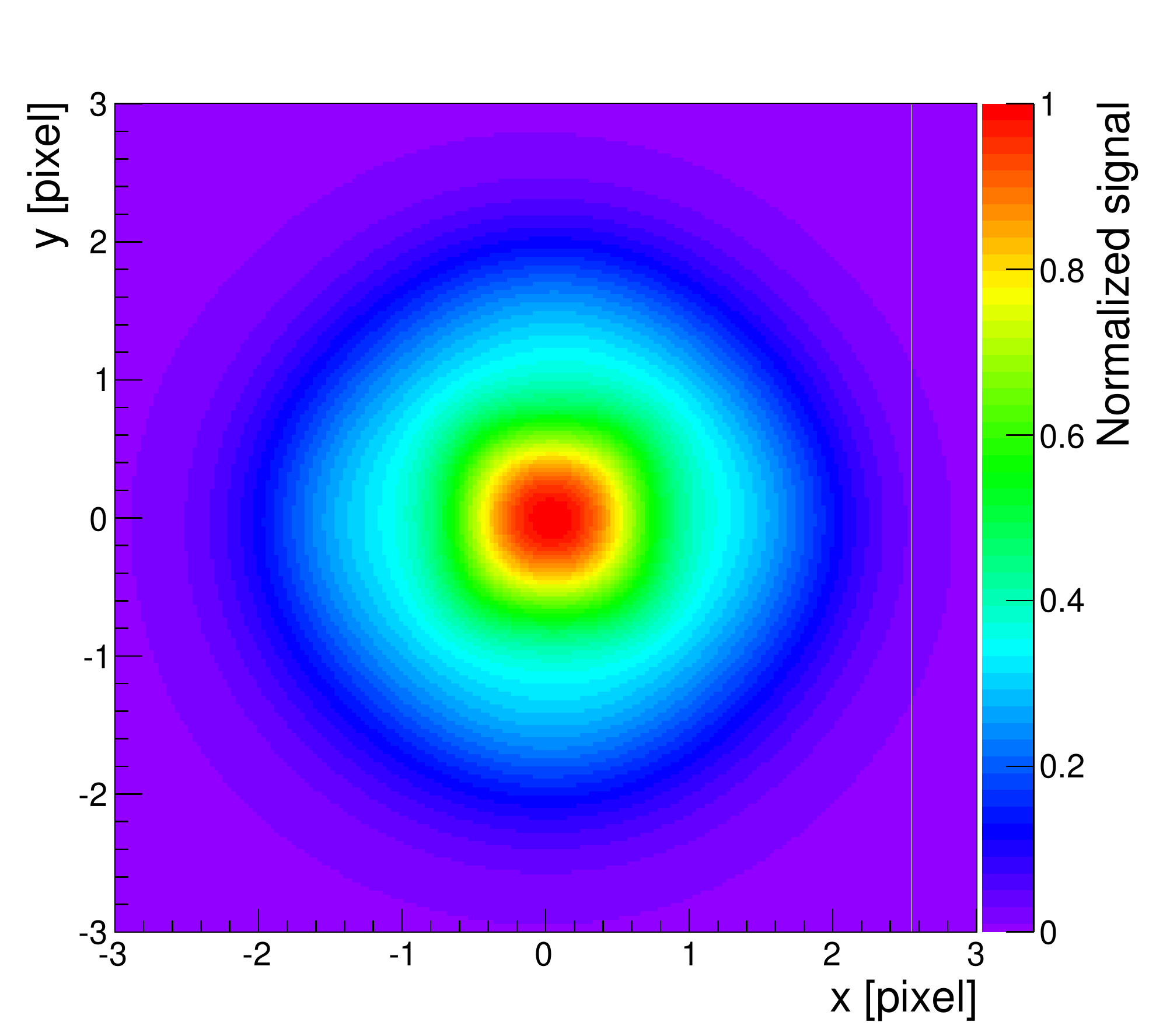}
	\includegraphics[width=0.32\textwidth]{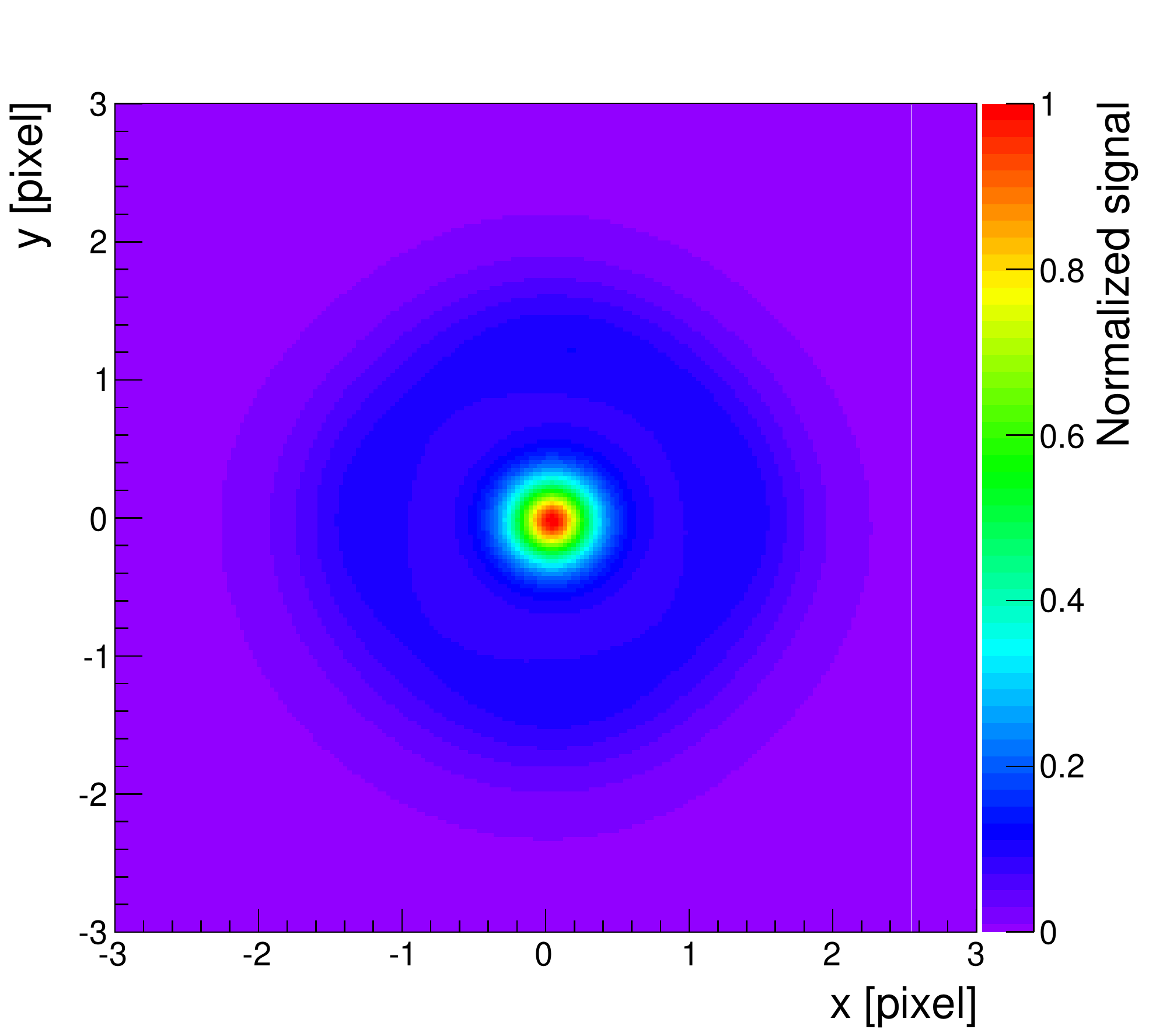}
}
\subfigure[200 pixels from the frame centre]{
	\includegraphics[width=0.32\textwidth]{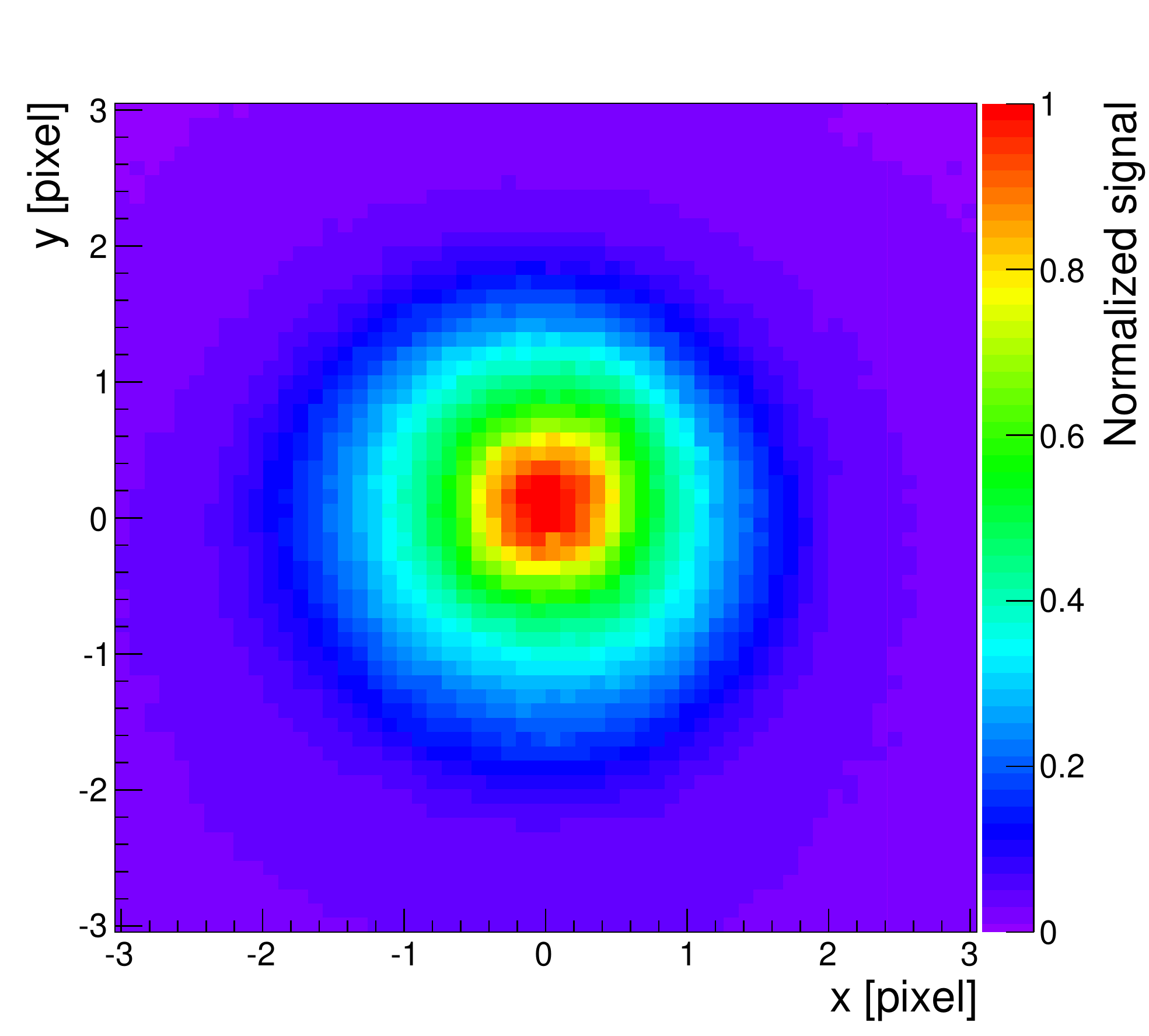}
	\includegraphics[width=0.32\textwidth]{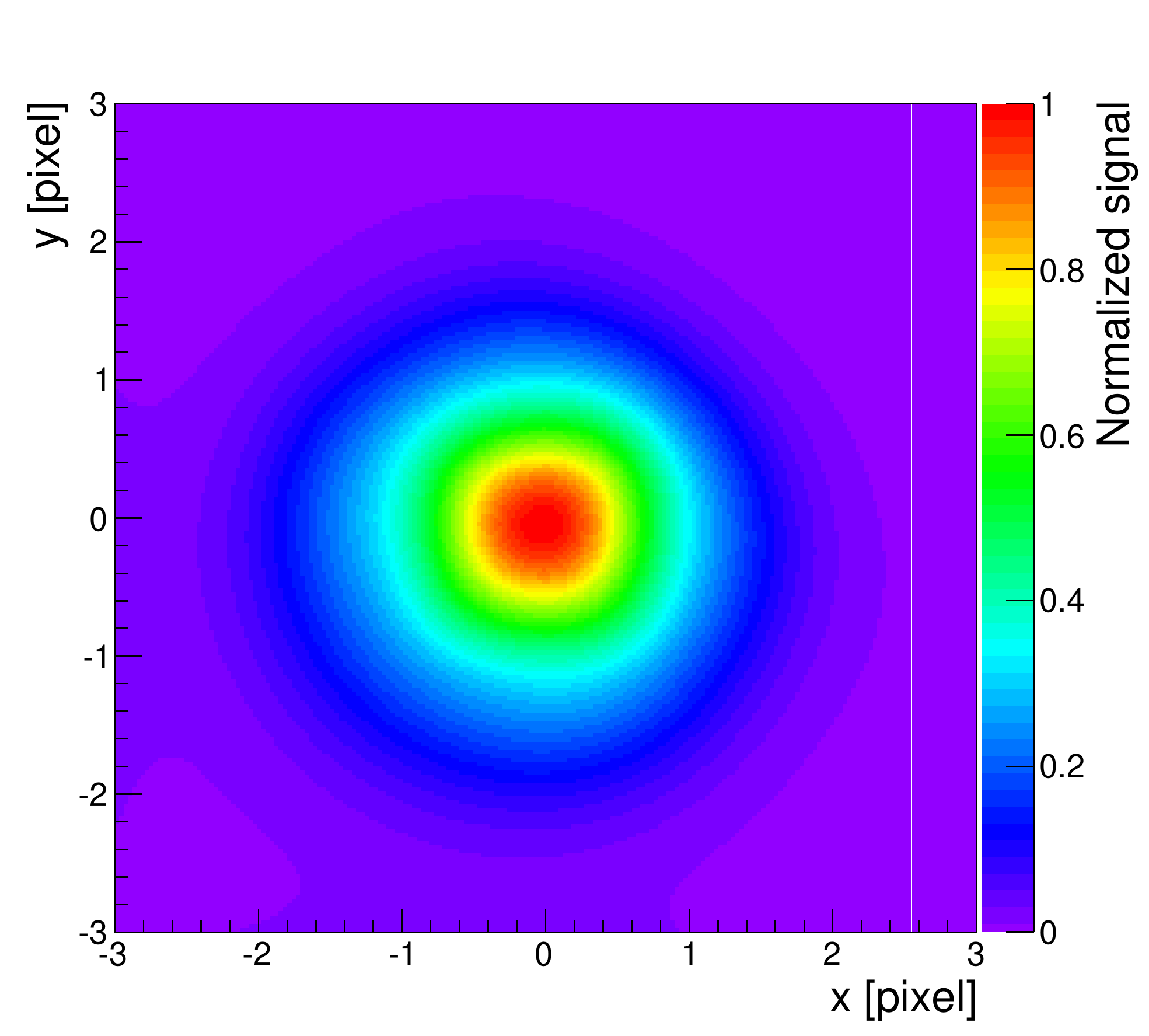}
	\includegraphics[width=0.32\textwidth]{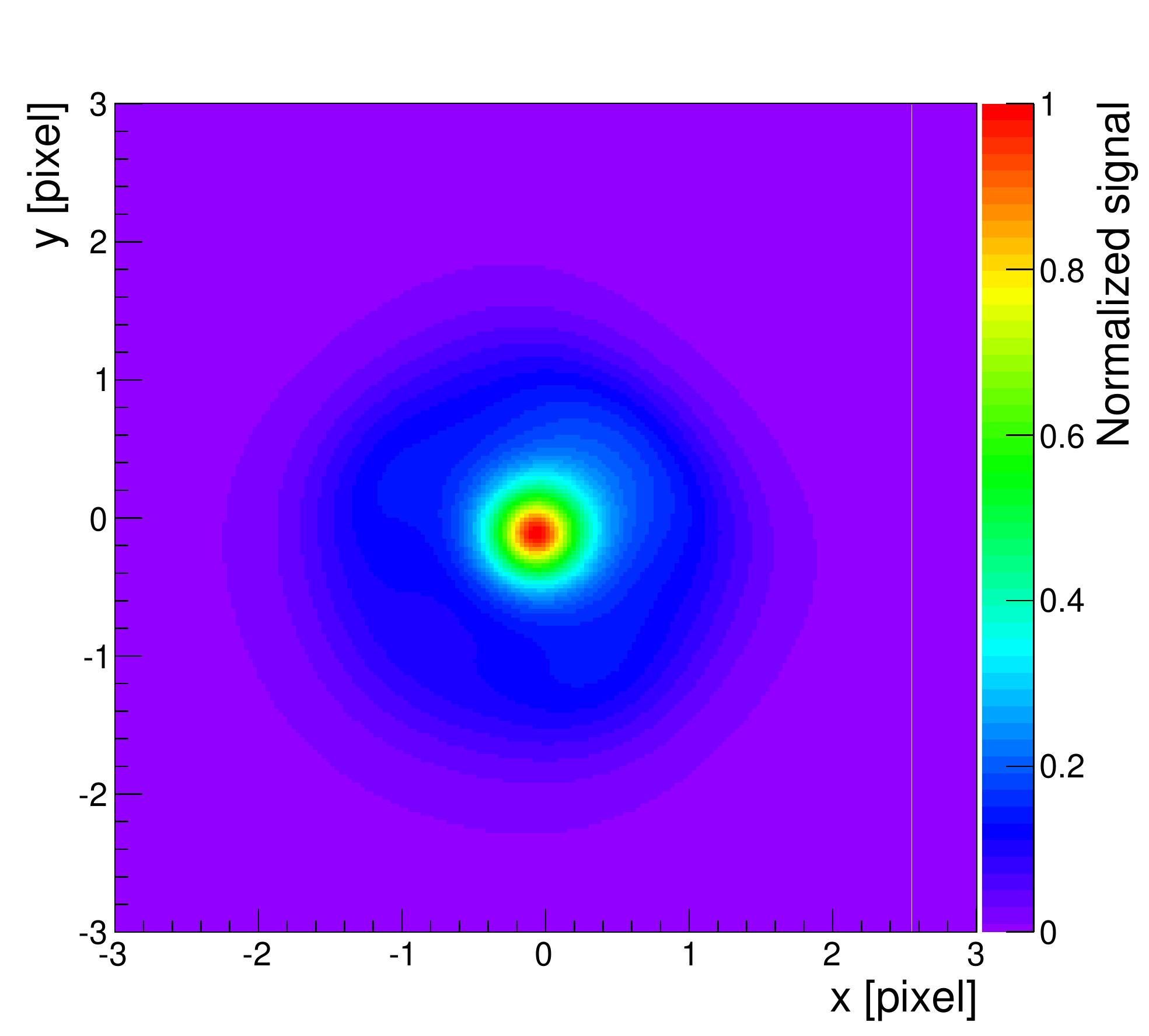}
}
\subfigure[400 pixels from the frame centre]{
	\includegraphics[width=0.32\textwidth]{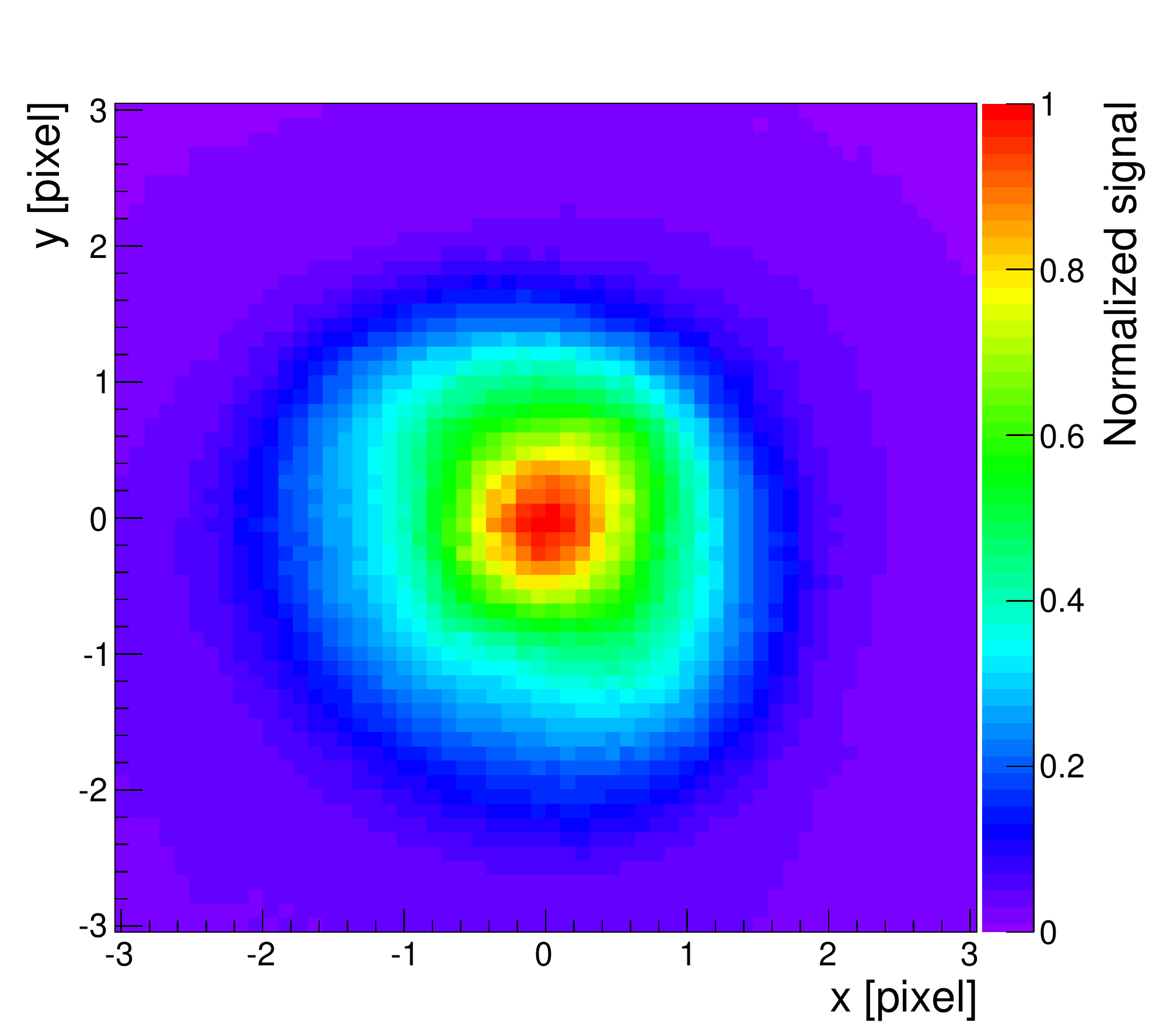}
	\includegraphics[width=0.32\textwidth]{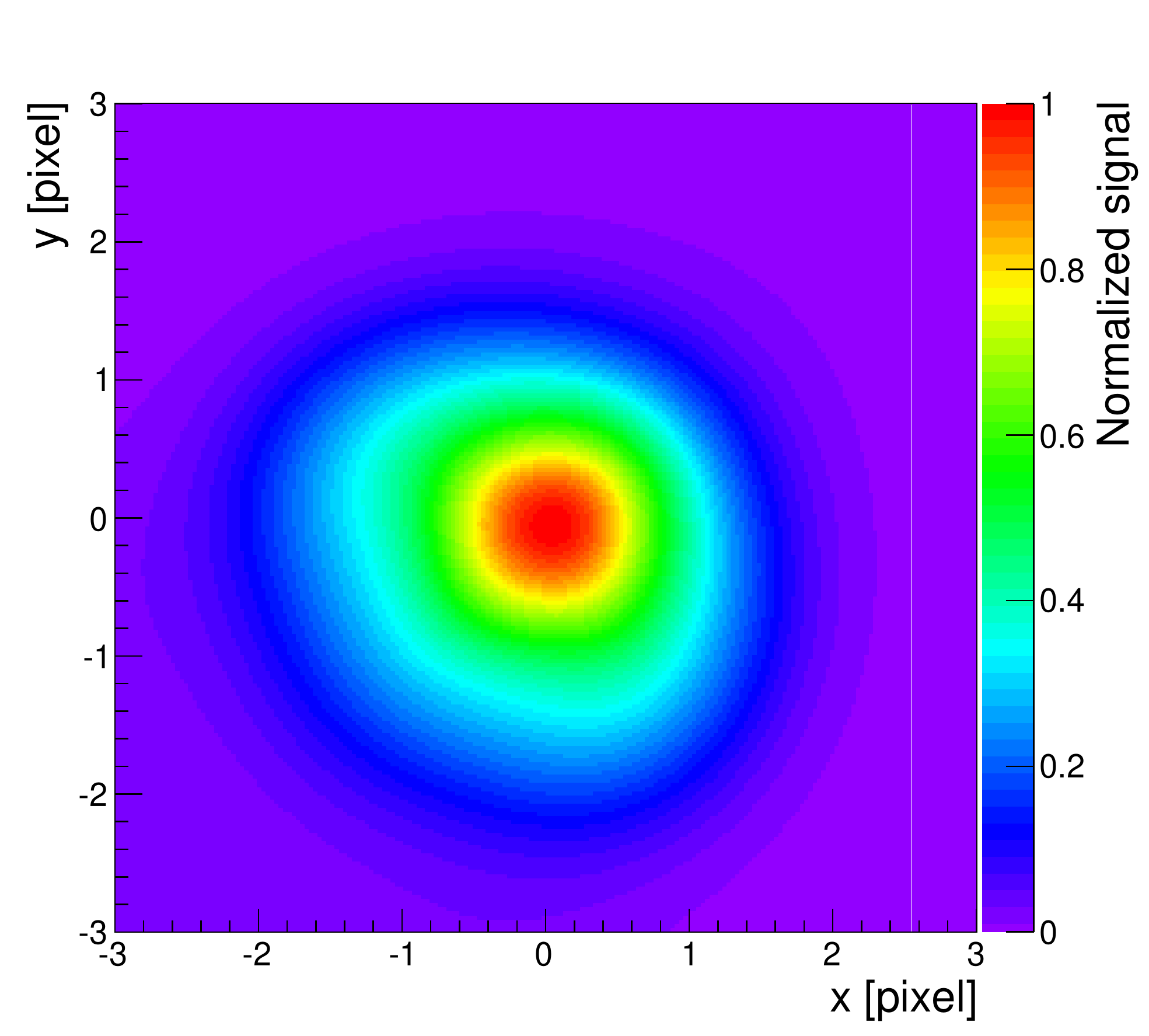}
	\includegraphics[width=0.32\textwidth]{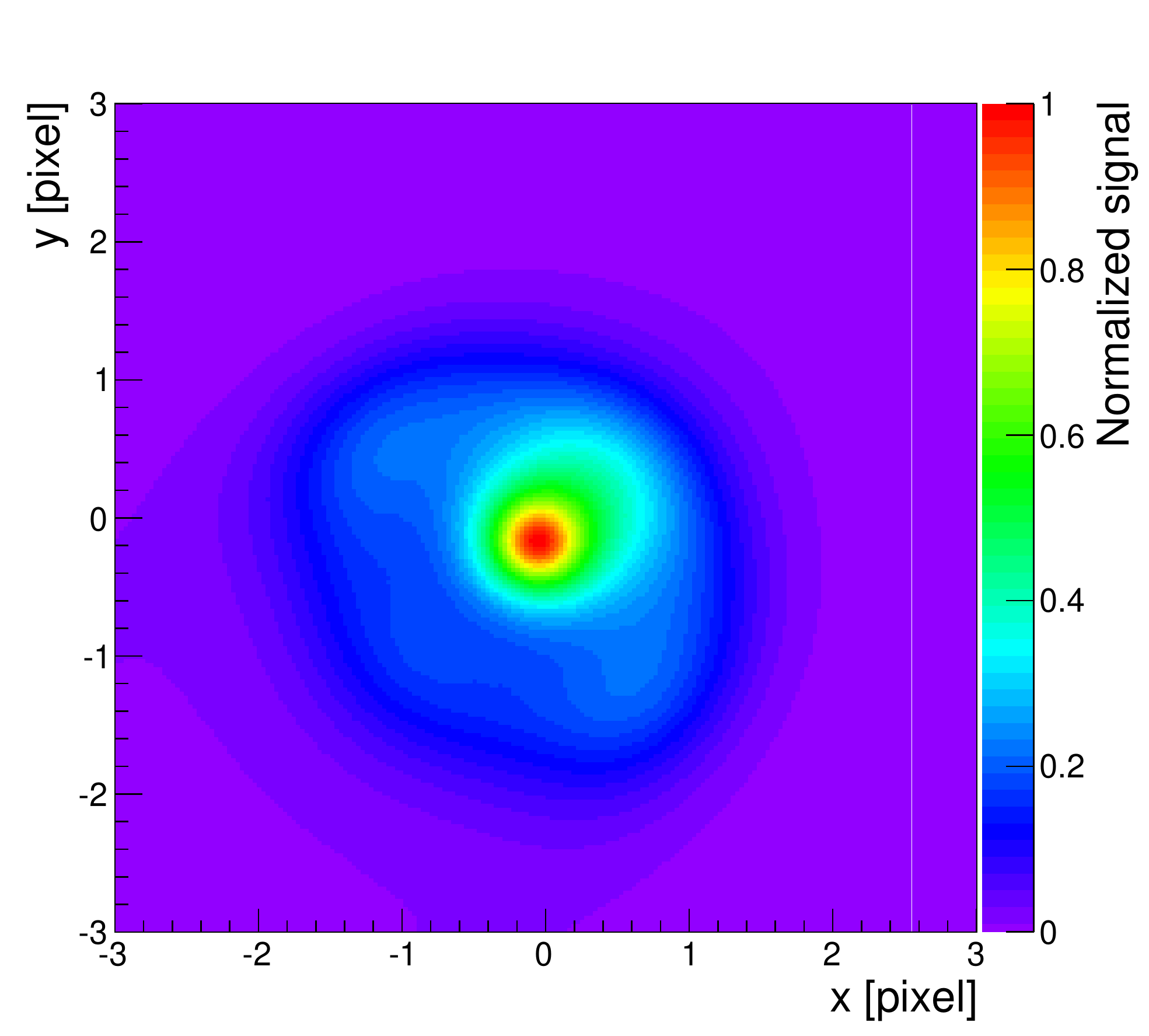}
}
\subfigure[600 pixels from the frame centre]{
	\includegraphics[width=0.32\textwidth]{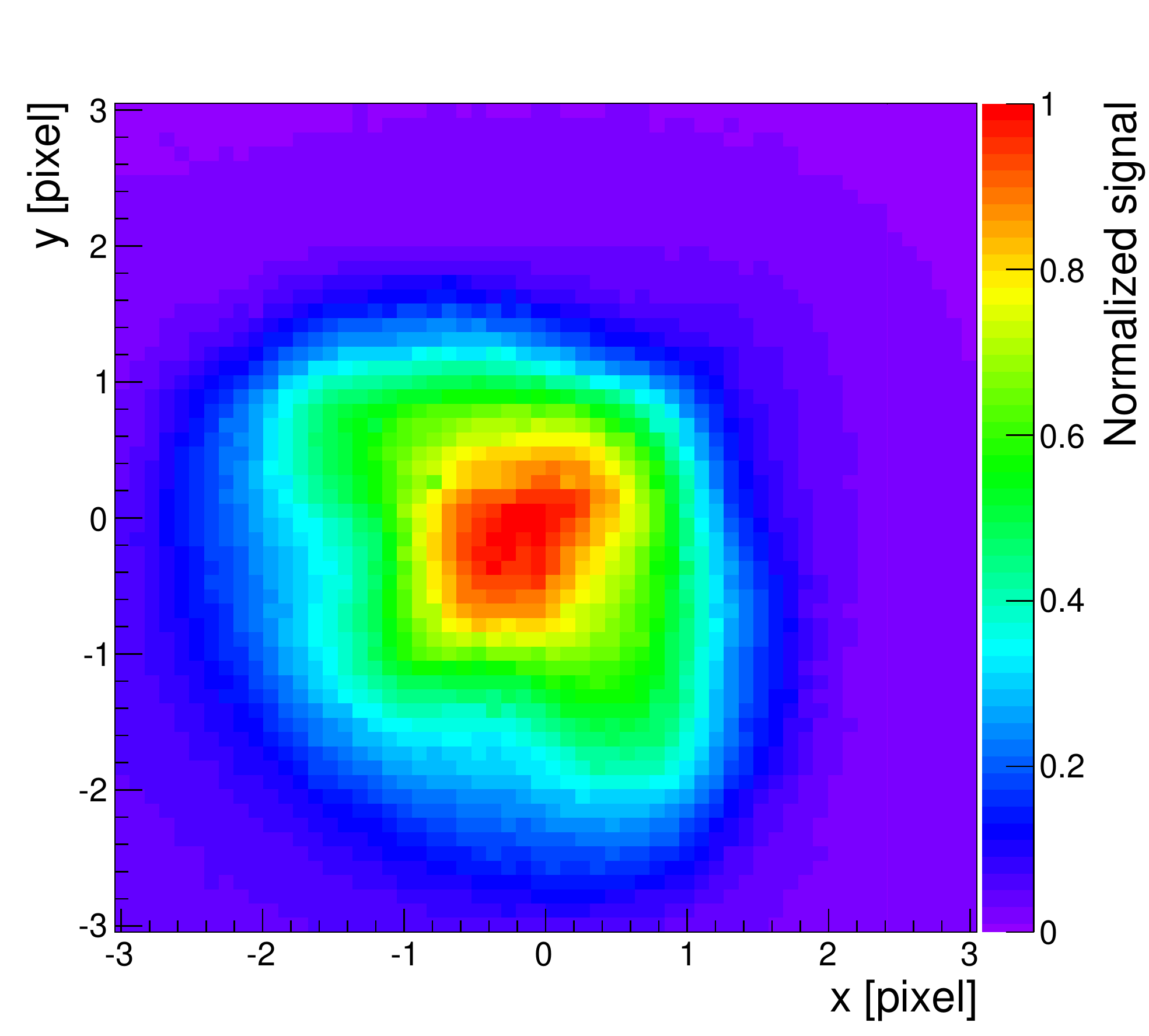}
	\includegraphics[width=0.32\textwidth]{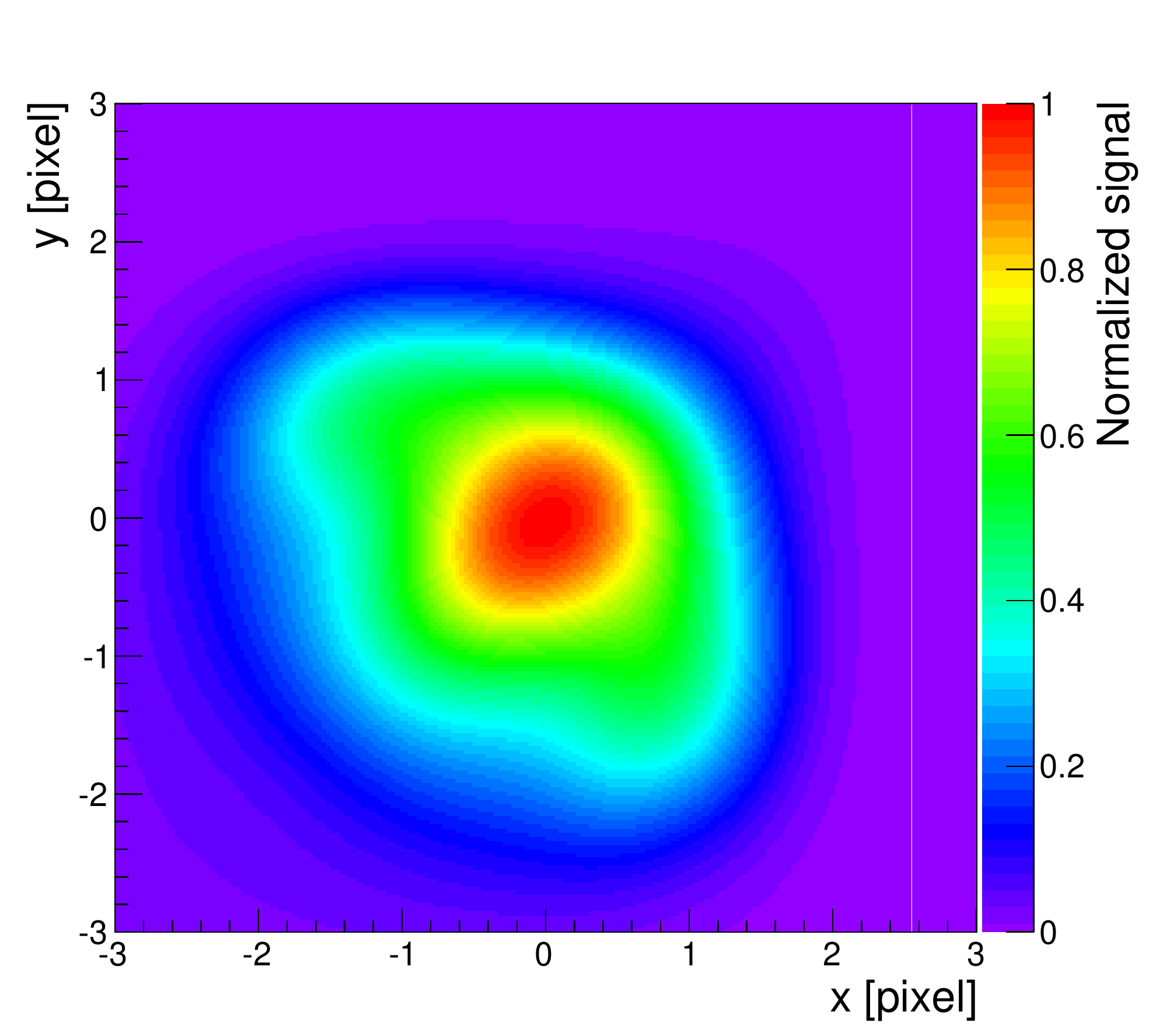}
	\includegraphics[width=0.32\textwidth]{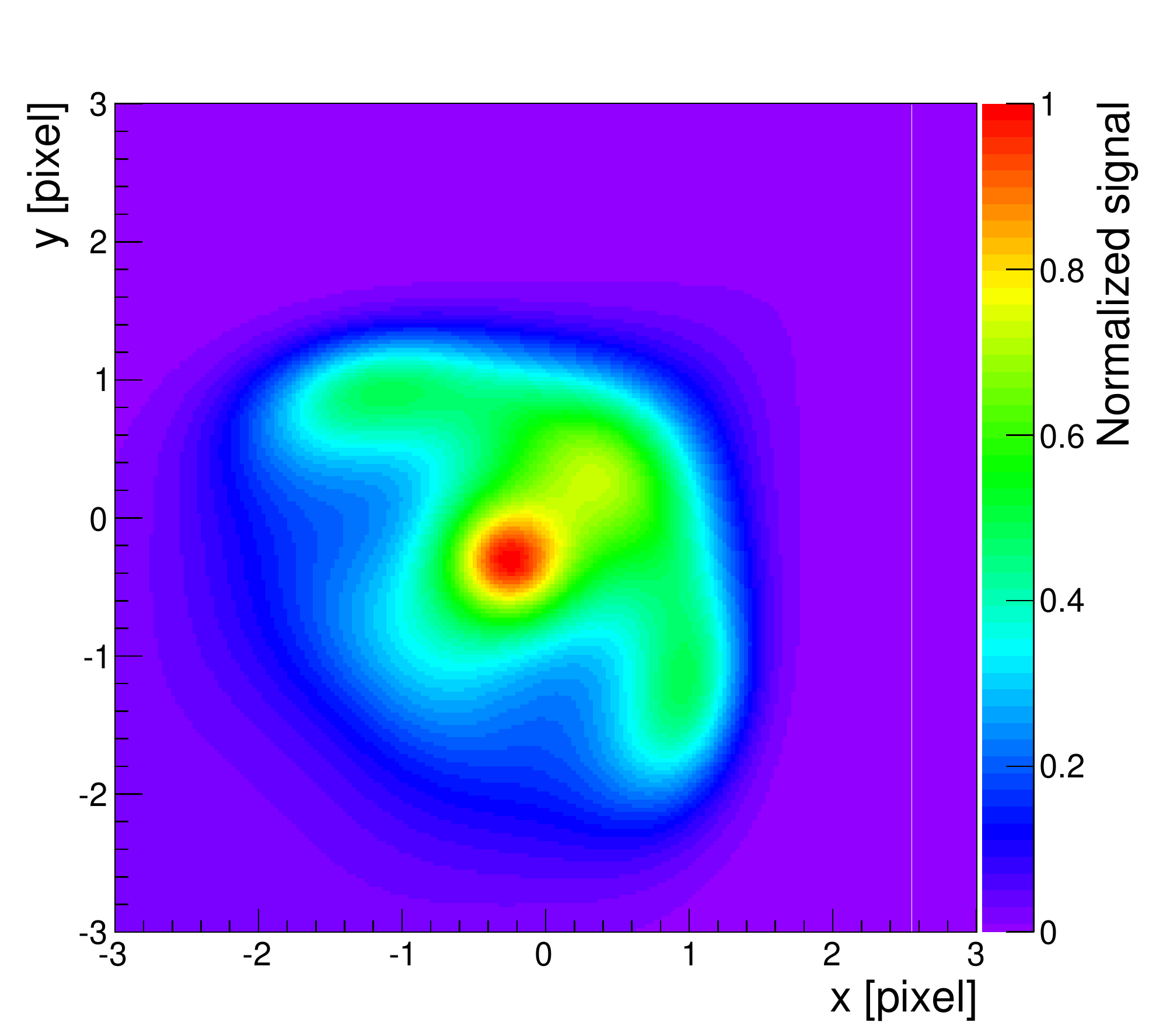}
}
\end{center}
\caption{Final results for the PSF modelling. Shown for 0-600 pixels from the frame centre are: measured profile (left), polynomial model of the PSF convoluted with the CCD pixel response (centre), and the obtained shape of the optical PSF before convolution with the CCD pixel structure (right).}
\label{fig_fin_psf}
\end{figure}

\begin{figure}
\begin{center}

\subfigure[800 pixels from the frame centre]{
	\includegraphics[width=0.32\textwidth]{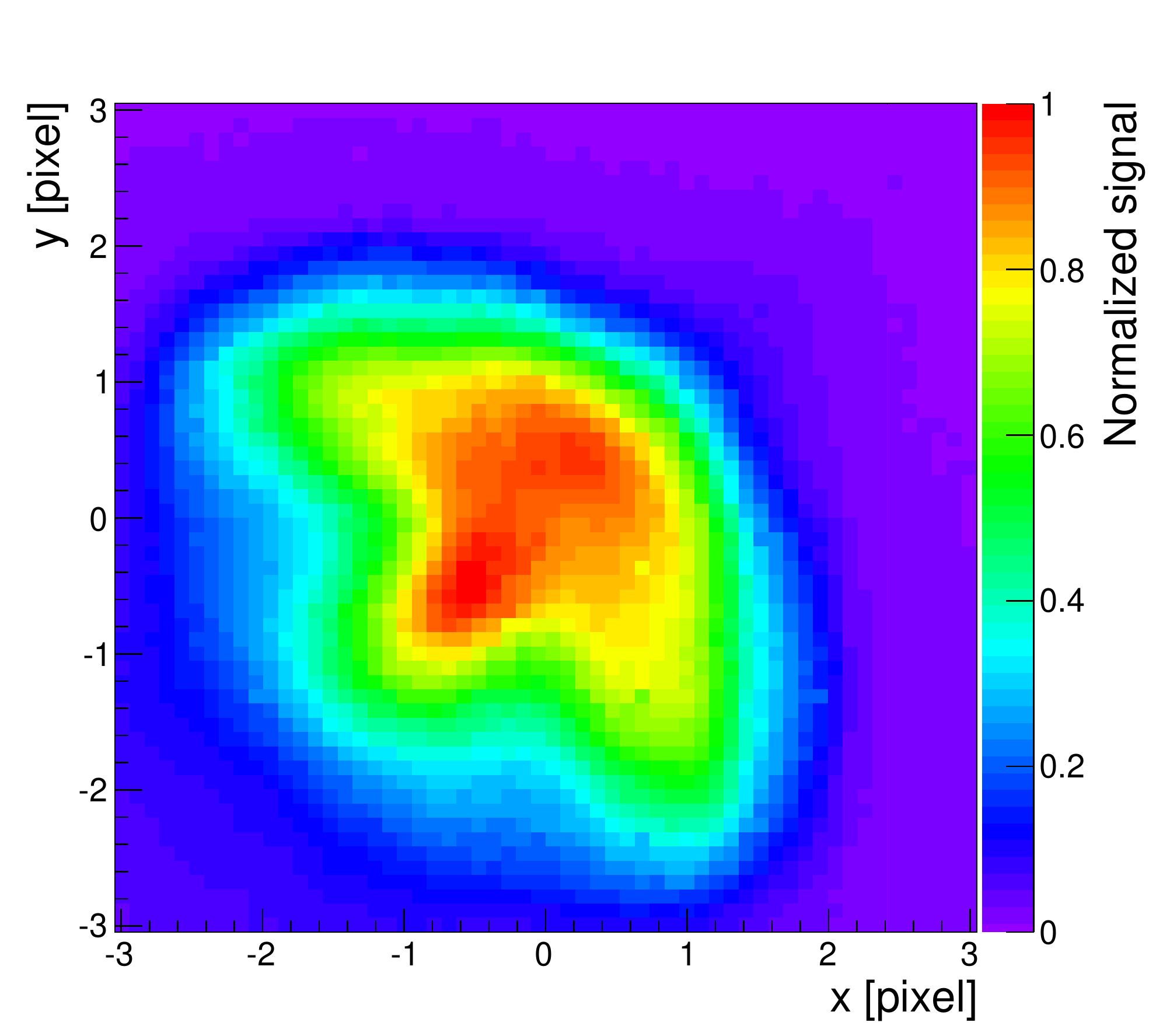}
	\includegraphics[width=0.32\textwidth]{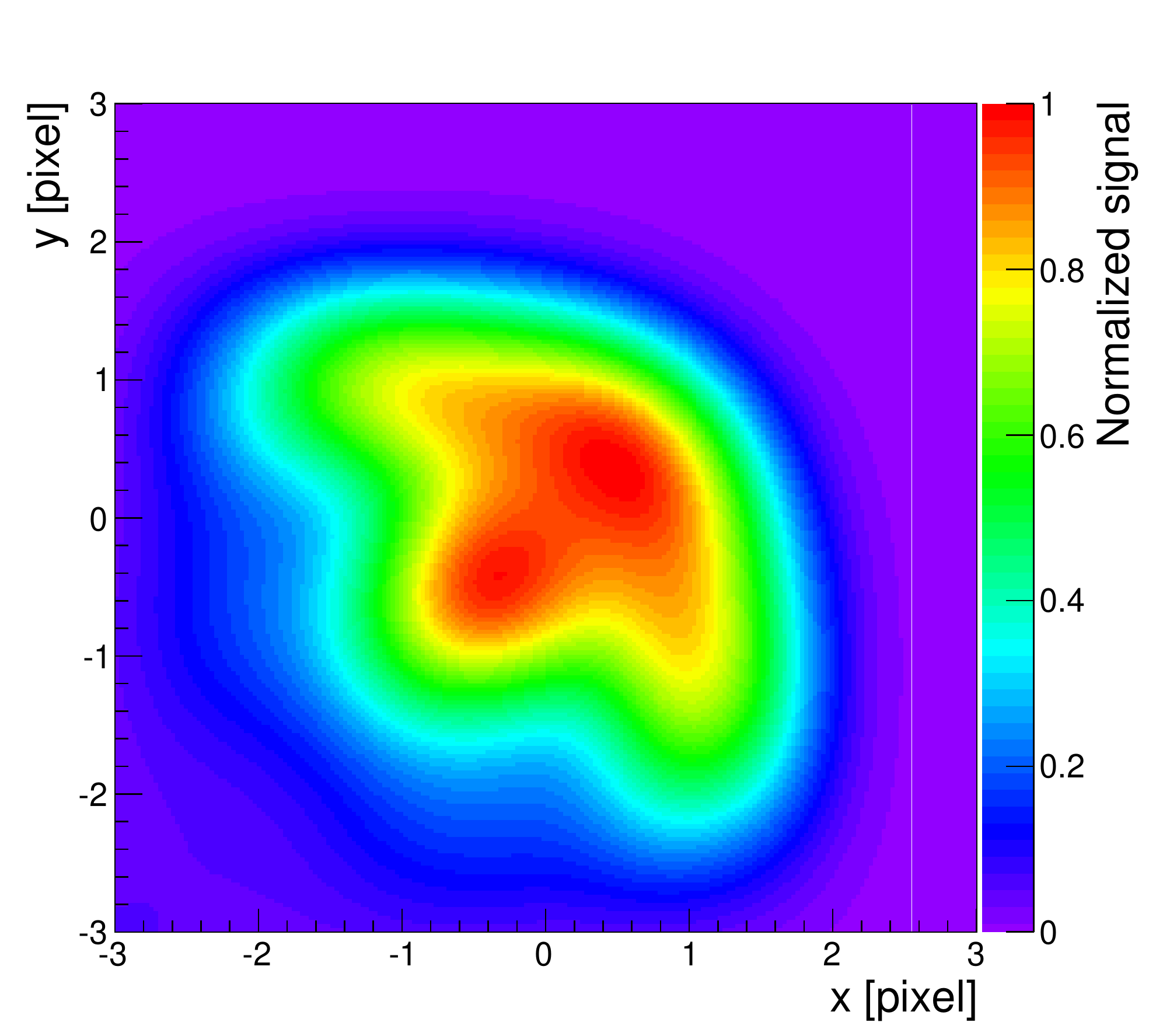}
	\includegraphics[width=0.32\textwidth]{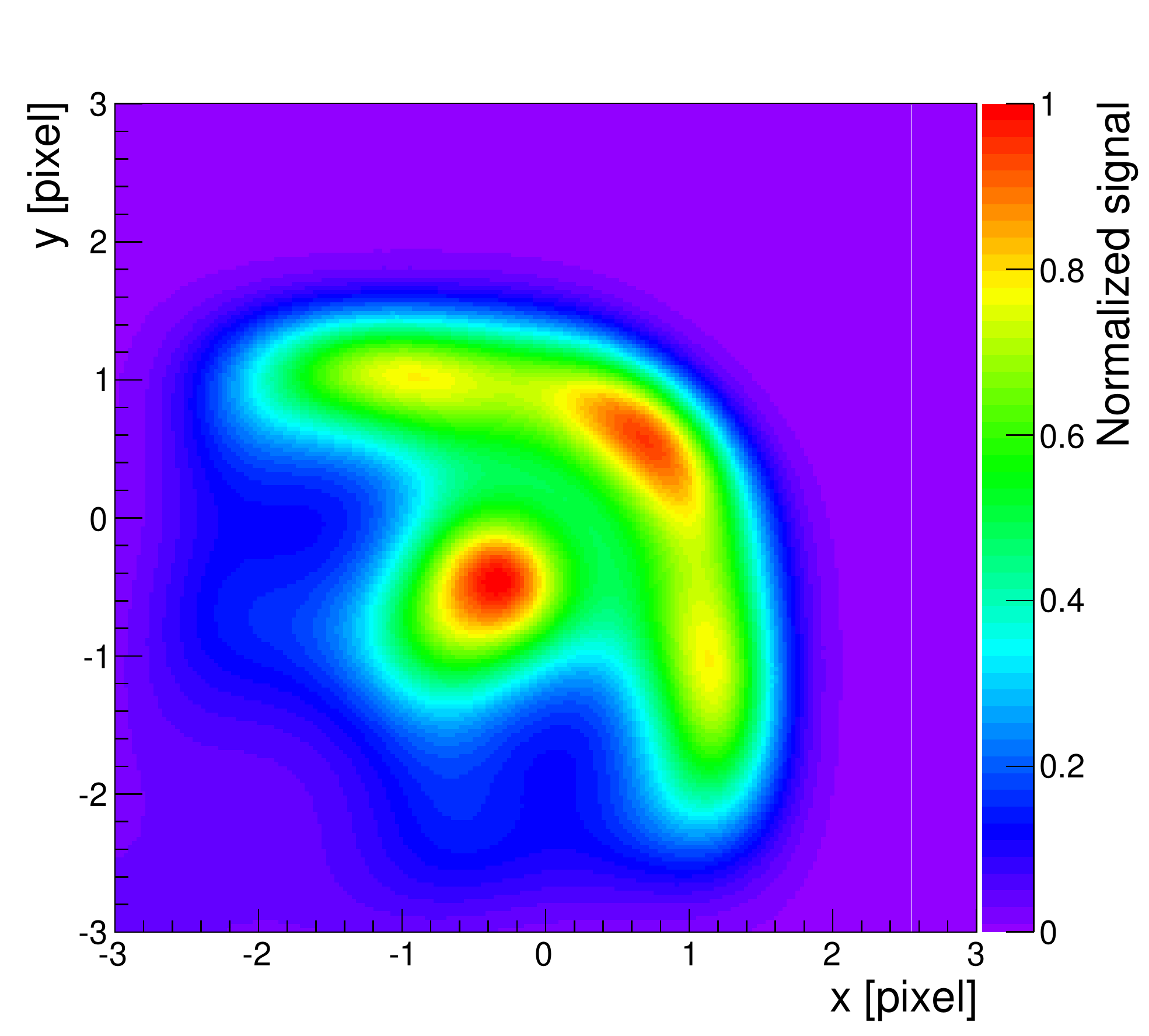}
}
\subfigure[1000 pixels from the frame centre]{
	\includegraphics[width=0.32\textwidth]{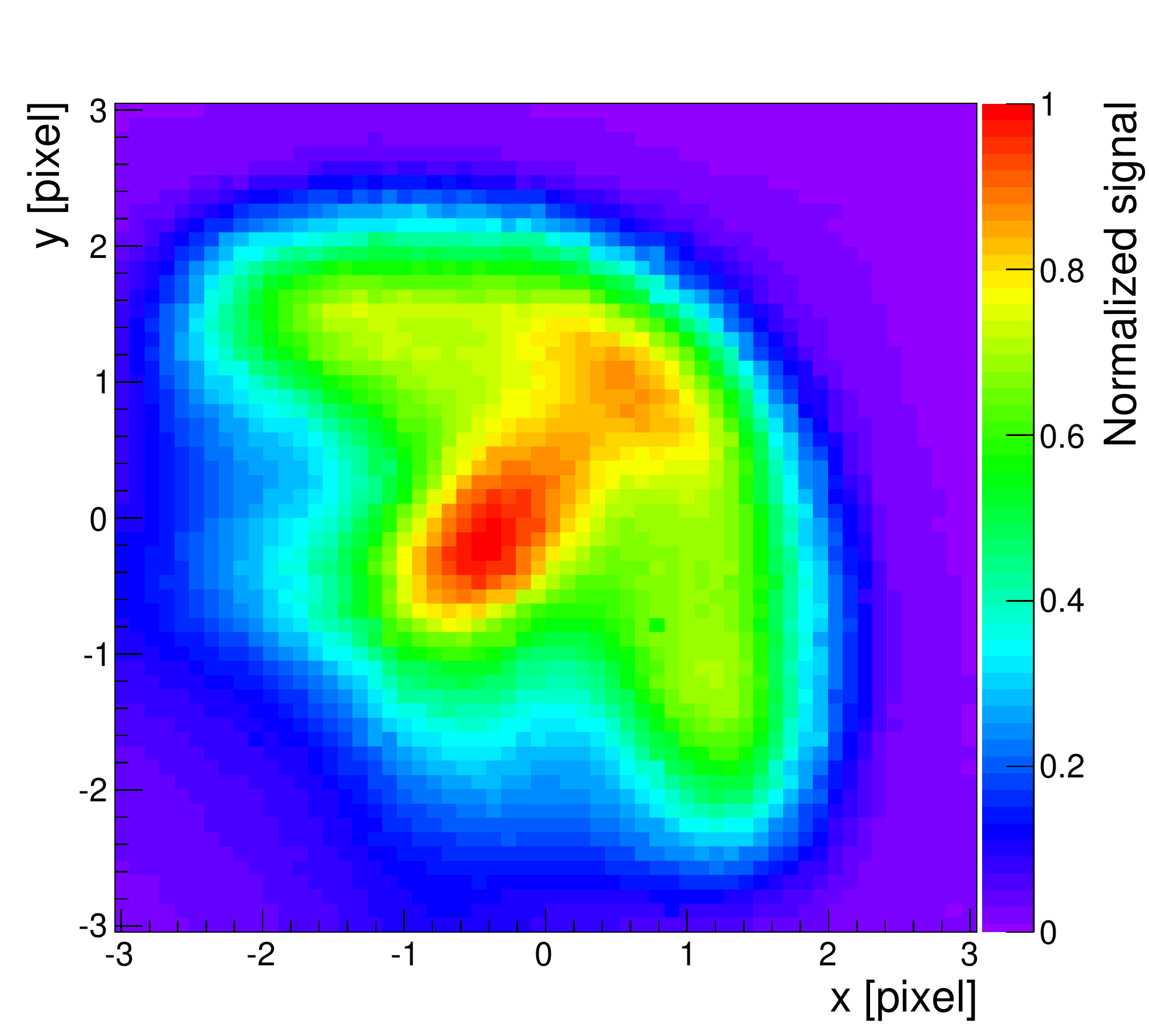}
	\includegraphics[width=0.32\textwidth]{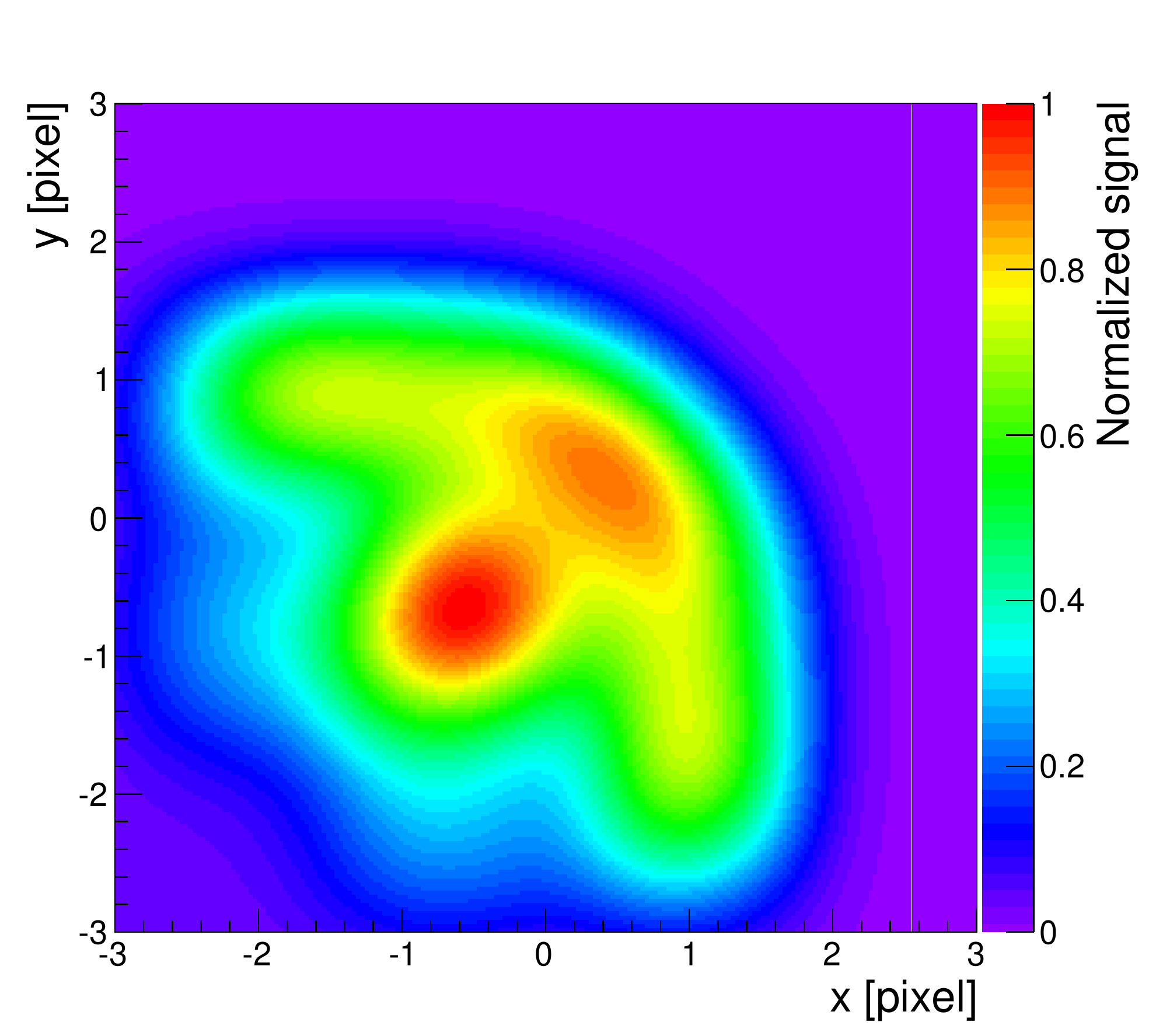}
	\includegraphics[width=0.32\textwidth]{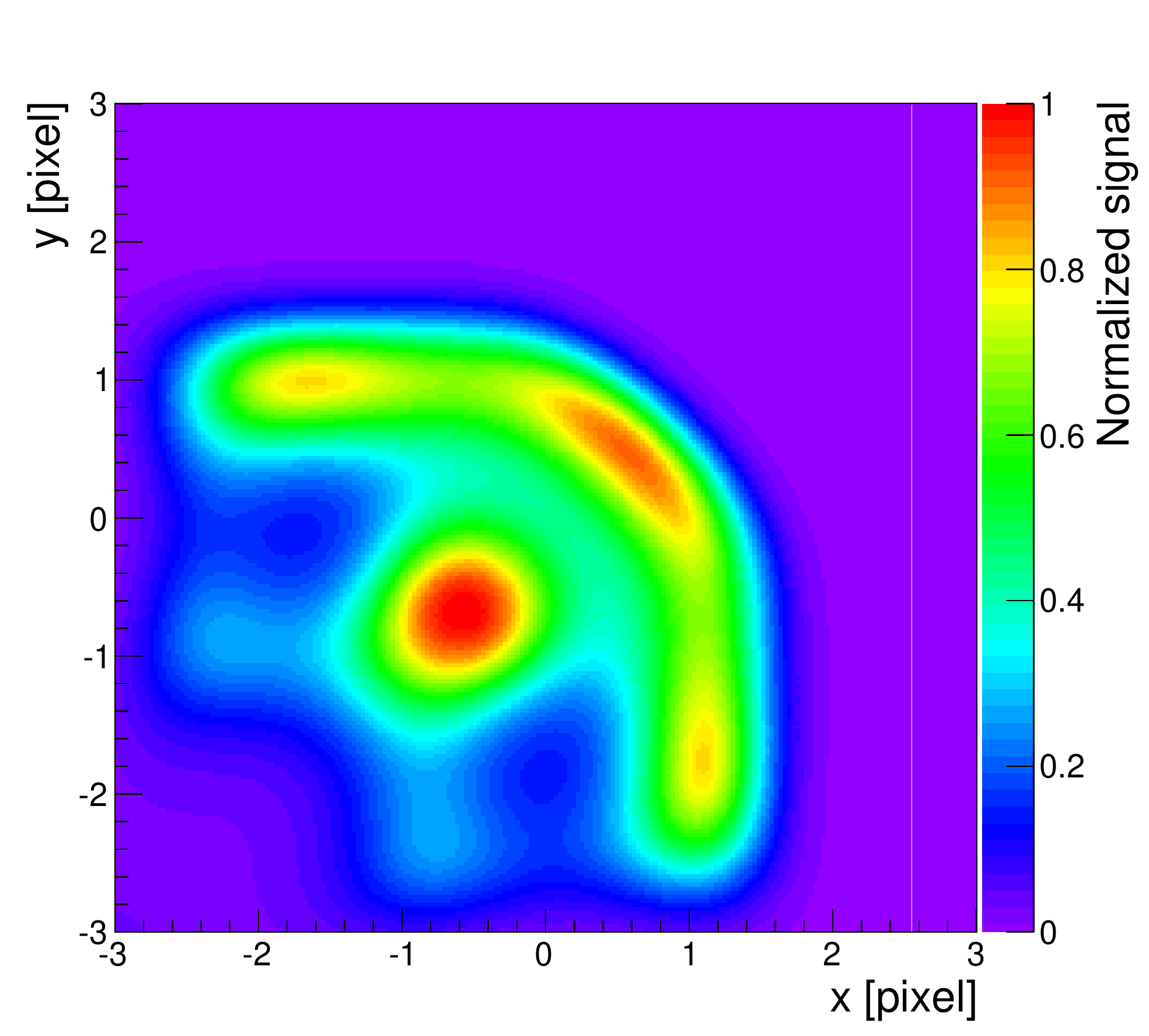}
}
\subfigure[1200 pixels from the frame centre]{
	\includegraphics[width=0.32\textwidth]{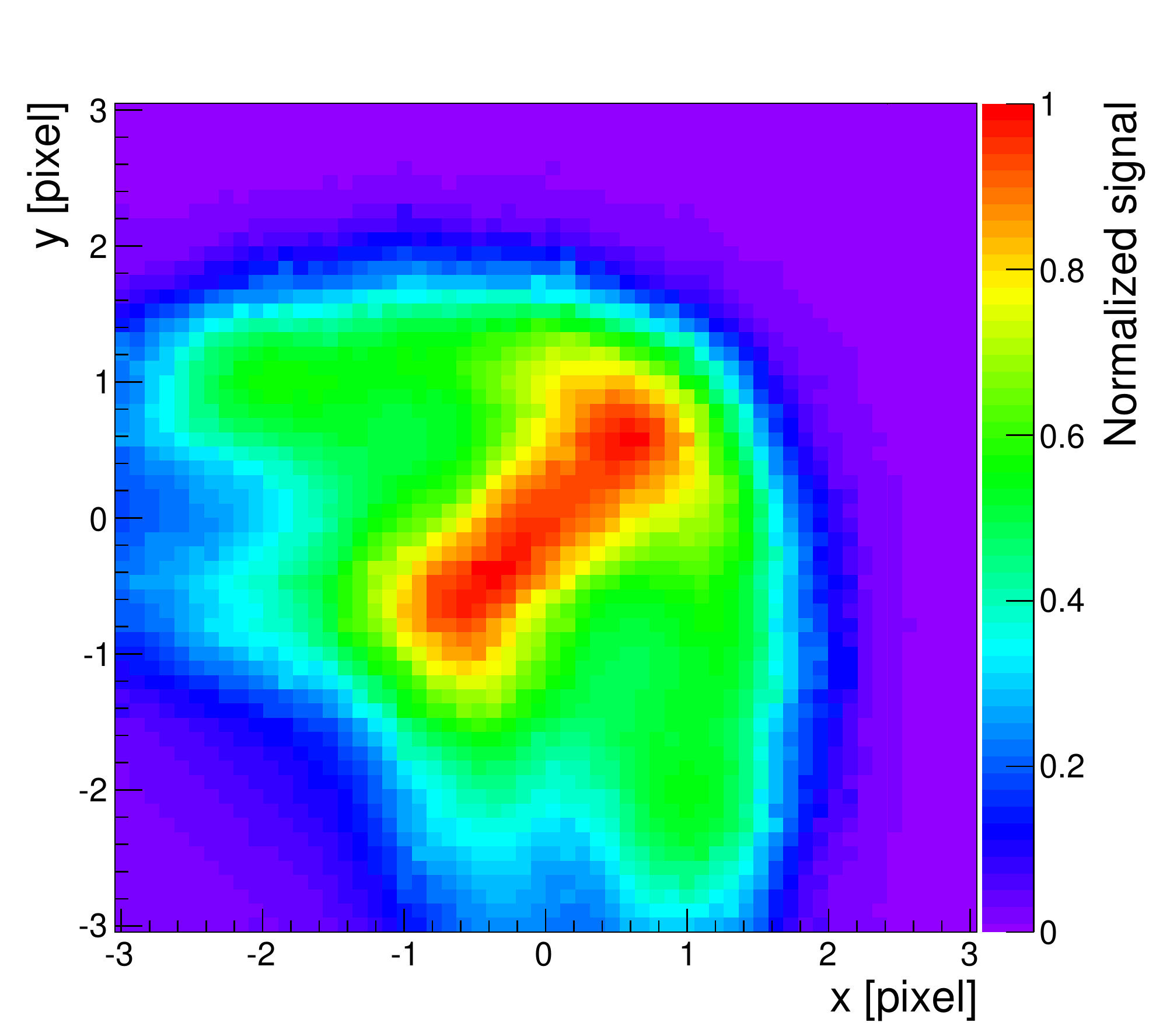}
	\includegraphics[width=0.32\textwidth]{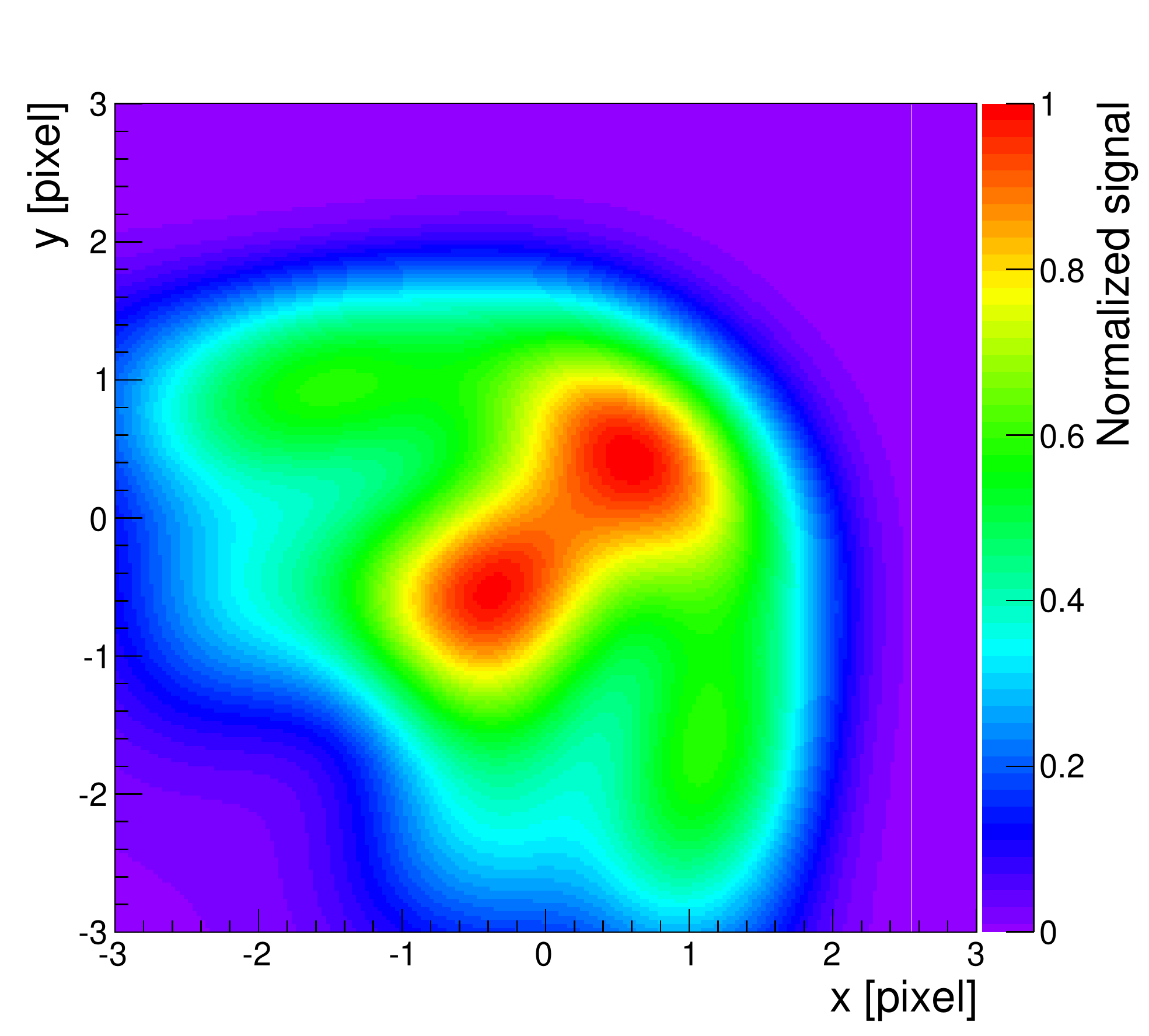}
	\includegraphics[width=0.32\textwidth]{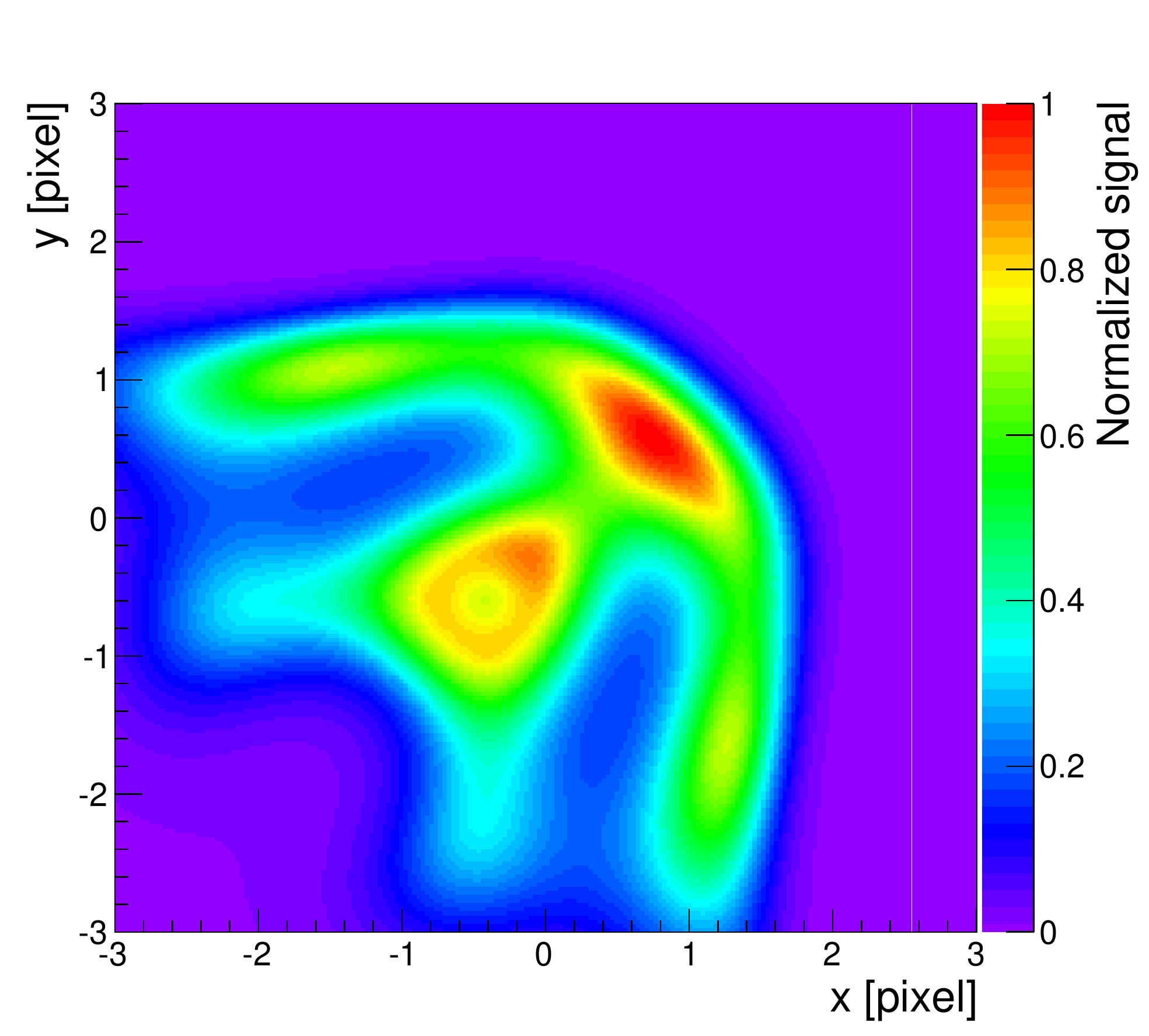}
}
\subfigure[1400 pixels from the frame centre]{
	\includegraphics[width=0.32\textwidth]{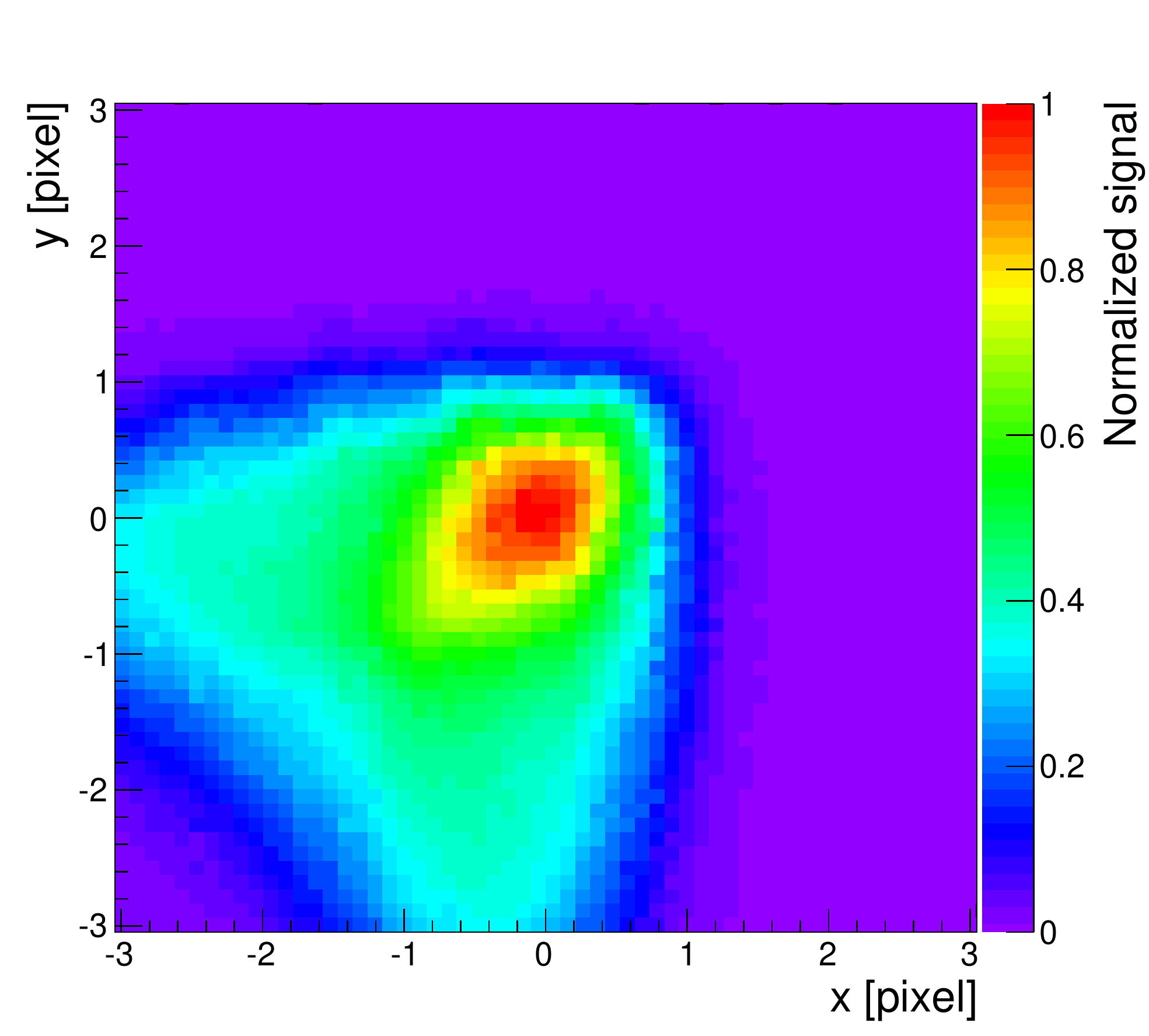}
	\includegraphics[width=0.32\textwidth]{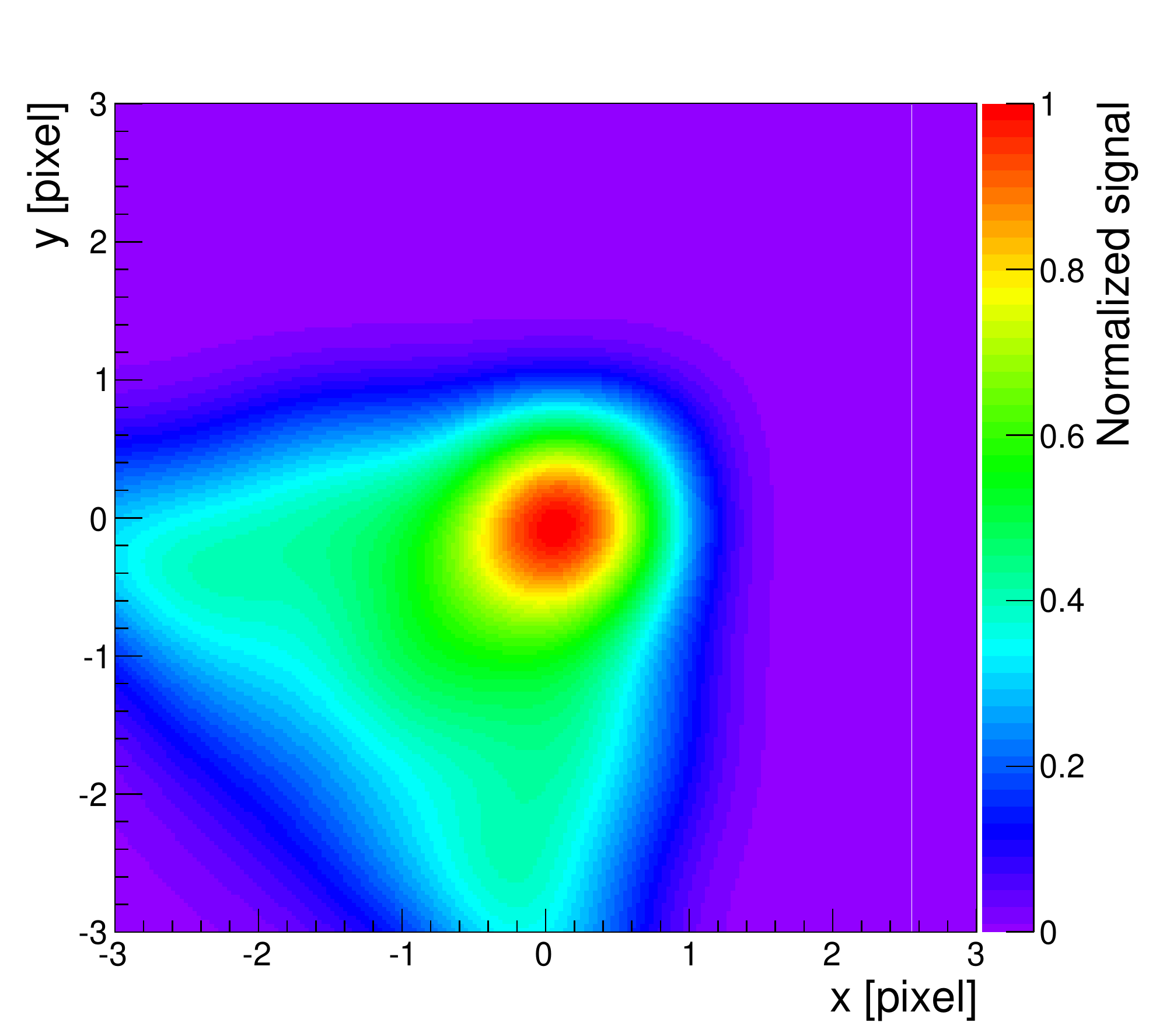}
	\includegraphics[width=0.32\textwidth]{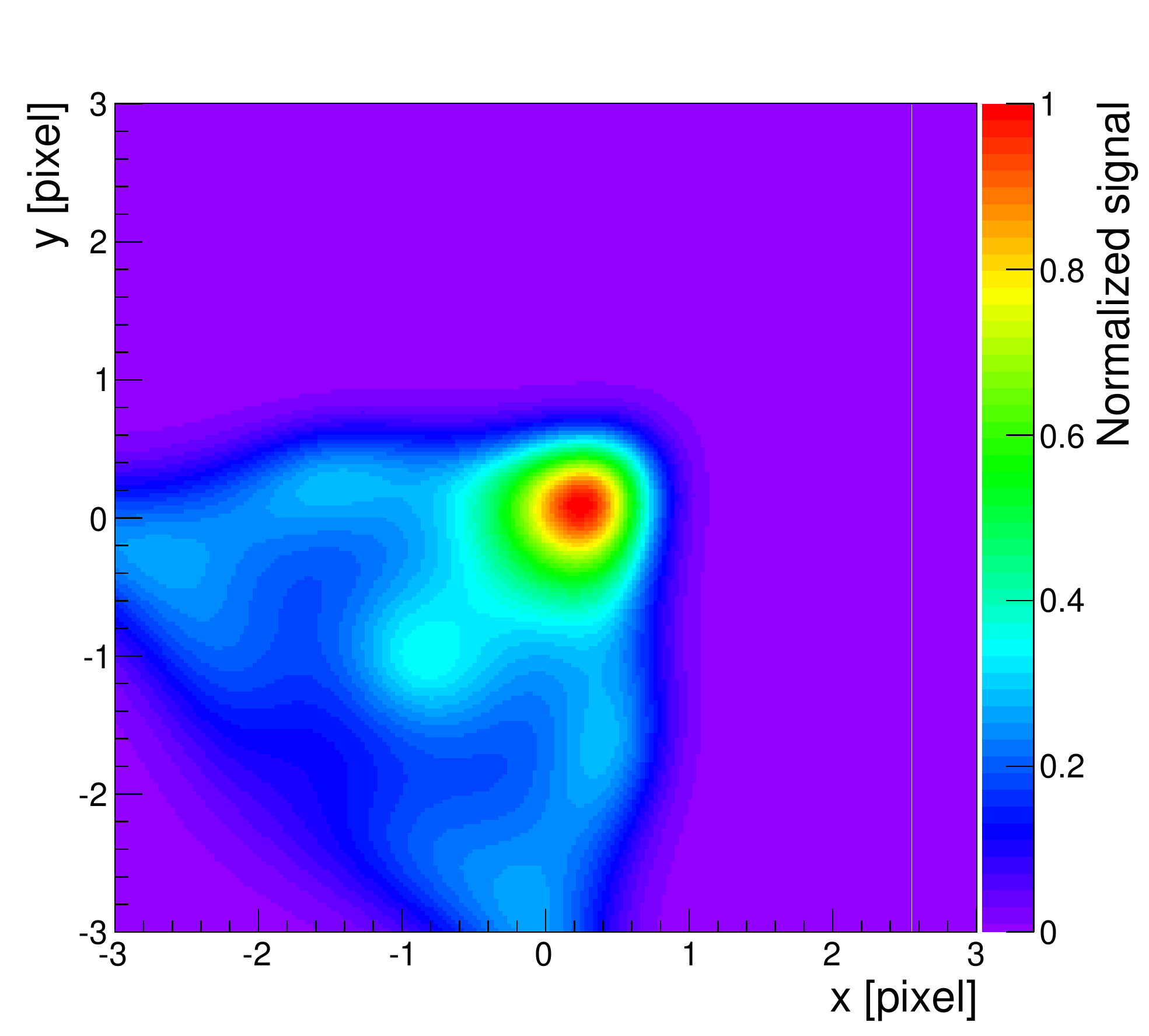}
}
\end{center}
\caption{Final results for the PSF modelling. Shown for 800-1400 pixels from the frame centre are: measured profile (left), polynomial model of the PSF convoluted with the CCD pixel response (centre), and the obtained shape of the optical PSF before convolution with the CCD pixel structure (right).}
\label{fig_fin_psf2}
\end{figure}

As shown in figures \ref{fig_fin_psf} and \ref{fig_fin_psf2}, the chosen parametrisation describes the central $6\times6$ pixels of the measured PSFs quite well. For comparison, results for the optical $PSF_L$ shape, before convolution with the CCD response function, are also shown. The residues (fig. \ref{fig_res}) range from $5.7\%$ of maximum signal for the PSF 200 pixels from the frame centre, up to $12.8\%$ for 1400 pixels (tab. \ref{res_tab}). However, maximum residue is highly dependent on the individual fluctuations of data points and does not describe the overall deviation of the model from data well. The general comparison between fits to different profiles is described better by the relative root mean square deviations (RRMSD), which is lowest for the profile 200 pixel from the frame centre and highest for profiles 1000 and 1200 from the frame centre, which are respectively 2.6 and 2.7 times higher (tab. \ref{res_tab}). The RRMSD seems to follow the apparent complexity of the PSF\footnote{We also considered other statistics for the purpose of model testing,
including the standard and reduced $\chi^2$. However, the $\chi^2$ value is very dependent on the signal-to-noise ratio. Therefore it is not difficult to get a relatively low value of these statistics for a bad model using very noisy data or data that includes significant amount of background, where the white noise dominates. Moreover, it is not easy to define the number of degrees of freedom for the reduced $\chi^2$ for the non-linear model, such as the one presented in this paper.}.

Residues, as well as RRMSD, clearly show that the obtained model does not fully describe the PSF measured in our system. The considered uncertainties are purely statistical and are further lowered by the averaging of multiple profiles. However, in the considered data systematic uncertainties, which can be related to the high-resolution profile reconstruction or PRF reconstruction methods, turn out to be significant. Therefore the actual RRMSD cannot quantitatively describe the quality of the fit (except the fact that large systematic differences exists) and can be used mainly for comparison purposes.

We can point out two causes for the systematic deviation of the model from the data, visible after studying the residues (fig. \ref{fig_res}). The first one is an assumption of axially symmetrical PSFs, which is not true. The second is apparently the too few components used to describe such complicated shapes. Both issues could be treated introducing more components. However, according to our experience, this would in turn make the interpolation between fitted shapes on the frame unstable. The selected number of polynomials was the compromise between the quality of the individual fit and the smoothness of the interpolation.

For comparison we fitted the model based on polar shapelets \citep{shapelets1,shapelets2} convoluted with the PRF\footnote{The unconvoluted fit of shapelets was also attempted, but the results were significantly worse.}. The same procedure of choosing the most significant symmetric components as for the polynomial fit was followed (with exactly the same fitting method), and the parameter describing the width of the shapelet was chosen to minimise the $\chi^2$ for all the profiles. The final basis for the shapelets involved the same number of free parameters as for the polynomial model.

The highest residues for shapelets range from $12.8\%$ for the 200 pixels from the frame centre to $17.9\%$ for 1200 pixels and are significantly higher then the polynomial model residues. The highest residue of the polynomial model equals the lowest residue of the shapelet model. The RRMSD is bigger by 24\% for 800 pixel from the frame centre up to 62\% for 1200 pixels from the frame centre (tab. \ref{res_tab}). Therefore we conclude that the proposed polynomial basis describes highly deformed PSFs in very wide-field experiments significantly better. Additionally, we attempted to fit cheblets \citep{cheblets}, which should properly describe extended wings of PSF, but results of the procedure were even worse than for modified Zernike polynomials or shapelets.

\begin{figure*}
\begin{center}
	\includegraphics[width=0.328\textwidth]{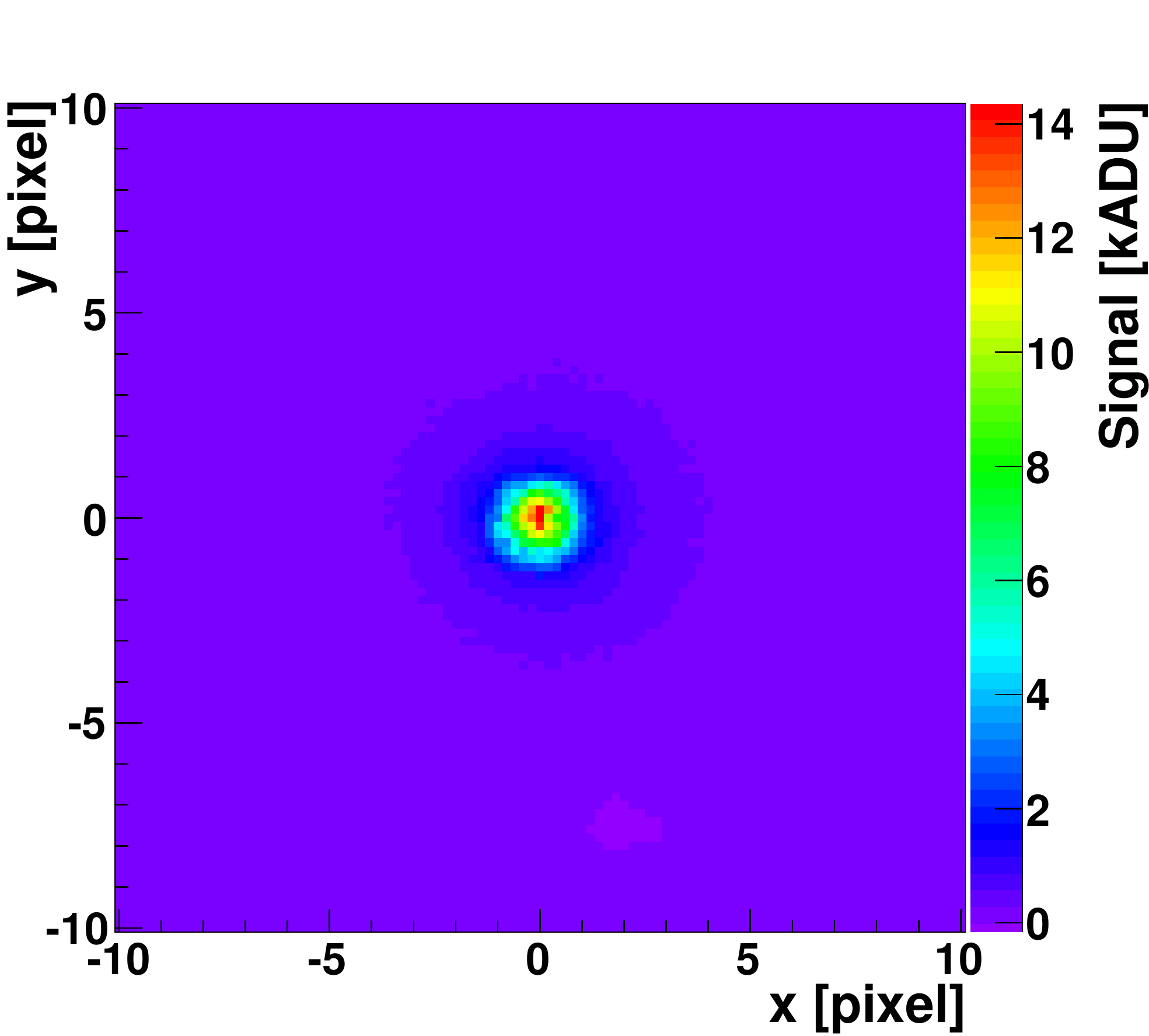}
	\includegraphics[width=0.328\textwidth]{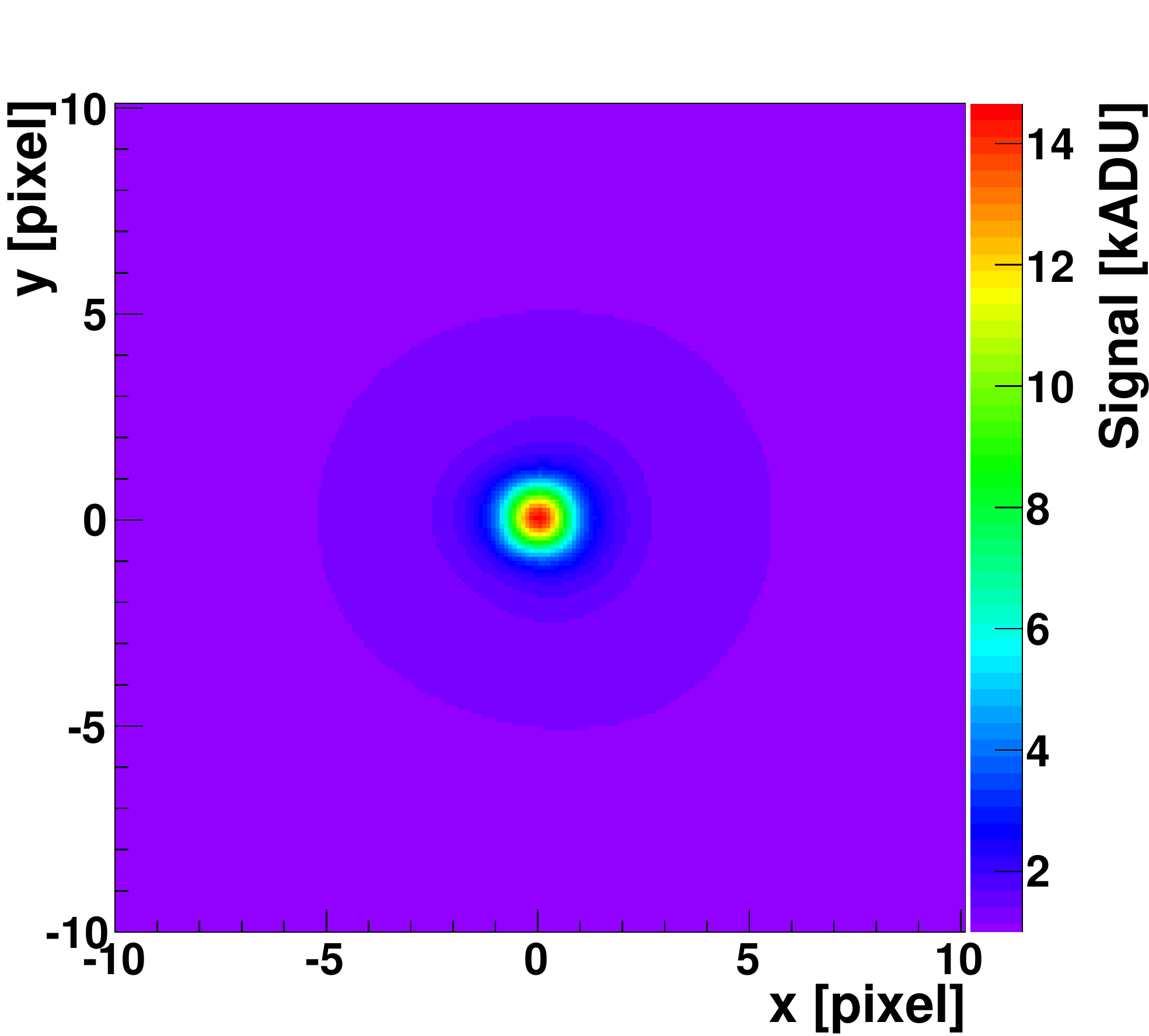}\\
	\includegraphics[width=0.328\textwidth]{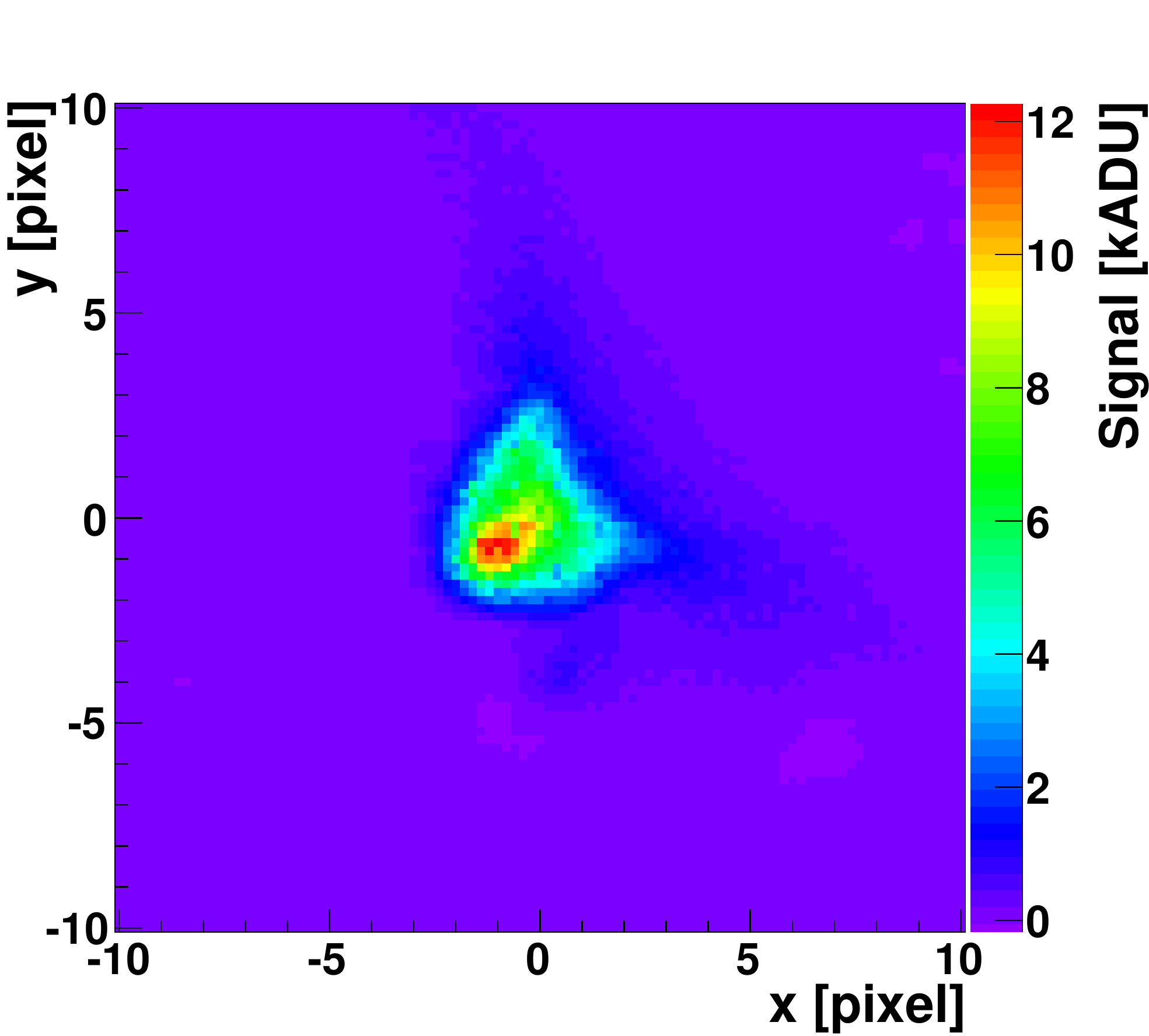}
	\includegraphics[width=0.328\textwidth]{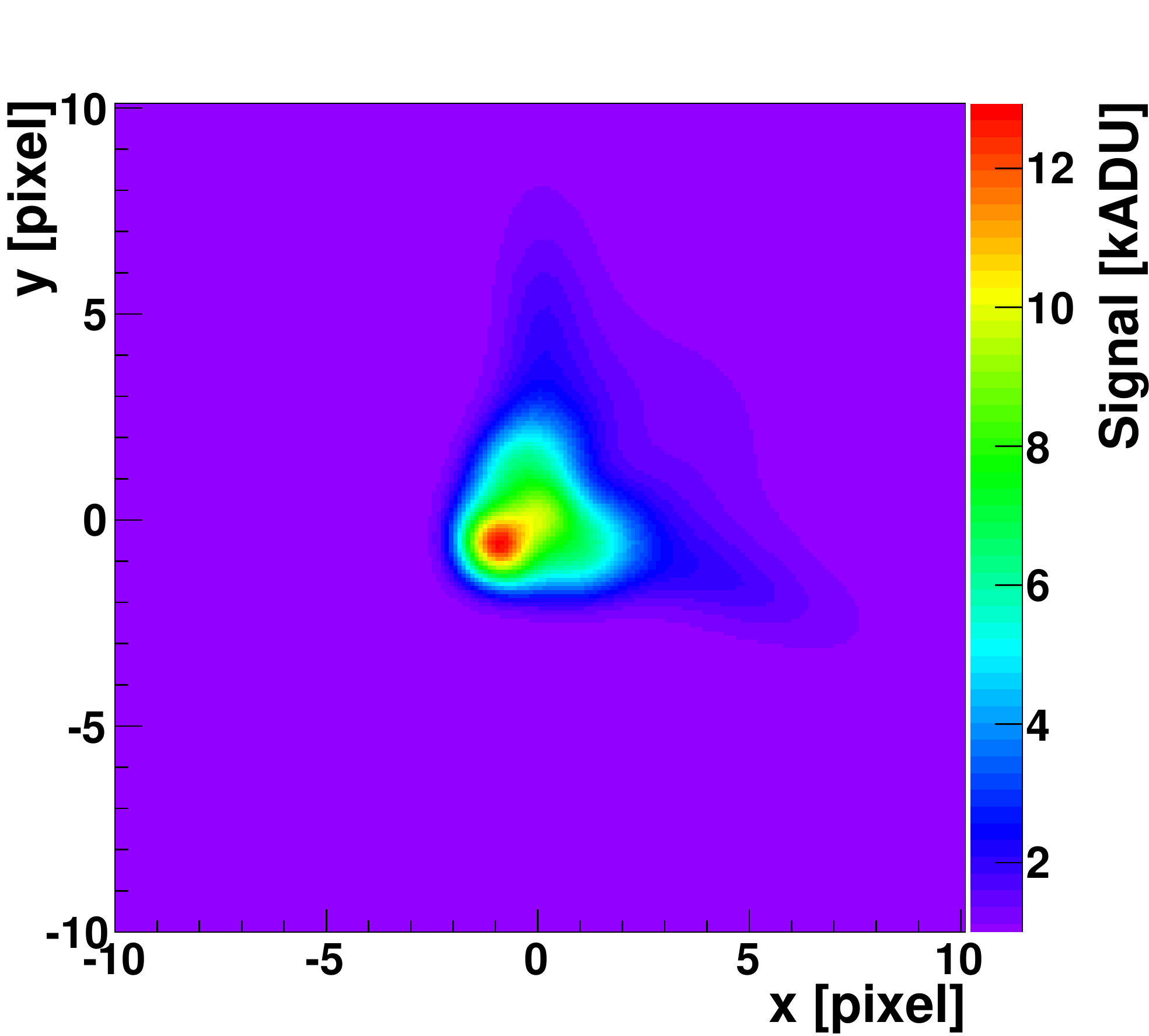}
\end{center}
\caption{Two sample reconstructed star profiles (on the left), used to obtain the final PSF model (on the right), for about 100 (top) and 1300 (bottom) pixels from the frame centre, respectively. The model reproduces profiles in high detail, although far tails and asymmetric parts are missing.}
\label{sample_PSF}
\end{figure*}

The popular requirement set on PSF modelling is for residues to be contained in the $1\%$ range of the maximum signal \citep{Stetson1992}. However, this is mostly the case of fitting a lower resolution profile. A single-fit to high-resolution PSF obtained from laboratory measurements corresponds to a combined fit of a single profile to 400 low resolution profiles. Performing the fit to low resolution data in our case would also result in lower residues and RRMSD for this particular fit, but the obtained model would depend on the actual PSF position with respect to the pixel grid. It also has to be noted that the requirement of $1\%$ was defined for seeing-limited instruments, with much smaller deviations of the PSF from well known shapes, and it simply does not apply to a very wide-field system with complicated optics such as ``Pi of the Sky''. Furthermore, residues of the very central profile and the widely used Moffat function were higher than $25\%$, as high as $14.5\%$ for modified Lorentzian+Gaussian, and $12.9\%$ for sum of two Moffats. The highest residues of our most deformed profiles are not higher than the residues for the simple central profile described by the double Moffat, which may be considered as proof of the quality of our fit.

Parametrisation used to describe the PSF measured for a white diode can also be used to model monochromatic or defocussed profiles. However, the results of such modelling were not as good as for the white PSFs.

\section{Sample model applications}

The time required to compute star profile from the polynomial model is low, although slightly higher than computing the two-dimensional Gaussian or Moffat functions. It becomes significant when calculating shapes for thousands of stars. For about 13000 stars\footnote{The number of stars on a frame from a sample field observed by ``Pi of the Sky''}, the computation time for profiles consisting of 36 pixels ($6\times6$ pixels) is about 30 seconds, on a single core of a 2.2 GHz processor running a 32 bit Linux. This gives approximately 2 ms calculation time per star. 

The processing time is an important issue in very wide-field, high time-resolution astronomical experiments, where the amount of data must be reduced in near real time to avoid exceeding the storage capacity. Therefore reduction algorithms have to be optimised to fulfil such a requirement. The current computing time is far too long for the ``Pi of the Sky'' on-line analysis, where single exposure is ten seconds. It should also be reduced significantly for a standard off-line analysis, where full reduction of the frame should take at most about two times the exposure time, to avoid the infinite growth of the amount of the data stored\footnote{The computation time is less relevant if data is reduced on a several fast computers, however, this is not the case of the ``Pi of the Sky'' experiment and, probably, most other small experiments}. It has to be noted that the full reduction process involves more time consuming steps than a simple profile or flux calculation, therefore we estimate that the calculation time should be reduced by at least an order of magnitude to be introduced into standard, all-star pipelines, at least for the computing power currently available in our experiment. However, the current calculation time makes the model very suitable for dedicated, special data-analysis applications.

A quick glance at data from real sky images taken with the prototype \pin detector in Chile\footnote{Camera used in laboratory experiments and those taking real sky images in Chile were identical in technical specification and were equipped with the same lenses.} shows that the star images are not identical in shape to the laboratory measurements, although their general properties are similar. This difference can result from different focussing of the lenses used in laboratory and sky measurements. Moreover, star images show that the real PSF does not change with distance from the frame centre, but also strongly depends on the azimuthal coordinate. This has to be due to mechanical differences appearing in the assembly procedure, because the dependence on azimuthal coordinate is most likely the result of a non-negligible tilt of the optical axis with respect to the CCD (sensor board) plane. This enforces a recalculation of model parameters for each set of equipment.

\subsection{Model for real sky data}

A set of 285 bright stars up to $9^\m$, without bright neighbours, scattered over the full frame was chosen. For every star scale (signal), the position and background level were fitted with the polynomial model described above, on each of the consecutive 172 frames of a chosen test field. Using obtained positions, a high resolution profile of each star was prepared, and then model parameters were refitted to each profile. The procedure was repeated iteratively, using the last iterations' parameters for calculating star positions in each reconstruction phase. In our case the final model was obtained after six iterations. A well known practise for such PSF iterative modelling (although for cases with much simpler PSF) is to perform three repetitions of the procedure.

As in the intermediate model, PSF has to be described for every position on the CCD, while fits were performed only for a finite set of stars with specific positions on the sensor. Thus each parameter value has been plotted in Cartesian coordinates and a Delaunay triangulation \citep{delaunay_tri} used for interpolation between measured points. Results for most of the parameters show general rotational symmetry, and it seemed that the polar coordinate system would be the most natural for the interpolation purposes.

The final model recalculated for the real sky data seems to reproduce collected images quite well, both for slightly asymmetric profiles close to the frame centre and for very deformed ones close to the frame edge, as shown in fig. \ref{sample_PSF}. However, as in the laboratory measurements, the far ``tails'' of deformed PSFs are not properly described. Moreover, one can see that the reconstructed profile is slightly asymmetric -- an effect that was not visible during laboratory measurements and that is probably induced by the slight tilt of the CCD surface relative to the lenses' axes.

\begin{figure}
\begin{center}
	\includegraphics[width=0.49\textwidth]{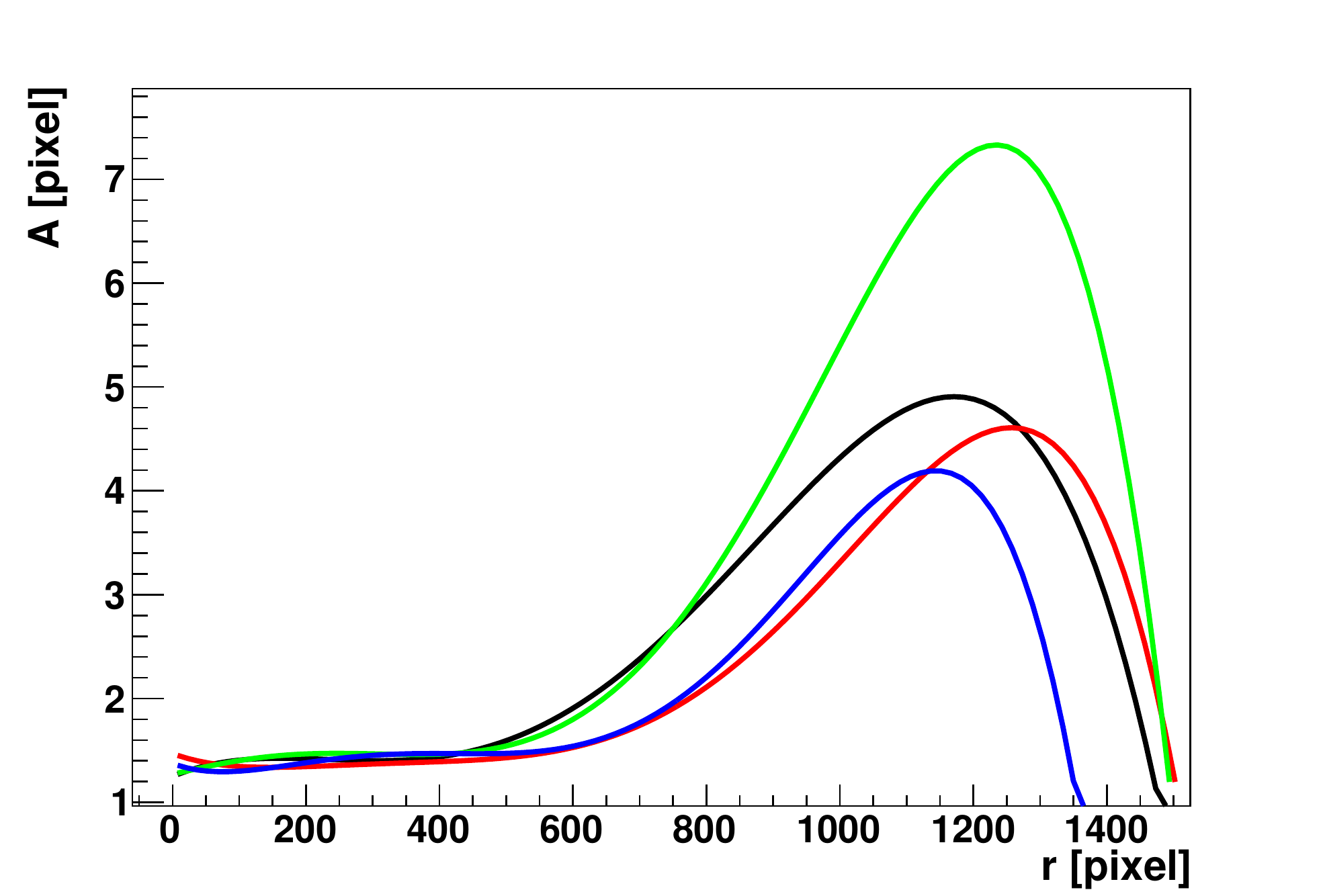}
\end{center}
\caption{Sensor area A covered by the signal larger than $50\%$ of the maximum signal for the final model of PSF as a function of distance from the frame centre r. Different curves' colours represent results obtained along diagonals of different quarters of the CCD.}
\label{model_fwhm}
\end{figure}

As mentioned before, the general shape of the PSF and its development on the CCD is similar for the artificial point source (laboratory measurements) and the real sky data. The similarity of the PSF shape evolution is visible in the plots showing area A above $50\%$ of the maximum signal as a function of distance from the frame centre r for the final model (fig. \ref{model_fwhm}) and the laboratory model (fig. \ref{fig_fwhm}). The plots cannot be directly compared, for in the former case a mathematical function is integrated, while in the latter real measurements are analysed. Still, the behaviour is the same: the signal is contained in a small number of pixels close to the frame centre and near the frame corner and spread over an area that is even few times larger for intermediate positions. The maximum position here, however, is oscillating around 1200 pixels from the frame centre, while it was about 1000 pixels for the laboratory measurements. The difference between maxima for different quarters of the CCD is most likely due to the sensor tilt described previously.

In general, there were enough star measurements (number of frames) considered for profile reconstruction and model fit. However, bright stars in the very corners of the frame seem to be somewhat scarce compared to the rapid changes in the PSF shape in this area. This obstacle cannot be easily solved, for there are simply very few stars in the corner of the frame compared to the other parts of the frame because of the sensor's geometry.

Additional systematic uncertainties in the model could come from the selected method of interpolation. However, results of photometry based on models using interpolation in Cartesian and polar coordinates for Delaunay triangles \citep{delaunay_tri} and Shephard's \citep{shepard_int} methods were very similar, so we conclude that the corresponding uncertainties are negligible.

\begin{figure*}
\begin{center}
	\includegraphics[width=0.49\textwidth]{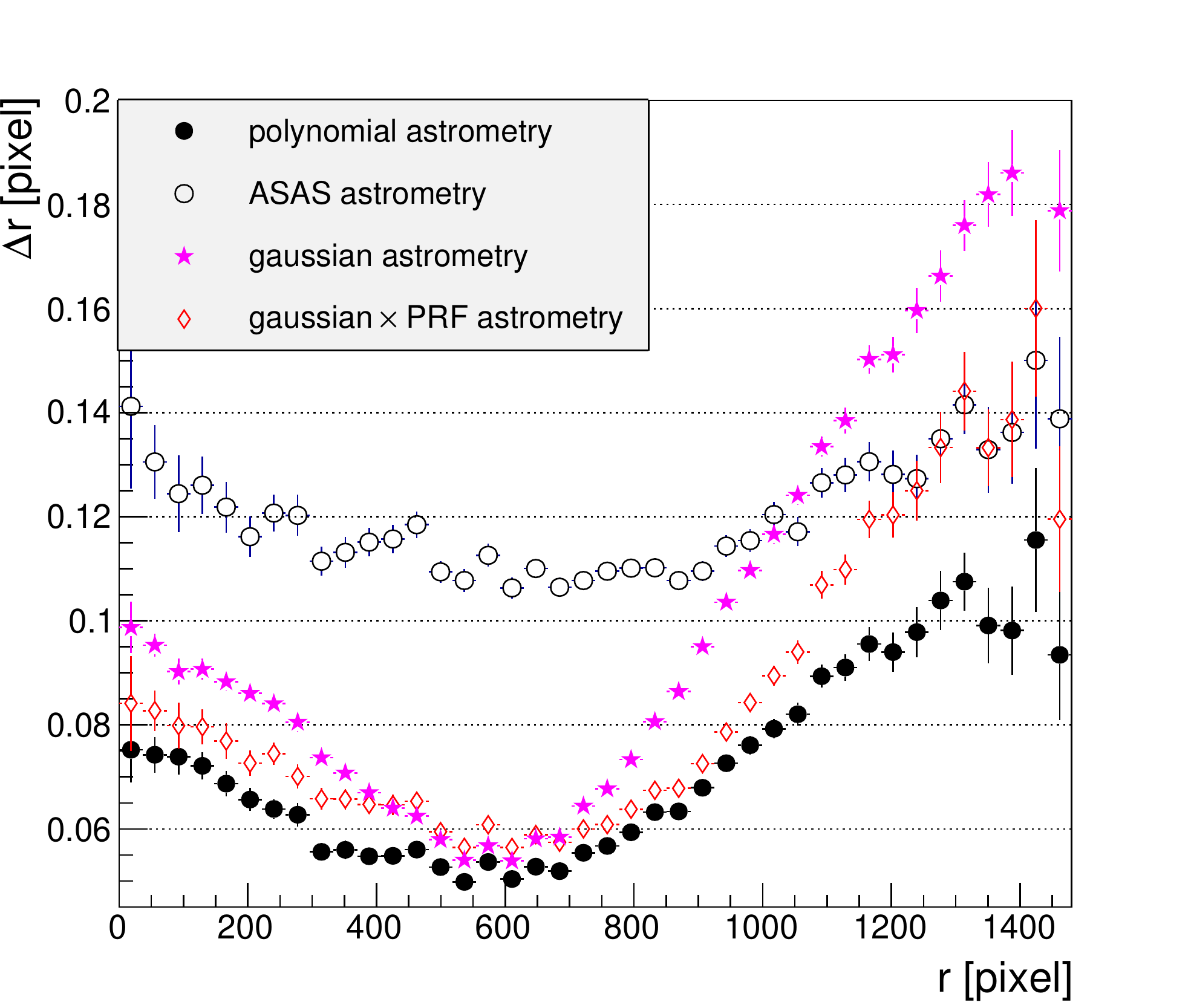}
	\includegraphics[width=0.47\textwidth]{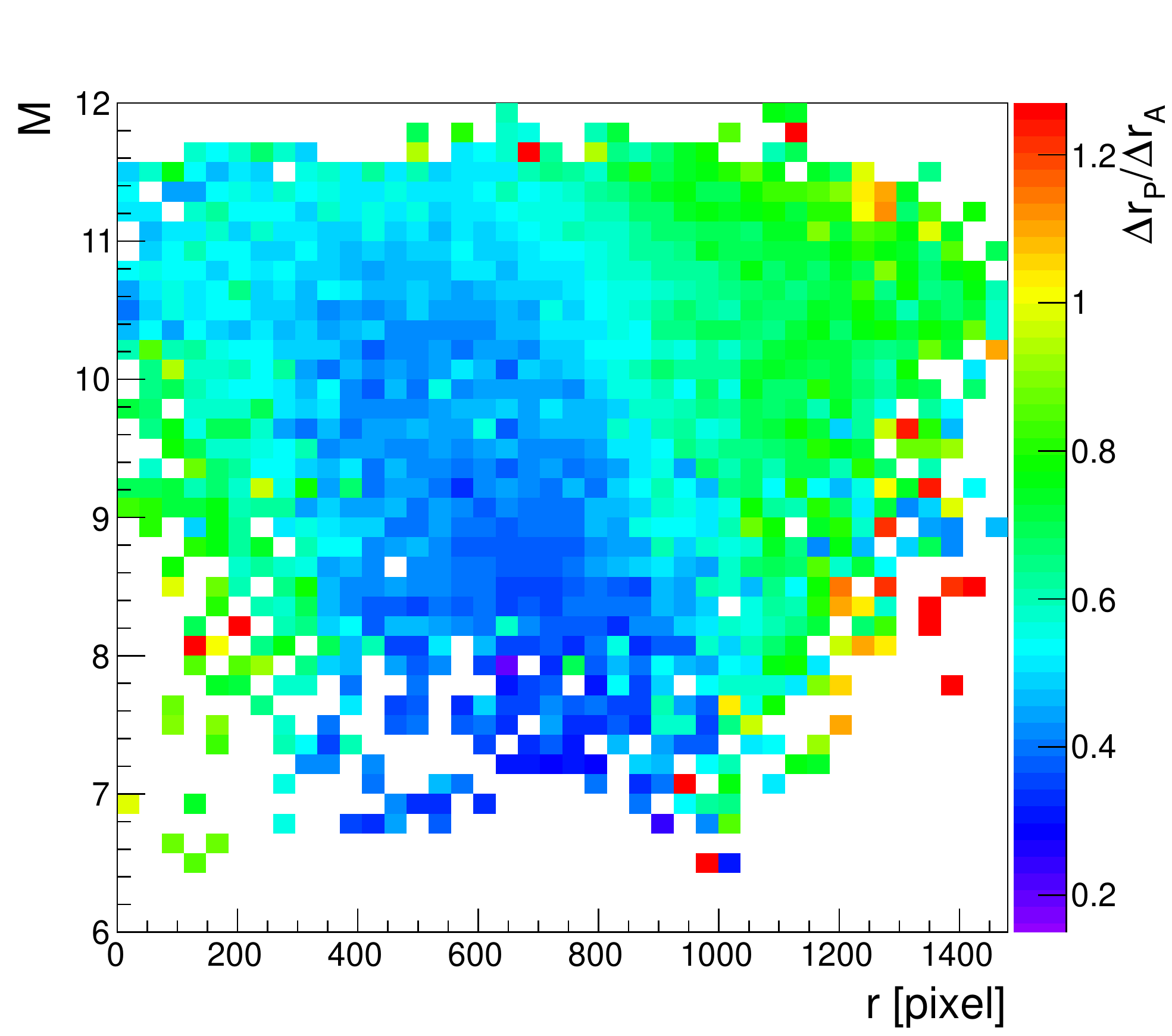}
\end{center}
\caption{Comparison between polynomial and ASAS astrometry results. Left: position spread $\mathrm{\Delta r}$  as a function of the distance from the frame centre r. Right: ratio $\frac{\mathrm{\Delta r_P}}{\mathrm{\Delta r_A}}$ of position spread of polynomial and ASAS astrometry as a function of star brightness and distance from the frame centre.}
\label{astro_multi}
\end{figure*}

\subsection{Astrometry and photometry}

The final model of the PSF (introducing 2D Delaunay triangles interpolation in polar coordinates) has been used as the basis for simple astrometry and photometry analysis. Each star on a frame has been fitted with a PSF specific to its coordinates, only the scale (signal level), background, and position on the pixel were treated as free parameters, not given by the model. Positions on the pixel in x and y coordinates were the only numerically fitted parameters, while the scale was being calculated analytically, according to the formula:

\begin{equation}
S = \frac{\displaystyle\sum_{x,y}f(x,y)\frac{s(x,y)+b}{\sigma(x,y)^2}-b*\displaystyle\sum_{x,y}\frac{f(x,y)}{\sigma(x,y)^2}}{\displaystyle\sum_{x,y}\frac{f(x,y)^2}{\sigma(x,y)^2}}
\label{scale_fit_formula}
\end{equation} 

\noindent where $x,y$ stand for pixels coordinates on the image, $f(x,y)$ denote the PSF model function value for centres of image pixels (for a given PSF position), $b$ is the background, and $\sigma(x,y)$ stands for the uncertainty of the signal in pixel $(x,y)$. We assume here that the signal measurement uncertainties $\sigma \simeq \sqrt{s(x,y)+b}$, where $s(x,y)$ is the signal in pixel $(x,y)$. This assumption, corresponding to the Gaussian regime of the Poisson distribution of photon statistics, resulted in the best fit compared to the other error models tested\footnote{We did not attempt to model the noise fully, which apart from the photon statistics noise of the source and background, should include factors such as gain, readout, and confusion noise. The optimal fit should include all those non-negligible factors, but then the analytic approach to the $\chi^2$ minimisation is no longer possible.}. The equation \ref{scale_fit_formula} is obtained from an analytic minimisation of a $\chi^2$ equation for the model function to real signal and background at a given PSF position. Although background can be fitted analytically as well, results turned up to be stablest if we set background to a trimmed median\footnote{The trimmed median in this case was median of values of all pixels, after removing pixel values outside mean $\pm 3\sigma$.} of all pixels within the area $40\times 40$ pixels around the star's centre.

We performed simple astrometry and photometry on 172 consecutive images of one sky field. To test the astrometry, we calculated (on consecutive frames) the spread of the distance between pairs of neighbouring stars $\rm{\Delta r}$. The distance between two stars should remain constant on all frames, thus any spread of measured distance between stars should be a direct result of imprecisions in position determination\footnote{It has to be noted that the real position of stars on the analysed frame series was not constant. While this reduces the quality of data, it serves the astrometry testing purposes well.}. The polynomial astrometry performs much better than the ASAS \citep{ASAS} algorithm, based on aperture calculations. The average spread improvement ranges from $\sim 1.4$ times ($\sim 23\%$) in the corner of the frame to $\sim 2.2$ times ($\sim 54\%$) for r=$400-700$ pixels from the frame centre, where the distance uncertainty is $\sim 0.05$~pixel for the polynomial and $\sim 0.11$~pixel for the ASAS astrometry (see fig. \ref{astro_multi}, left). It should be noted that the ratio of the estimated uncertainties is the lowest for the part of the frame containing the most stars\footnote{The number of stars is highest for this part of the frame simply for geometrical reasons.}. The spread increases for the distances very close to and far from the frame centre, where the PSF of a star becomes smaller.

It is clear that the polynomial astrometry outperforms profile astrometry made with a Gaussian profile (with the width optimal for the central, least deformed PSF), a behaviour expected for a heavily under-sampled data. Adjusting the function width to fit would probably improve the situation, but it can be concluded that the area occupied by the model is not the main factor responsible for the precision of the astrometry. If it was, the astrometry should improve close to the corner of the frame, where the area becomes smaller, as indicated in fig. \ref{model_fwhm}. Instead, the polynomial advantage over the Gaussian is highest for the centre of the frame (better up to $\sim 0.025$ of the pixel), for which the Gaussian width was optimised, and for the corners of the frame, where the size of the core of the PSF goes to similar values (being $\sim 0.08$ pixels better). A significant share in the astrometric precision has to be attributed to the PRF. The Gaussian convoluted with the PRF (with the width again optimised for central stars) gives much closer results to the polynomial fit, but shows the same behaviour as the pure Gaussian fit. It is again worst than the fit in the centre and the corner of the frame, but this time by $\sim 0.01$ and $\sim 0.04$ pixel, respectively. It has to be noted that all the presented astrometric algorithms have the highest precision close to 600 pixels from the frame centre. We assume that PSF in this region exhibits the best compromise between the size of the PSF and its deformation.

Comparison of the two astrometry methods as a function of the distance to frame centre r and magnitudo M is shown in fig. \ref{astro_multi} (right). For nearly the whole magnitudo-position space, polynomial astrometry performs better than ASAS. The lowest values of ratio between methods results -- $\frac{\mathrm{\Delta r_P}}{\mathrm{\Delta r_A}}$, reaching 0.3 for bright stars slightly increases with magnitudo, but the range in the distance from the frame centre where the profile astrometry performs much better becomes wider. The brightness dependency shows that ASAS astrometry gives better results for some points on the boundaries of the plot. This can be attributed to the fit instabilities and statistics for a small sample of stars in these regions.

\begin{figure*}
\begin{center}
	\includegraphics[width=0.47\textwidth]{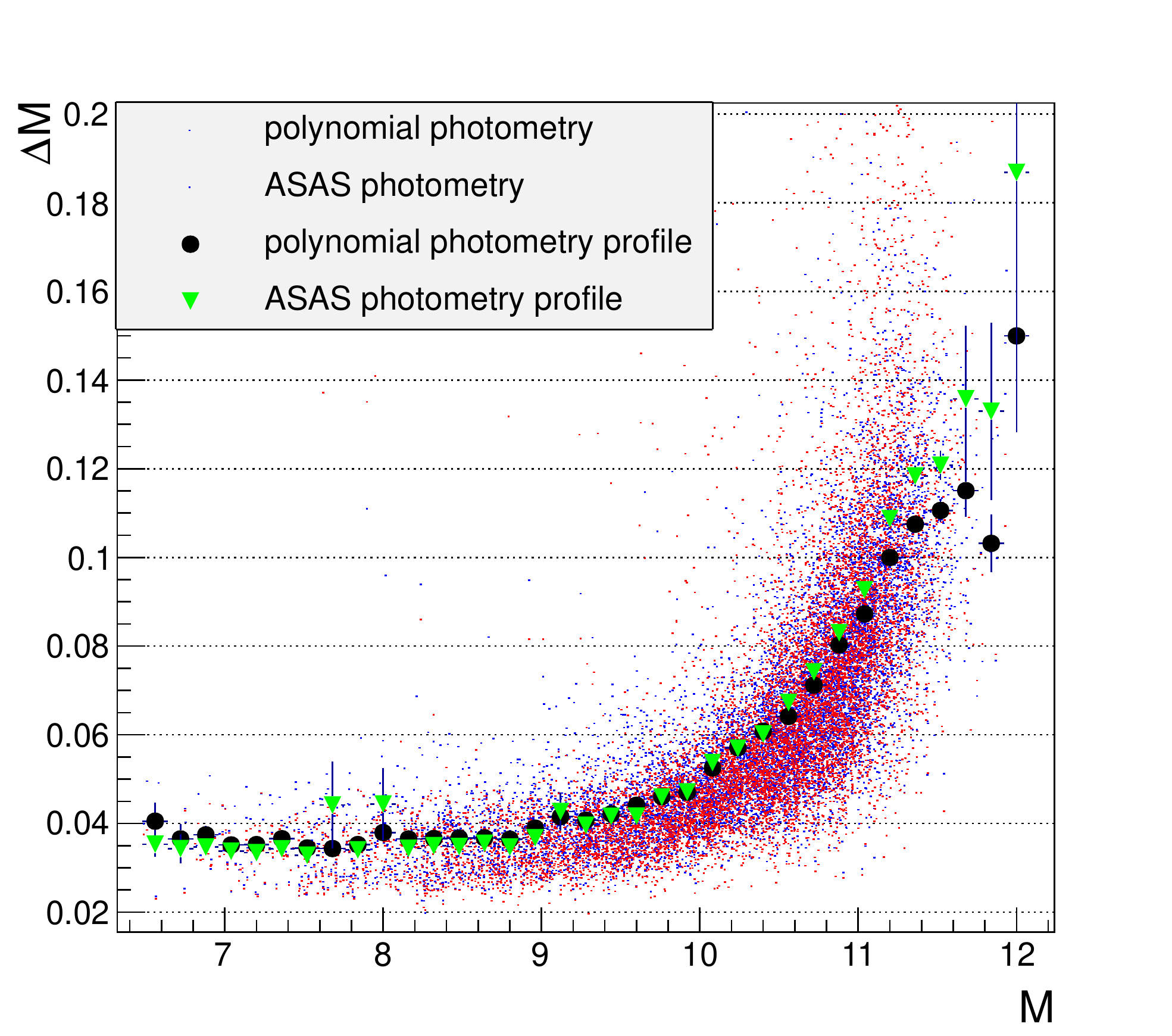}
	\includegraphics[width=0.47\textwidth]{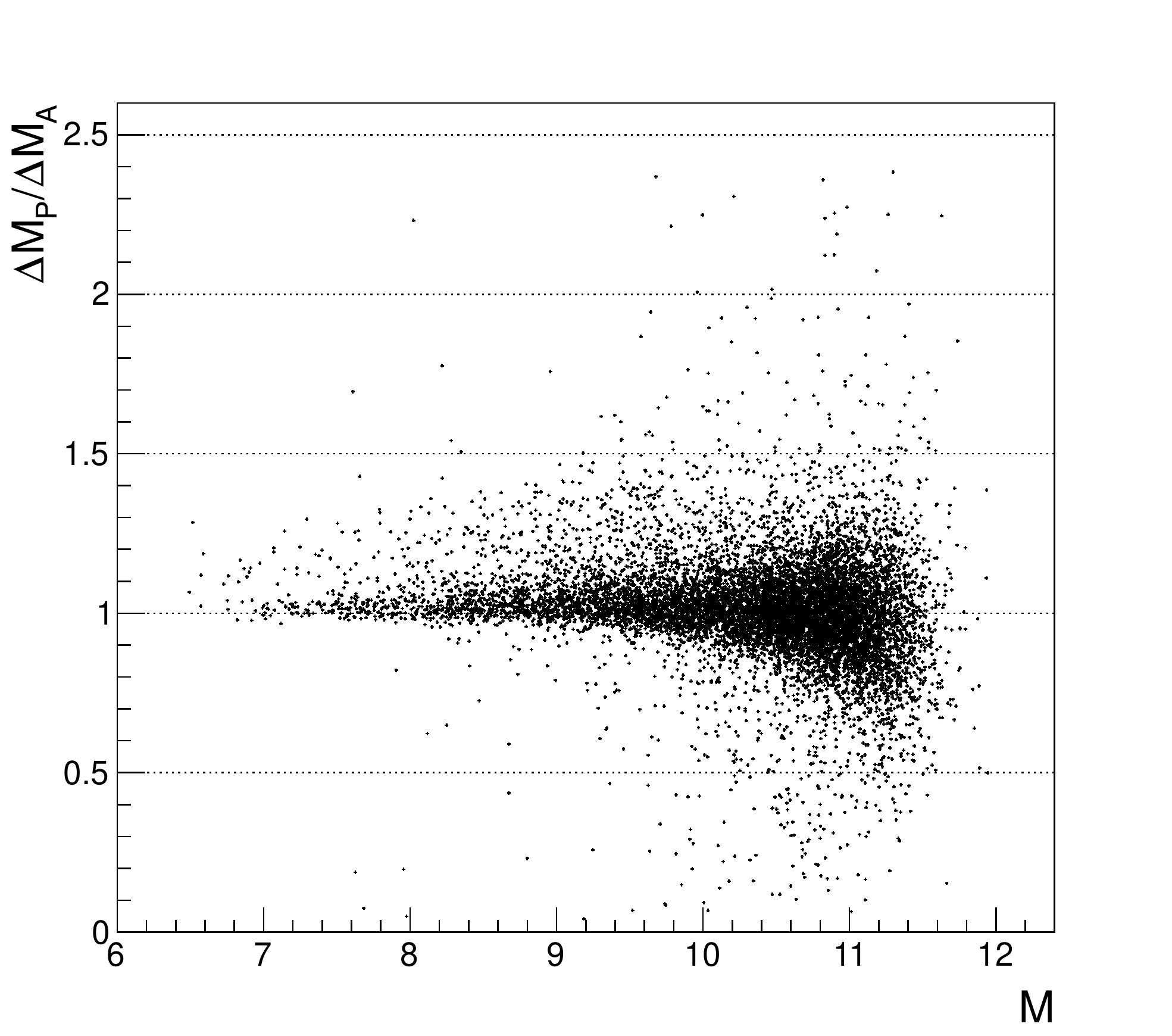}
\end{center}
\caption{Measured magnitudo spread $\mathrm{\Delta M}$ for polynomial (blue points) and ASAS photometry (red points) (left) and the ratio of the polynomial to ASAS photometry spread $\frac{\mathrm{\Delta M_P}}{\mathrm{\Delta M_A}}$ (right) as a function of the magnitudo M of the measured star.}
\label{fig_dmag_mag}
\end{figure*}

Even the lowest precision of $7''$, which is the result of recalculating the lowest precision shown in fig. \ref{astro_multi} to angular units with the ``Pi of the Sky'' pixel size of $36''$, is enough for following up on our observations with many large telescopes. However, the higher the precision, the lower the chance for accidental misidentification of the detected source with sources in star catalogues. The precision of current catalogues ranges between $0.1''$ and $0.2''$, being $0.15''$ for the APASS photometric survey \citep{APASS}. The APASS uses cameras with $2.6''$ angular pixel size. Simple rescaling of our maximum precision of $1.8''$ (0.05 pixel) to this pixel size would give $0.13''$ precision, and $\sim 0.2''$ precision on average. Our precision can also be compared to, for example, the All Sky Astrometric Surver (ASAS), which is around 0.2 pixel ($3''$) \citep{ASAS_cat}. The PSF deformation in both of these experiments is close to negligible compared to ours. Therefore we conclude that the resulting astrometry of the polynomial model is highly satisfactory.

Results of the fitted magnitudo spread $\rm{\Delta M}$\footnote{The magnitudo spread is obtained from the fitted signal, without applying any corrections based on reference stars, due to the correction calculation time and its low relevance for comparing photometry methods.} vs magnitudo M, as obtained from the polynomial profile photometry and compared to ASAS photometry are shown in fig. \ref{fig_dmag_mag}. The smallest obtained brightness spread is about $0.02^\m$ for nearly the whole available brightness range (fig. \ref{fig_dmag_mag}, left). Up to $9^\m$ the spread rarely exceeds $0.05^\m$, but for dimmer stars it steeply ascends to reach close to $0.2^\m$ for $\sim11^\m$. The best range for polynomial photometry over the whole frame is around $7^\m-9^\m$. For brighter stars fits are less stable probably due to saturation, the CCD nonlinearity, and more evident ``tails'' of PSF not properly described by the model. For dimmer stars the reason is simply the descending signal-to-noise ratio and neighbouring brighter stars disturbing the measurement, with the number of these increasing with magnitudo.

The polynomial photometry is comparable to ASAS aperture photometry, similar to or slightly worse for most stars up to $\sim8.7^\m$, and slightly improving with magnitudo (fig.~\ref{fig_dmag_mag}, right). However, for most stars, the ratio of brightness-measurement uncertainties $\frac{\mathrm{\Delta M_P}}{\mathrm{\Delta M_A}}$ remain close to 1 over the full magnitudo range. This is expected behaviour when apertures with optimal diameters are used \citep{Irwin}. It is very likely that the tail towards values higher than 1 is due to instabilities in fitting the model to some stars. We have to keep in mind that the very basic profile photometry performed automatically on a large number of stars is compared here with sophisticated and mature aperture photometry. Most of the stars have the same magnitudo spread in both types of photometry, which suggest that the spread is due to real fluctuations in measurements. Therefore, to compare differences between the algorithms, data of better photometric quality should be tested.

The level of photometric uncertainties below $0.01^{\rm m}$ would allow for discoveries of extrasolar planets, however it requires hardware dedicated to this task. Very wide-field experiments such as ``Pi of the Sky'' can aim at observing variable stars, which in general have amplitudes of variability between few magnitudos and hundredths of a magnitudo. The ASAS experiment claims that it has precision below $0.05^{\rm m}$ in general, and it allowed for discoveries and analysis of a substantial amount of variable stars \citep{ASAS_cat}. Reaching below $0.01^{\rm m}$ could allow us to analyse the variability of more stars, such as short-period Beta Cephei ($0.015^{\rm m}$ to $0.025^{\rm m}$ amplitude), Delta Scuti ($0.003^{\rm m}$ to $0.9^{\rm m}$), and ZZ Ceti stars ($0.001^{\rm m}$ to $0.2^{\rm m}$) \citep{amplitudes}. On the other hand, a photometric survey, such as APASS, gives roughly $0.01^{\rm m}-0.05^{\rm m}$ average precision for the brightest unsaturated stars (preliminary data), and the upper boundary on precision extends above $0.1^{\rm m}$ about $3.5^{\rm m}$ above saturation level \citep{APASS}, as in our experiment. We conclude that this photometric precision is satisfactory, especially for a non-photometric data and for only the first step in reduction (direct, unfiltered flux from the measurements). Nevertheless, we should aim at getting as close to the $0.01^{\rm m}$ precision as possible. The full reduction performed on 20 stacked exposures allows us to reach precision of $0.018^{\rm m}$-$0.024^{\rm m}$ \citep{photo_pi}. However, if this precision is achievable with 10 s exposures, which is much shorter than for nearly all astronomical experiments, remains to be seen as well as what requirements are there for still fairly unexplored regime of variability which is of order of seconds.

\subsection{Other applications}

Applications of the obtained PSF model are not limited to photometry and astrometry. One of the other possible utilisations is a dedicated search for signals at specific coordinates on the frame. An example may be the search for an optical precursor \citep{optic_precursor} to ``the naked eye'' burst GRB080319B, which, due to its unprecedented brightness suits the task perfectly. With the polynomial model we managed to set limits on precursor emission to $12^\m$, much better than the previous limits of $11.5^\m$ obtained with aperture photometry \citep{lwp_precursor}. Additionally, the developed PSF model allowed us to determine the position on the CCD where the object astrometry and photometry results are most precise. This can be used to define the optimal pointing policy for follow-up observations performed by ``Pi of the Sky'' telescopes.

The precise polynomial PSF model also allows for generating artificial sky frames that are very similar to the real ones and to obtain star images in ``controlled'' conditions, managing their coordinates, brightness, variability, background fluctuations, etc. Such simulation allows for testing future hardware and future algorithms and for determining experiment parameters, such as star separation distance, and parallax uncertainties. \citep{simul}

\section{Conclusions}

In this work, we have discussed the precise reconstruction of a very wide-field camera's PSF. The effective modelling performed with modified Zernike polynomials reproduces even the most deformed star images in high detail, where residues are below $13\%$ of the maximum signal for high resolution profiles, significantly better than any other method that has been tested.

The polynomial model should be easily adaptable to other very wide-field experiments with a simple procedure of refitting polynomial coefficients, as shown in the case of real sky data from the \pin project. We have shown that the model can significantly increase the astrometry precision. Additionally, the very simple profile photometry utilising the model gives similar results to a sophisticated aperture photometry, proving the model's quality. This opens numerous possibilities, such as performing photometry on dense fields of highly deformed stars, which would result in significant uncertainties with traditional algorithms, or very sensitive searching for weak signals or simulating real sky frames.

Other approaches to the modelling of very wide-field experiments' PSF can still be tested, such as PCA, which is very successful in the weak lensing experiments. However, the PCA, along with other popular methods such as shapelets, was tested mainly on small (compared to ours) deformations of PSF, where linear combination of model components is justified. We suspect that in the case of high deformation, the quality of our model lies in the non-linear combination (this could be rendered plausible by the simplified Rayleigh-Sommerfeld equation of PSF generation (eq. \ref{eq_aber_diff})). Therefore it remains to be seen if PCA performs better on a very wide-field experiment's data\footnote{It should be noted that deriving a set of principal components from a large set of heavily under-sampled data, such as in the ``Pi of the Sky'' experiment, may require non-standard steps, and the procedure may be quite complicated. Clearly, this requires a dedicated analysis, independent of the presented one.}.

On the other hand, we are aware that the chosen basis, as well as most of the other bases that are popular in the PSF modelling, describe deviations from a rotationally symmetric shape (or elliptic shape), while the shown PSFs are far from circular and tend to extend the signal towards the centre of the frame and reduce it in the opposite direction. Therefore a model that accounts for this complicated, multidirectional deformation, could prove more successful with fewer components. Inventing such a model with a dedicated basis is a mathematical challenge.

The presented model is probably the first development of the parametrisation of the highly deformed PSF in very wide-field experiments, and the future may (hopefully) show increasing interest in this area. Nevertheless, we conclude that our model is a significant advance for experiments such ``Pi of the Sky'' and may in future prove to be a crucial tool for data analysis, as well as algorithms and hardware development.

\begin{acknowledgements}
This work was supported by the Polish Ministry of Science and Higher Education in 2009-2012 as a research project.
\end{acknowledgements}

\bibliographystyle{aa}
\bibliography{paper}

\clearpage
\onecolumn

\end{document}